\documentclass[11pt,a4paper]{article}
\pdfoutput = 1
\usepackage{jheppub}
\usepackage{color}
\usepackage{amsmath,bm}
\usepackage{graphicx}
\usepackage{pgfplots,pgfplotstable}
\usetikzlibrary{intersections}
\pgfplotsset{compat=newest}

\pgfplotsset{every axis/.style={
    width=12cm,
    height=10cm,
    grid=both,
    scaled ticks=false,
    yticklabel style={/pgf/number format/.cd, fixed,precision=5}
  }
}

\newcommand{\as}{\alpha_{\mathrm{s}}}
\newcommand{\LA}{\mathrm{A}}

\newcommand{\LB}{\mathrm{B}}

\newcommand{\LF}{\mathrm{F}}
\newcommand{\LI}{\mathrm{I}}

\newcommand{\LL}{\mathrm{L}}

\newcommand{\LP}{\mathrm{P}}
\newcommand{\LR}{\mathrm{R}}

\newcommand{\LT}{\mathrm{T}}

\newcommand{\LZ}{\mathrm{Z}}
\newcommand{\La}{\mathrm{a}}
\newcommand{\Lb}{\mathrm{b}}
\newcommand{\Lc}{\mathrm{c}}
\newcommand{\Ld}{\mathrm{d}}
\newcommand{\Le}{\mathrm{e}}
\newcommand{\Lf}{\mathrm{f}}
\newcommand{\Lg}{\mathrm{g}}

\newcommand{\Lp}{\mathrm{p}}
\newcommand{\Lr}{\mathrm{r}}

\newcommand{\Lu}{\mathrm{u}}

\newcommand{\GeV}{\ \mathrm{GeV}}
\newcommand{\TeV}{\ \mathrm{TeV}}

\newcommand{\Vvirt}{{\cal S}}
\newcommand{\Vreal}{{\cal V}}

\definecolor{red}{rgb}{1,0,0}

\def\mi{{\mathrm i}}
\newcommand{\MSbar}{\overline {\text{MS}}}

\def\ket#1{\big|{#1}\big\rangle}
\def\bra#1{\big\langle{#1}\big|}
\def\<>#1{\big\langle{#1}\big\rangle}
\def\[]#1{\big[{#1}\big]}

\def\sket#1{\big|{#1}\big)}
\def\sbra#1{\big({#1}\big|}
\def\sbrax#1{\big({#1}}        % No "|"

\newbox\charbox
\newbox\slabox
\def\s#1{{      % Feynman slash
        \setbox\charbox=\hbox{$#1$}
        \setbox\slabox=\hbox{$/$}
        \dimen\charbox=\ht\slabox
        \advance\dimen\charbox by -\dp\slabox
        \advance\dimen\charbox by -\ht\charbox
        \advance\dimen\charbox by \dp\charbox
        \divide\dimen\charbox by 2
        \raise-\dimen\charbox\hbox to \wd\charbox{\hss/\hss}
        \llap{$#1$}
}}

\title{Summing threshold logs in a parton shower}

\author[a]{Zolt\'an Nagy}

\author[b]{and Davison E.\ Soper}

\affiliation[a]{
DESY\\
Notkestrasse 85\\
22607 Hamburg, Germany
}

\affiliation[b]{
Institute of Theoretical Science\\
University of Oregon\\
Eugene, OR  97403-5203, USA
}

\emailAdd{Zoltan.Nagy@desy.de}
\emailAdd{soper@uoregon.edu}

\abstract{
When parton distributions are falling steeply as the momentum fractions of the partons increases, there are effects that occur at each order in $\as$ that combine to affect hard scattering cross sections and need to be summed. We show how to accomplish this in a leading approximation in the context of a parton shower Monte Carlo event generator.
}


\begin{document}
\maketitle

%-------------------------------------------------------------------
\section{Introduction}%%%%%%%%%%%%%%%%
\label{sec:intro}

One often wants to calculate the cross section for a hard process in hadron-hadron collisions under the circumstance that one or both of the required momentum fractions $\eta_\La$ and $\eta_\Lb$ for the initial state partons is large. Then the corresponding parton distribution functions $f_{a/A}(\eta_\La,\mu^2)$ or $f_{b/B}(\eta_\Lb,\mu^2)$  will be steeply falling as a function $\eta_\La$ and $\eta_\Lb$. In this case, the cross section calculated beyond the leading order is enhanced compared to the leading order cross section. The effect is said to be due to ``threshold logarithms.''\footnote{For the emission of soft gluons from, say, parton ``a'', we have an integration over the fraction $(1-z)$ of $\eta_\La$ that is taken by the gluons. There is a parton factor $f_{a/A}(\eta_\La/z,\mu^2)$ and a singular function of $(1-z)$. The integration is limited by the falloff of $f_{a/A}(\eta_\La/z,\mu^2)$ as $(1-z)$ increases. This integration produces logarithms of the effective upper bound on $(1-z)$. For details, see section \ref{sec:comparison}.} The amount of enhancement is proportional to how fast $f_{a/A}(\eta_\La,\mu^2)$ or $f_{b/B}(\eta_\Lb,\mu^2)$ is falling. The threshold enhancement grows as $\eta_\La$ and $\eta_\Lb$ increase, but $\eta_\La$ and $\eta_\Lb$ do not have to be close to 1 for the threshold logarithms to start to be significant. One can count $\eta_\La > 0.1$ and $\eta_\Lb > 0.1$ as being part of the threshold region. 

In the threshold region, it is useful to sum the enhanced contributions. Already in 1987, Sterman \cite{Sterman1987} pointed out the issue and showed how to understand it and sum the threshold logarithms. Since then, there has been a substantial development in the field, including both treatments following the traditional direct approach  \cite{AppellStermanMackenzie, Catani32, CataniWebberMarchesini, Magnea1990, Korchemsky1993, CataniManganoNason32, SudakovFactorization, KidonakisSterman, KidonakisOderdaSterman, KidonakisOderdaStermanColor, LaenenOderdaSterman, CataniManganoNason, LiJointR, KidonakisOwensPhotons, StermanVogelsang2001, KidonakisOwensJets, LSVJointR, KSVJointR, Kidonakis2005, MukherjeeVogelsang, Bolzoni2006, deFlorianVogelsang, RavindranSmithvanNeerven, RavindranSmith, BonviniForteRidolfi, BFRSaddlePt, deFlorianJets, Catani:2014uta, Ahmed:2014uya, BonviniMarzani, Bonvini:2015ira} and treatments using soft-collinear effective theory \cite{ManoharSCET, IdilbiJiSCET, BecherNeubert, BecherNeubertPecjak, BecherNeubertXu, BecherSchwartz, Stewart:2009yx, Beneke:2010da, Bauer:2011uc, Wang:2014mqt, Li:2016axz, Lustermans:2016nvk}.  Comparisons of these approaches can be found in refs.~\cite{Bonvini:2012az, StermanZeng, Bonvinietalcompare}.

Even before the study of threshold logarithms began, Sj\"ostrand \cite{EarlyPythia} and Marchesini and Webber \cite{EarlyHerwig} had introduced parton shower event generators that were important in the mid 1980s and whose direct descendants are still essential tools today \cite{dipolesG, lambdaR, dipolesGP, Herwig1992, Pythia1994, SherpaAlpha, SjostrandSkands, NSRingberg, DinsdaleTernickWeinzierl, SchumannKrauss, Herwig, Sherpa, PlatzerGiesekeI, PlatzerGiesekeII, HochePrestel, Pythia}. These programs are based on iterating parton splittings derived from perturbative quantum chromodynamics (QCD). Because of this iterative structure based on the soft and collinear singularities of the theory, they sum many sorts of logarithms in a natural way. For instance, parton shower event generators can sum logarithms of $k_\LT^2/M_\LZ^2$, where $k_\LT$ is the transverse momentum of a Z-boson produced in the Drell-Yan process \cite{NSpT}. Also, the analysis of ref.~\cite{CataniWebberMarchesini} connects the summation of threshold logarithms to the structure of a parton shower.

Current standard parton shower event generators do not sum threshold logarithms. However, it is possible to rearrange them so that they do sum the leading threshold logarithms. In this paper, we show how to do that.

Our analysis is based on our own parton shower event generator, \textsc{Deductor} \cite{NSI, NSII, NSspin, NScolor, Deductor, ShowerTime, PartonDistFctns, ColorEffects}. We make this definite choice because we need to be specific with respect to shower kinematics, the choice of shower ordering variable, splitting functions, and the treatment of color. For this paper, we use a new version, 2.0.2, of \textsc{Deductor} \cite{DeductorCode}.

Even though our analysis uses definitions specific to \textsc{Deductor}, we think that it will be apparent that the results could be adapted to other hardness based parton shower generators, such as \textsc{Pythia} \cite{Pythia} and \textsc{Sherpa} \cite{Sherpa}. With somewhat more restructuring, it should be possible to adapt the results to the angle ordered parton shower generator \textsc{Herwig} \cite{Herwig}.

We present some of the features of the theoretical framework in section \ref{sec:ordering}. Then, before we begin the analysis, we exhibit in section \ref{sec:mainfeatures} some of the main features of the result and of where this result comes from. Inevitably, we leave a lot out. The details are presented later in the paper. Section \ref{sec:pdfs} discusses the role of parton distribution functions. Section \ref{sec:Vreal} analyzes the Sudakov exponent conventionally used in parton shower algorithms. Section \ref{sec:Vvirt} presents a revised Sudakov exponent to be used if one wants to include threshold logarithms. Sections \ref{sec:enhancementintegrand} and \ref{sec:enhancementLPplus} analyze the exponential that contains the threshold logarithms. Section \ref{sec:comparison} presents a comparison of the leading behavior of the threshold factor in this paper to the factor in previous treatments of threshold logarithms. Section \ref{sec:numerical} presents numerical results obtained with the \textsc{Deductor} parton shower. There are three concluding sections. In section \ref{sec:summary}, we summarize the main steps of the analysis that has been presented, leaving out the supporting arguments. The reader may wish to read section \ref{sec:summary} first, then return to it after absorbing the supporting arguments. In section \ref{sec:choices}, we describe some of the choices available in running \textsc{Deductor}. In section \ref{sec:outlook}, we say something about the advantages and disadvantages of putting threshold enhancements into a parton shower. We have tried to keep  the main presentation relatively short by relegating calculational details to three appendices.

\section{The \textsc{Deductor} framework}%%%%%%%%%%%%%%%%
\label{sec:ordering}

Our analysis is based on the parton shower event generator, \textsc{Deductor} \cite{NSI, NSII, NSspin, NScolor, Deductor, ShowerTime, PartonDistFctns, ColorEffects}, which uses specific choices with respect to shower kinematics, the shower ordering variable, the parton splitting functions, and the treatment of color. In this section, we outline some of these choices that play a role in the analysis of this paper.

An exact treatment of leading threshold logarithms requires an exact treatment of color, which is available in the general formalism 
of Refs.~\cite{NSI, NSII, NSspin, NScolor, Deductor, ShowerTime, PartonDistFctns, ColorEffects}. The exact color treatment is not implemented in the code of \textsc{Deductor}. Rather, we are able to use only an approximation, the leading-color-plus (LC+) approximation. However, in this paper we develop the theory using exact color. Then the LC+ approximation consists of simply dropping some terms. 

In \textsc{Deductor}, we order splittings according to decreasing values of a hardness parameter $\Lambda^2$. The hardness parameter is based on virtuality. For massless partons, the definition is
\begin{equation}
\begin{split}
\label{eq:Lambdadef}
\Lambda^2 ={}& \frac{(\hat p_l + \hat p_{m+1})^2}{2 p_l\cdot Q_0}\ Q_0^2
\hskip 1 cm {\rm final\ state},
\\
\Lambda^2 ={}&  -\frac{(\hat p_\La - \hat p_{m+1})^2}{2 p_\La \cdot Q_0}\ Q_0^2
\hskip 1 cm {\rm initial\ state}.
\end{split}
\end{equation}
Here the mother parton in a final state splitting has momentum $p_l$ and the daughters have momenta $\hat p_l$ and $\hat p_{m+1}$. For an initial state splitting in hadron A, the mother parton has momentum $p_\La$, the new initial state parton has momentum $\hat p_\La$ and the final state parton created in the splitting has momentum $\hat p_{m+1}$.\footnote{Here $p_\La$ is the momentum of the mother parton in the sense of shower evolution that moves from hard interactions to softer interactions. With initial state splittings, the shower development moves backwards in physical time.} For hadron B, we replace ``a'' $\to$ ``b.'' We denote by $Q_0$ a fixed vector equal to the total momentum of all of the final state partons just after the hard scattering that initiates the shower. We often use a dimensionless ``shower time'' variable given by
\begin{equation}
t = \log(Q_0^2/\Lambda^2)
\;.
\end{equation}
Then $t$ increases as the shower develops.

One could use some other hardness parameter to order the shower. For instance, various measures of the transverse momentum in a splitting are popular choices. In this paper, we use $\Lambda^2$ because that is what we have implemented in \textsc{Deductor}.

In an initial state splitting, parton distribution functions enter the splitting functions. Then we need a scale parameter for the parton distribution functions. We use the virtuality of the splitting:
\begin{equation}
\begin{split}
\label{eq:musqamusqb}
\mu_\La^2(t) ={}& 2 p_\La\cdot Q_0 \,e^{-t}
\;,
\\
\mu_\Lb^2(t) ={}& 2 p_\Lb\cdot Q_0 \,e^{-t}
\;.
\end{split}
\end{equation}
Similarly, for a final state splitting of parton $l$, the virtuality parameter is 
\begin{equation}
\label{eq:musql}
\mu_l^2(t) = 2 p_l\cdot Q_0 \,e^{-t}
\;.
\end{equation}
It proves convenient to use a dimensionless virtuality variable $y$. For an initial state splitting,
\begin{equation}
\label{eq:ydef0}
y = \frac{\mu_\La^2(t)}{2 p_\La\cdot Q} = \frac{\mu_\La^2(t)}{Q^2}
\;,
\end{equation}
where $Q = p_\La + p_\Lb$ is the total momentum of the final state particles in the shower state before the splitting. We collect these and other notations used throughout the paper in appendix \ref{sec:notation}.

\textsc{Deductor} 1.0 uses non-zero charm and bottom quark masses, both in the final and the initial states. However, we have simplified the code in \textsc{Deductor} 2.0 \cite{DeductorCode} that we use here by setting the charm and bottom quark masses to zero in the parton kinematics (although charm and bottom quark thresholds in the evolution of parton distribution functions remain).  This is evidently an approximation, but this approximation does not matter for the physics investigated in this paper, the threshold logarithms at very large parton collision energies. 

The general formalism includes parton spins, but the current implementation in \textsc{Deductor} simply eliminates spin by averaging over the mother parton spin before each splitting and summing over the daughter parton spins. In this paper we keep the analysis as simple as we can by doing the same spin averaging, so that spin dynamics does not play a role.

\section{Some main features of the threshold enhancement}
\label{sec:mainfeatures}

As in our previous work \cite{NSI, NSII, NSspin, NScolor, Deductor, ShowerTime, PartonDistFctns, ColorEffects}, we find it useful to think of parton shower evolution as describing the evolution of a statistical state vector $\sket{\rho(t)}$. The shower time $t$ represents the hardness scale of the current state. As $t$ increases, parton emissions get softer and softer. At shower time $t$, $\sket{\rho(t)}$ represents the distribution of partonic states in an ensemble of runs of the Monte Carlo style program. Without color or spin, $\sket{\rho(t)}$ would simply represent a probability distribution in the number $m+2$ of partons ($m$ final state partons and two initial state partons), their flavors $f$, and their momenta $p$. With the inclusion of color, $\sket{\rho(t)}$ is a little more complicated: it represents the probability distribution of parton flavors and momenta and the density matrix in the quantum color space. As stated in the introduction, we ignore spin in this paper. If spin were included, $\sket{\rho(t)}$ would represent the density matrix in the combined color and spin space.

In the notation we use here, a measurement of some property $F$ of the partonic system, like the number of jets with transverse momenta bigger than some given value, is represented by a vector $\sbra{F}$. The average cross section corresponding to measurement $F$ at time $t$ is $\sbrax{F}\sket{\rho(t)}$. Of particular importance is the completely inclusive measurement, which we denote by $\sbra{1}$. Thus the total cross section represented by state $\sket{\rho(t)}$ is $\sbrax{1}\sket{\rho(t)}$. To compute $\sbrax{1}\sket{\rho(t)}$, we sum over the number of partons and their flavors, integrate over the parton momenta, and take the trace of the density matrix in color space.

We represent the evolution of the shower by the equation
\begin{equation}
\label{eq:evolution}
\frac{d}{dt}\,\sket{\rho(t)} = [{\cal H}_I(t) - \Vvirt(t)]\sket{\rho(t)}
\;.
\end{equation}
Here ${\cal H}_I(t)$ represents real parton emissions and increases the number of final state partons by one. It contains splitting functions derived from QCD and a ratio of parton distribution functions. The operator $\Vvirt(t)$ leaves the number of partons, their momenta, and their flavors unchanged. It is, however, an operator on the color space. The operator $\Vvirt(t)$ contains terms representing two effects. It contains a term derived from one loop virtual graphs. It also contains a term proportional to the derivative with respect to scale of the parton distribution functions. This term is needed because parton distribution functions appear at the hard interaction and then (with a different scale) at each parton splitting. Thus one needs a term in $\Vvirt(t)$ that cancels the parton distributions at the old, harder, scale and inserts parton distributions at the new, softer, scale.

It is useful to solve the shower evolution equation (\ref{eq:evolution}) in the form
\begin{equation}
\label{eq:rhoevolution}
\sket{\rho(t)} = {\cal U}_\Vvirt(t,t_0)\sket{\rho(t_0)}
\;,
\end{equation}
where
\begin{equation}
\label{eq:evolutionsolution}
{\cal U}_\Vvirt(t,t_0) = {\cal N}_\Vvirt(t,t_0)
+ \int_{t_0}^t\!d\tau\ 
{\cal U}_\Vvirt(t,\tau)
{\cal H}_I(\tau)
{\cal N}_\Vvirt(\tau,t_0) 
\;.
\end{equation}
Here ${\cal N}_\Vvirt(\tau,t_0)$ is the no-splitting operator, given by a time-ordered exponential of the integral of $\Vvirt(\tau')$,
\begin{equation}
{\cal N}_\Vvirt(\tau,t_0) = \mathbb T \exp\left[
-\int_{t_0}^{\tau} d\tau'\ \Vvirt(\tau')\right]
\;.
\end{equation}
Thus, the number of partons, their flavors, and their momenta remain unchanged from time $t_0$ to some intermediate time $\tau$. The probability for the state can be multiplied by a numerical factor or, in general, by a matrix in color space. The no-splitting operator is usually called the Sudakov factor. In eq.~(\ref{eq:evolutionsolution}), there is a splitting at the intermediate time $\tau$, as specified by the splitting operator ${\cal H}_I(\tau)$. After that, there is more evolution according to the full evolution operator ${\cal U}_\Vvirt(t,\tau)$.

The evolution equation eq.~(\ref{eq:evolution}) is what one gets in a rather direct way from first order QCD perturbation theory. It is, however, not exactly what is computed in \textsc{Deductor~1.0} or in other standard parton shower event generators. The reason is that this evolution equation does not exactly conserve the total cross section:
\begin{equation}
\frac{d}{dt}\,\sbrax{1}\sket{\rho(t)} \ne 0
\;.
\end{equation}
We can conserve the total cross section if we change the evolution equation to
\begin{equation}
\label{eq:evolution2}
\frac{d}{dt}\,\sket{\hat\rho(t)} = [{\cal H}_I(t) - \Vreal(t)]\sket{\hat\rho(t)}
\;,
\end{equation}
where 
\begin{equation}
\label{eq:Vrealdef}
\sbra{1}[{\cal H}_I(t) - \Vreal(t)] = 0
\;,
\end{equation}
so that
\begin{equation}
\frac{d}{dt}\,\sbrax{1}\sket{\hat\rho(t)} = 0
\;.
\end{equation}
That is, we {\it define} $\Vreal(t)$ from ${\cal H}_I(t)$ so that
\begin{equation}
\label{eq:Vrealdef2}
\sbra{1}\Vreal(t) = \sbra{1}{\cal H}_I(t)
\;.
\end{equation}
With this definition, the color trace of the revised Sudakov factor,
\begin{equation}
{\cal N}_{\Vreal}(t_2,t_1) = \mathbb T \exp\left[
-\int_{t_1}^{t_2} d\tau\ \Vreal(\tau)\right]
\;,
\end{equation}
represents the probability for the parton system not to split between shower time $t_1$ and shower time $t_2$. It is standard to use Sudakov factors that are exponentials of $\Vreal(t)$ defined by eq.~(\ref{eq:Vrealdef}) to define a parton shower. That is the choice in \textsc{Deductor 1.0}.

When one constructs a parton shower using $\Vreal(t)$, the cross section associated with $\sket{\hat\rho(t)}$ is what it was at the hard interaction, say $t = 0$. That is to say, the total cross section is the Born cross section. What the parton shower does is to distribute the fixed cross section into cross sections for the different final states that the starting partons could evolve into.

The idea of the summation of threshold logarithms is that the total cross section is {\em not} just the Born cross section. Rather it contains corrections from higher orders of perturbation theory. To see this effect, we should use $\Vvirt(t)$ instead of $\Vreal(t)$. When we do that, we have a Sudakov factor that we can write as
\begin{equation}
\label{eq:SudakovFactor}
{\cal N}_\Vvirt(t_2,t_1) = \mathbb T \exp\left[
\int_{t_1}^{t_2} d\tau\ (-\Vreal(\tau) 
+ [\Vreal(\tau) - \Vvirt(\tau)])\right]
\;.
\end{equation}
Compared to the standard Sudakov factor made using $\Vreal(\tau)$, there is then an extra term in the exponent, $[\Vreal(\tau) - \Vvirt(\tau)]$. The most important parts of this are those associated with initial state evolution, $[\Vreal_\La(\tau) - \Vvirt_\La(\tau)]$ and $[\Vreal_\Lb(\tau) - \Vvirt_\Lb(\tau)]$.

There is a good deal of calculation needed to find $[\Vreal_\La(\tau) - \Vvirt_\La(\tau)]$. However, the most important term in the result is pretty simple to understand. For a partonic state with $m+2$ partons with momenta $p$, flavors $f$, and colors $c',c$, we find that the contribution is
\begin{equation}
\begin{split}
\label{eq:K2bis}
[\Vreal_{\La}(t)&
- \Vvirt_\La(t)]\sket{\{p,f,c',c\}_{m}}
\\
={}& 
\Bigg\{
\frac{\as}{2\pi}\,
\int_0^{1/(1+y)}\!dz\, 
\left(
\frac{
f_{a/A}(\eta_{\La}/z,\mu_\La^2(t))}
{zf_{a/A}(\eta_{\La},\mu_\La^2(t))}\,
P_{a a}(z)
- \frac{2C_a}{1-z}
\right)
[1\otimes 1]
\\&-
\frac{\as}{2\pi}\,
\int_0^{1}\!dz 
\left(
\frac{f_{a/A}(\eta_{\La}/z, \mu_\La^2(t))}
{zf_{a/A}(\eta_{\La}, \mu_\La^2(t))}\,
P_{a a}\!\left(z\right)
 -
\frac{2 C_a }{1-z}
\right)
[1\otimes 1]
\\ & + \cdots
\Bigg\}
\sket{\{p,f,c',c\}_{m}}
\;.
\end{split}
\end{equation}
The omitted terms here need not concern us at the moment. The function $P_{a a}\!\left(z\right)$ is the flavor conserving part of the evolution kernel for the parton distributions $f_{a/A}(\eta_{\La},\mu^2)$. The factors $[1\otimes 1]$ represent unit operators on the color space. The index $a$ represents the flavor of parton ``a'' in the state $\sket{\{p,f,c',c\}_{m}}$. The factors $C_a$ are color factors, $C_\LF$ for quark or antiquark flavors, $C_\LA$ for gluons. The parameter $y$, defined in eq.~(\ref{eq:ydef0}), is a dimensionless virtuality variable for a splitting at shower time $t$. For large $t$, the splitting has small virtuality and $y \ll 1$.

The integration variable $z$ is the momentum fraction in the splitting: parton ``a'' carries momentum fraction $\eta_\La$ before the splitting and the new parton ``a'' carries momentum fraction $\eta_\La/z$ after the splitting.

The first term in eq.~(\ref{eq:K2bis}) comes from $\Vreal_{\La}(t)$, which is given in eq.~(\ref{eq:calRa2encore}) and involves parton distribution functions and the splitting functions that describe real emissions. The second term comes from $\Vvirt_\La(t)$, which is given in eq.~(\ref{eq:calAa}) and involves the evolution of the parton distribution functions together with contributions from virtual graphs. The full $\Vreal_{\La}(t) - \Vvirt_\La(t)$ is given in eq.~(\ref{eq:VminusSa0}).

In the first term in eq.~(\ref{eq:K2bis}), the upper limit $z < 1/(1+y)$ comes from the kinematics of parton splitting: for a finite virtuality, $(1-z)$ cannot be too small. See eq.~(\ref{eq:zlimit}). In the second term, the upper limit $z < 1$ comes from the upper limit in the evolution equation for parton distribution functions. We see that the two terms are almost identical. They differ only in the upper limits of the $z$-integrations. Adding the two terms, we have
\begin{equation}
\begin{split}
\label{eq:K3bis}
[\Vreal_{\La}(t)&
- \Vvirt_\La(t)]\sket{\{p,f,c',c\}_{m}}
\\
={}& 
\Bigg\{
-\frac{\as}{2\pi}\,
\int_{1/(1+y)}^1\!dz\,
\left(
\frac{
f_{a/A}(\eta_{\La}/z,\mu_\La^2(t))}
{z f_{a/A}(\eta_{\La},\mu_\La^2(t))}
P_{a a}(z)
- \frac{2C_a}{1-z}
\right)
[1\otimes 1]
\\ & + \cdots
\Bigg\}
\sket{\{p,f,c',c\}_{m}}
\;.
\end{split}
\end{equation}
Only a small integration range, corresponding to soft gluon emission, remains. In this range, we can use
\begin{equation}
P_{a a}(z) \approx \frac{2zC_a}{1-z}
\;.
\end{equation}
This gives
\begin{equation}
\begin{split}
\label{eq:K3bisagain}
[\Vreal_{\La}(t)&
- \Vvirt_\La(t)]\sket{\{p,f,c',c\}_{m}}
\\
={}& 
\Bigg\{
\frac{\as}{2\pi}\,
\int_{1/(1+y)}^1\!dz\,
\left(
1 -
\frac{
f_{a/A}(\eta_{\La}/z,\mu_\La^2(t))}
{f_{a/A}(\eta_{\La},\mu_\La^2(t))}
\right)
\frac{2C_a}{1-z}\,
[1\otimes 1]
\\ & + \cdots
\Bigg\}
\sket{\{p,f,c',c\}_{m}}
\;.
\end{split}
\end{equation}
We integrate over a small range of $z$ near $z = 1$. There is a $1/(1-z)$ singularity, the ``threshold singularity.'' The ratio of parton distribution functions is 1 at $z = 1$. Thus the integral is not singular. But the integral is large if $f_{a/A}(\eta_{\La},\mu^2)$ is a fast varying function of $\eta_{\La}$. When $\eta_{\La}$ is not small, say $\eta_{\La} > 0.1$, $f_{a/A}(\eta_{\La},\mu^{2})$ is in fact a fast varying function of $\eta_{\La}$. For that reason, we retain this contribution. The difference $[\Vreal_{\La}(t) - \Vvirt_\La(t)]$ appears in the exponent of the Sudakov factor, so by keeping this contribution, we sum the effects that come from the threshold singularity.

This ``shower'' version of the summation of threshold logs looks rather different from the direct QCD or the SCET descriptions of the same physics. We will discuss the connections in section \ref{sec:comparison}.

With this introduction to provide hints about where we are going, we now proceed to a detailed examination of how the leading threshold logarithms can be summed within a parton shower event generator, at least after we apply the LC+ approximation \cite{NScolor} to simplify the color structure.

\section{The parton distribution functions}%%%%%%%%%%%%%%%%
\label{sec:pdfs}

As was noted already in ref.~\cite{Sterman1987}, the precise definition of parton distributions matters for the question of threshold logarithms. In this paper, we are dealing with a parton shower algorithm. As we will see in more detail in the following sections, it is important that the behavior of the parton distribution functions match the structure of the parton shower algorithm. We discuss the needed definitions in this section.

\subsection{Parton evolution equations}
\label{sec:partonevolution}

Begin with the $\MSbar$ parton distribution functions with one loop evolution. Call the scale variable $\mu_\perp^2$. We have
\begin{equation}
\begin{split}
\label{eq:DGLAP0}
\frac{d f_{a/A}^{\MSbar}(\eta_\La,\mu_\perp^2)}{d\log \mu_\perp^2}  
={}& 
\int_0^1\!dz \sum_{\hat a} \frac{\as(\lambda_\LR \mu_\perp^2)}{2\pi}\,
\theta(\mu_\perp^2 > m_\perp^2(a))
\\&\times
\Bigg\{ 
\frac{1}{z}\, P_{a\hat a}(z)\,
f_{\hat a/A}^{\MSbar}(\eta_\La/z,\mu_\perp^2)
- \delta_{a, \hat a}\left(\frac{2 C_a}{1-z} - \gamma_a
\right)
f_{a/A}^{\MSbar}(\eta_\La,\mu_\perp^2)
\Bigg\}
\;.
\end{split}
\end{equation}
In the case of a bottom or charm quark, evolution of $f_{a/A}(\eta_\La,\mu_\perp^2)$ occurs only for  $\mu_\perp^2 > m_\perp^2(a)$, where $m_\perp(a)$ is the quark mass. Below this scale, $f_{a/A}(\eta_\La,\mu_\perp^2) = 0$.\footnote{This is to be understood as a one loop matching condition between five flavor $\MSbar$ renormalization and four flavor renormalization and then between four and three flavor renormalization.} For a light quark or gluon, we define $m_\perp^2(f) = m_\perp^2({\rm start})$ to be a starting scale for shower evolution and for parton distribution function evolution. In \textsc{Deductor}, we take $m_\perp^2({\rm start})$ near $1 \GeV^2$. In eq.~(\ref{eq:DGLAP0}), $P_{a\hat a}(z)$ are the familiar DGLAP kernels and $C_a$ and $\gamma_a$ are the standard flavor dependent constants recorded in eqs.~(\ref{eq:Cf}) and (\ref{eq:gammaf}).  We take the argument of $\as$ to be $\lambda_\LR \mu_\perp^2$, where
\begin{equation}
\label{eq:lambdaR}
\lambda_\LR = \exp\left(- [ C_\LA(67 - 3\pi^2)- 10\, n_\Lf]/[3\, (33 - 2\,n_\Lf)]\right) \approx 0.4
\;.
\end{equation}
That is, we use the $\MSbar$ scale $\mu_\perp^2$ for $\as$ in the evolution equation and we insert an extra factor $\lambda_\LR$. (In $\lambda_\LR$,  the number $n_\Lf$ of active flavors depends on $\mu_\perp^2$.) The factor $\lambda_\LR$ in the argument of $\as$ is, strictly speaking, beyond the order of perturbation theory that we control in a leading order shower, but it is helpful in generating certain next-to-leading logarithms \cite{lambdaR, NSpT}. 

To make the structure of the $z \to 1$ singularities in this equation clearer, we write
\begin{equation}
\label{Pregdef}
P_{a\hat a}(z) = \delta_{a \hat a}\,\frac{2 z C_a}{1-z}
+ P_{a\hat a}^{\rm reg}(z)
\;,
\end{equation}
where $P_{a\hat a}^{\rm reg}(z)$ is nonsingular as $z \to 1$. Then we can write
\begin{equation}
\begin{split}
\label{eq:DGLAP0alt}
&\frac{d f_{a/A}^{\MSbar}(\eta_\La,\mu_\perp^2)}{d\log \mu_\perp^2}  
=\\&  \qquad 
f_{a/A}^{\MSbar}(\eta_\La,\mu_\perp^2) 
\Bigg\{ \frac{\as(\lambda_\LR \mu_\perp^2)}{2\pi}\,
\theta(\mu_\perp^2 > m_\perp^2(a))\,\gamma_a + 
\int_0^1\!dz\ \frac{\as(\lambda_\LR \mu_\perp^2)}{2\pi}\,
\theta(\mu_\perp^2 > m_\perp^2(a))
\\& \qquad\times
\Bigg( 
-\frac{2 C_a}{1-z}
\left[
1 - 
\frac{f_{a/A}^{\MSbar}(\eta_\La/z,\mu_\perp^2)}
{f_{a/A}^{\MSbar}(\eta_\La,\mu_\perp^2)}
\right]
+
\sum_{\hat a} P_{a\hat a}^{\rm reg}(z)
\frac{f_{\hat a/A}^{\MSbar}(\eta_\La/z,\mu_\perp^2)}
{z f_{a/A}^{\MSbar}(\eta_\La,\mu_\perp^2)}
\Bigg)
\Bigg\}
\;.
\end{split}
\end{equation}

Now we look at the parton evolution that matches shower evolution in \textsc{Deductor}, which uses the virtuality based ordering variable $\Lambda^2$ defined in eq.~(\ref{eq:Lambdadef}). It thus needs parton distribution functions $f_{a/A}(\eta,\mu_\Lambda^2)$, where $\mu_\Lambda^2$ represents the virtuality in an initial state splitting, as in eq.~(\ref{eq:musqamusqb}). In the conventional $\MSbar$ parton distribution functions \cite{CSpartons}, the scale variable is the renormalization scale for the functions with $\MSbar$ renormalization. In $\MSbar$ renormalization at one loop, we have a dimensionally regulated integration,
\begin{equation}
\int_0^1\! dz\ \mu_\perp^{2\epsilon}\int 
\frac{d^{2 - 2\epsilon}\bm k_\perp}{(2\pi)^{2 - 2\epsilon}}
\cdots
\;,
\end{equation}
over the momentum fraction $z$ and the transverse momentum $\bm k_\perp$ of the splitting. The integration is ultraviolet divergent at $\epsilon = 0$. The $\MSbar$ prescription is to subtract the $1/\epsilon$ pole along with certain conventional finite terms. This gives us a finite result in which the scale of the squared transverse momentum $\bm k_\perp^2$ is set by $\mu_\perp^2$.  This is not quite the same as setting the scale of the virtuality $|k^2|$ in the splitting. The two variables are related,{\footnote{For a derivation, see eq (3.23) of ref.~\cite{ShowerTime} with all of the masses equal to zero. Following the discussion in appendix \ref{sec:notation}, we adopt $(1-z) |k^2|$ as the definition of $\bm k_\perp^2$ for an initial state splitting.} for small angle splittings, by $\bm k_\perp^2 = (1-z) |k^2|$.

Using $\bm k_\perp^2 = (1-z)|k^2|$, we see that the scale $\mu_\perp^2$ of $\bm k_\perp^2$ is related to the scale of $\mu_\Lambda^2$ of $|k^2|$ by $\mu_\perp^2 = (1-z)\mu_\Lambda^2$. This works for $\mu_\perp^2 > m_\perp^2(a)$. Below $\mu_\perp^2 = m_\perp^2(a)$, evolution for partons of flavor $a$ stops. We want parton evolution and shower evolution to match. Thus we define 
\begin{equation}
\label{eq:muperpsq}
\mu_\perp^2(z,\mu_\Lambda^2) = \max[(1-z)\mu_\Lambda^2, m_\perp^2(a)]
\;.
\end{equation}
With this definition of the scales, under under an infinitesimal change of $\mu_\Lambda^2$ we have
\begin{equation}
\delta\log\mu_\perp^2 = 
\begin{cases}
\delta\log\mu_\Lambda^2 & (1-z)\mu_\Lambda^2 > m_\perp^2(a) \\
0                       & (1-z)\mu_\Lambda^2 < m_\perp^2(a) 
\end{cases}
\;.
\end{equation}
Thus the one loop evolution equation for virtuality based parton distribution functions is almost the same as the evolution equation for $\MSbar$ parton distribution functions. We need theta functions that turn off the evolution for flavor $a$ unless $(1-z) \mu_\Lambda^2 > m_\perp^2(a)$ and we should use $(1-z) \lambda_\LR \mu_\Lambda^2$ in the argument of $\as$. We simplify this a bit and use
\begin{equation}
\begin{split}
\label{eq:DGLAPLambda}
\frac{d f_{a/A}(\eta_\La,\mu_\Lambda^2)}{d\log \mu_\Lambda^2}  
={}&
f_{a/A}(\eta_\La,\mu_\Lambda^2) 
\Bigg\{ \frac{\as(\lambda_\LR \mu_\Lambda^2)}{2\pi}\,
\theta(\mu_\Lambda^2 > m_\perp^2(a))\,
\gamma_a 
\\&+ 
\int_0^1\!dz\ 
\theta((1-z)\mu_\Lambda^2 > m_\perp^2(a))
\\& \qquad\times
\Bigg( 
-\frac{\as((1-z)\lambda_\LR \mu_\Lambda^2)}{2\pi}\,
\frac{2 C_a}{1-z}
\left[
1 - 
\frac{f_{a/A}(\eta_\La/z,\mu_\Lambda^2)}
{f_{a/A}(\eta_\La,\mu_\Lambda^2)}
\right]
\\&\qquad\qquad +
\frac{\as(\lambda_\LR \mu_\Lambda^2)}{2\pi}\,
\sum_{\hat a} P_{a\hat a}^{\rm reg}(z)
\frac{f_{\hat a/A}(\eta_\La/z,\mu_\Lambda^2)}
{z f_{a/A}(\eta_\La,\mu_\Lambda^2)}
\Bigg)
\Bigg\}
\;.
\end{split}
\end{equation}
In the first term, there is no change from the $\MSbar$ evolution equation. In the remaining two terms, there is an integration over $z$ and we identify $\mu_\perp^2$ with $(1-z) \mu_\Lambda^2$ in the theta function that provides an infrared cutoff on $\mu_\perp^2$. In the second term, which has a $1/(1-z)$ singularity from soft gluon emission, we use $(1-z) \lambda_\LR \mu_\Lambda^2$ as the argument of $\as$, while in the third term, with no $1/(1-z)$ singularity, we simplify the evolution by omitting the factor $(1-z)$ in the argument of $\as$.

With these choices, parton evolution matches the conventions that we use in the shower splitting kernels in \textsc{Deductor}. The parton distributions $f_{a/A}(\eta_\La,\mu_\Lambda^2)$ that we use are obtained by solving the evolution equation (\ref{eq:DGLAPLambda}) using an $\MSbar$ parton distribution set as the initial condition at the starting scale $m_\perp^2({\rm start})$ for the shower. For the starting distributions, we use the CT14 NLO parton set \cite{CT14}.

\subsection{Approximate analytic result}

In section \ref{sec:comparison}, we will present an analytical comparison of the results of this paper for the Drell-Yan cross section to standard analytical results in the leading log approximation. For this purpose, we will need an approximate analytical relation between the parton distribution functions $f_{a/A}(\eta_\La,\mu_\Lambda^2)$ obtained by solving eq.~(\ref{eq:DGLAPLambda}) and the $\MSbar$ parton distribution functions $f_{a/A}^{\MSbar}(\eta_\La,\mu_\perp^2)$ obtained by solving eq.~(\ref{eq:DGLAP0alt}). Our aim is to understand just the leading contributions to threshold behavior, so we do not specify the argument of $\as$ and we make some approximations that correspond to including only the effect of soft gluon emissions.

We write eqs.~(\ref{eq:DGLAPLambda}) and (\ref{eq:DGLAP0alt}) as equations for the Mellin moments of the parton distributions,
\begin{equation}
\tilde f(N) = \int_0^1\!\frac{d\eta}{\eta}\ \eta^{N} f(\eta)
\;.
\end{equation}
We introduce a parameter $\lambda$ with $\lambda = 0$ corresponding to $\MSbar$ evolution, eq.~(\ref{eq:DGLAP0alt}), and $\lambda = 1$ corresponding to $\Lambda^2$ evolution, eq.~(\ref{eq:DGLAPLambda}). We denote by $\mu_0^2$ the appropriate scale parameter, $\mu_\perp^2$ for $\lambda = 0$ or $\mu_\Lambda^2$ for $\lambda = 1$. Keeping only the leading singular terms in the kernel, we have
\begin{equation}
\begin{split}
\label{eq:DGLAPsoft}
\frac{d \tilde f_{a/A}(N,\mu_0^2,\lambda)}{d\log \mu_0^2}  
\approx{}& 
\int_0^1\!dz\  \frac{\as}{2\pi}\,
\theta((1-z)^\lambda \mu_0^2 > m_\perp^2(a))
\\&\times 
\frac{2 C_a}{1-z}
\left[z^N - 1\right]
\tilde f_{a/A}(N,\mu_0^2,\lambda)
\;.
\end{split}
\end{equation}
The solution of this is
\begin{equation}
\begin{split}
\label{eq:DGLAPsoftsoln}
\tilde f_{a/A}(N,\mu_0^2,\lambda)  
\approx{}& 
\tilde f_{a/A}(N,m_\perp^2({\rm start}),0)
\\&\times
\exp\bigg\{
\int_{m_\perp^2({\rm start})}^{\mu_0^2}\!\frac{d\bar \mu^2}{\bar \mu^2}
\int_0^1\!dz\  \frac{\as}{2\pi}\,
\theta((1-z)^\lambda \bar\mu^2 > m_\perp^2(a))
\\&\qquad\times 
\frac{2 C_a}{1-z}
\left[z^N - 1\right]
\bigg\}
\;.
\end{split}
\end{equation}
At the starting scale, $m_\perp^2({\rm start})$, we use the $\MSbar$ parton distributions ($\lambda = 0$) as a boundary value.

Now we can regard $\lambda$ as a continuous variable. Then eq.~(\ref{eq:DGLAPsoftsoln}) gives us a differential equation for the $\lambda$ dependence of $\tilde f_{a/A}(N,\mu_0^2,\lambda)$:
\begin{equation}
\begin{split}
\label{eq:lambdadiffeqnN}
\frac{d\tilde f_{a/A}(N,\mu_0^2,\lambda)}{d\lambda}  
\approx{}& 
\tilde f_{a/A}(N,\mu_0^2,\lambda)
\\&\times
\int_0^1\!dz\  \frac{\as}{2\pi}\,
\theta((1-z)^\lambda \mu_0^2 > m_\perp^2(a))
\log(1-z)
\frac{2 C_a}{1-z}
\left[z^N - 1\right]
\;.
\end{split}
\end{equation}
When we take the inverse Mellin transform of this, we find
\begin{equation}
\begin{split}
\label{eq:lambdadiffeqneta}
\frac{df_{a/A}(\eta_\La,\mu_0^2,\lambda)}{d\lambda}  
\approx{}& 
-f_{a/A}(\eta_\La,\mu_0^2,\lambda)
\int_0^1\!dz\  \frac{\as}{2\pi}\,
\theta((1-z)^\lambda \mu_0^2 > m_\perp^2(a))
\log(1-z)
\\&\qquad\times 
\frac{2 C_a}{1-z}
\left[1
-\frac{f_{a/A}(\eta_\La/z, \mu_0^2,\lambda)}
{f_{a/A}(\eta_\La, \mu_0^2,\lambda)}\right]
\;.
\end{split}
\end{equation}

Notice the factor
\begin{equation}
R = \frac{f_{a/A}(\eta_\La/z,\mu_0^2,\lambda)}
{f_{a/A}(\eta_\La,\mu_0^2,\lambda)}
\;.
\end{equation}
In the kinematic region of interest for threshold logarithms, $R$ is approximately independent of $\lambda$. For instance, with the $\lambda = 1$ parton distributions obtained by solving eq.~(\ref{eq:lambdadiffeqneta}), we find for $a = \Lu$ and $\eta_a = 0.3$, $\mu = 2 \TeV$, that $R \approx z^{2.8}$ for $0.8 < z < 1$. For $\lambda = 0$, we solve the ordinary first order $\MSbar$ evolution equation. Then $R \approx z^{2.9}$ for $0.8< z < 1$. There is a difference, but it is small. 

We can get an instructive analytic result if we make the approximation that the $R$ is independent of $\lambda$. Then the solution of eq.~(\ref{eq:lambdadiffeqneta}) is
\begin{equation}
\label{eq:Zadef}
f_{a/A}(\eta_\La,\mu_0^2) = Z_a(\eta_\La,\mu_0^2)\,f^{\MSbar}_{a/A}(\eta_\La,\mu_0^2)
\;,
\end{equation}
where $f_{a/A}(\eta_\La,\mu_0^2) = f_{a/A}(\eta_\La,\mu_0^2,1)$ and $f^{\MSbar}_{a/A}(\eta_\La,\mu_0^2) = f_{a/A}(\eta_\La,\mu_0^2,0)$ and where
\begin{equation}
\begin{split}
\label{eq:Za0}
Z_a(\eta_\La,\mu_0^2) 
={}& 
\exp\!\Bigg( - 
\int_0^1\!dz 
\int_0^1\!d\lambda\ 
\frac{\as}{2\pi}\,
\theta((1-z)^\lambda\mu_0^2 > m_\perp^2(a))\,
\log(1-z)
\\& \qquad\times 
\frac{2 C_a}{1-z} \left\{
1 - 
\frac{f_{a/A}(\eta_\La/z,\mu_0^2)}{f_{a/A}(\eta_\La,\mu_0^2)}
\right\}
\Bigg)
\;.
\end{split}
\end{equation}
It will prove useful to change variables from $\lambda$ to $\mu_\perp^2 = (1-z)^\lambda \mu_0^2$. Then $\log(1-z)\, d \lambda = d\mu_\perp^2/\mu_\perp^2$ and $\lambda = 0$ corresponds to $\mu_\perp^2 = \mu_0^2$ while $\lambda = 1$ corresponds to  $\mu_\perp^2 = (1-z) \mu_0^2$. Thus (after also choosing a standard argument for $\as$)
\begin{equation}
\begin{split}
\label{eq:Za1}
Z_a(\eta_\La,\mu_0^2) ={}& 
\exp\!\Bigg( 
\int_0^1\!dz 
\int_{(1-z)\,\mu_0^2}^{\mu_0^2} \frac{d\mu_\perp^2}{\mu_\perp^2}\,
\frac{\as(\lambda_\LR \mu_\perp^2)}{2\pi}\,\theta(\mu_\perp^2 > m_\perp^2(a))
\\& \qquad \times
\frac{2 C_a}{1-z} \left\{
1 - 
\frac{f_{a/A}(\eta_\La/z,\mu_0^2)}{f_{a/A}(\eta_\La,\mu_0^2)}
\right\}
\Bigg)
\;.
\end{split}
\end{equation}
When does $Z_a(\eta_\La,\mu_0^2)$ differ significantly from 1? There is a factor $\as$ in the exponent, which generally makes the exponent small. However, the factor multiplying $\as$ can be large when the parton ratio $R$ is a fast varying function of $z$. We will use this result in section \ref{subsec:thecomparison}.

\section{The probability preserving integrand $\Vreal(t)$}
\label{sec:Vreal}

As outlined in section \ref{sec:mainfeatures}, we seek to calculate (with suitable approximations) the difference $(\Vreal(\tau) - \Vvirt(\tau))$ between the integrand of the Sudakov exponent $\Vreal(\tau)$ that preserves probabilities as the shower evolves and the Sudakov integrand $\Vvirt(\tau)$ that is based on virtual diagrams and the evolution of the parton distribution functions. 

We begin in this section with $\Vreal(\tau)$ as defined in \textsc{Deductor} 1.0, but with all quark masses set to zero. We sketch the needed calculations in appendix \ref{sec:CalculateV}. We will use the results in section \ref{sec:enhancementintegrand}. 

The operator $\Vreal(\tau)$ is used to define the Sudakov factor,
\begin{equation}
{\cal N}_{\Vreal}(t_2,t_1) = \mathbb T \exp\left[
-\int_{t_1}^{t_2}\! d\tau\ \Vreal(\tau)\right]
\;.
\end{equation}
The exponent $\int_{t_1}^{t_2} d\tau\ \Vreal(\tau)$, when we take its trace in color, is the total probability to have a real parton splitting between shower times $t_1$ and $t_2$. Thus ${\cal N}_{\Vreal}(t_2,t_1)$ is the probability not to have had a splitting. The operator $\Vreal(\tau)$ has, in general, a non-trivial color structure. For that reason, the exponential function is ordered in shower time $\tau$, as indicated by the $\mathbb T$ instruction. In \textsc{Deductor}, we apply an approximation, the LC+ approximation, that makes $\Vreal(\tau)$ diagonal in color. Then the $\mathbb T$ ordering instruction is not needed. In this paper, we keep the full color structure for all operators, but then at the end we can apply the LC+ approximation.

For many of our formulas for $\Vreal(\tau)$, it is convenient to define energies and angles in a reference frame in which the total momentum $Q$ of all of the final state partons has the form $Q = (E_Q,\vec 0)$. Then $\vec p_l$ is the momentum of parton $l$. The angle $\theta_{kl}$ between two parton momenta is defined in this same reference frame. Thus
\begin{equation}
\begin{split}
E_Q^2 ={}& Q^2
\;,
\\
1 - \cos\theta_{kl} ={}& \frac{p_l\cdot p_k\ Q^2}
{p_l\cdot Q\ p_k\cdot Q}
\;.
\end{split}
\end{equation}
It is convenient to define
\begin{equation}
\label{eq:alformulas}
a_l = \frac{Q^2}{2\,p_l\cdot Q} = \frac{E_Q}{2|\vec p_l|}
\;.
\end{equation}
See appendix \ref{sec:notation}, where $C_{f_l}$ and $\gamma_{f_l}$ are also defined.

In our analysis, we use a dimensionless virtuality variable $y$ given by eqs.~(\ref{eq:ydef0}) and (\ref{eq:ydef}). The variable $y$ is fixed by the shower time $\tau$ according to eq.~(\ref{eq:yfromt}). The structure of $\Vreal(\tau)$ is rather complicated, but we simplify it by taking the leading behavior as $y \to 0$. That is, we use the leading behavior of the splitting functions in the limits of soft or collinear splittings. 

The parton shower in \textsc{Deductor} has an infrared cutoff: no splittings with a transverse momentum smaller than $m_\perp({\rm start}) \sim 1 \GeV$ are generated. In terms of $y$ and the momentum fraction $z$ in the splitting, the cutoff is $y (1-z) > m_\perp^2({\rm start})/(2 p_\La\cdot Q)$ for an initial state splitting and $y z(1-z) > m_\perp^2({\rm start})/(2 p_l\cdot Q)$ for an final state splitting. However, in calculating the leading $y \ll 1$ behavior of $\Vreal(\tau)$, we assume that $y$ is not so small that we need to be concerned with this infrared cutoff. In fact, we will find that very tiny values of $y$ are not relevant for the threshold effects that we investigate. Thus we simply calculate $\Vreal(\tau)$ with no infrared cutoff.

Some of the terms in $\Vreal(\tau)$ depend on the angle $\theta_{kl}$ between the splitting parton $l$ and a parton $k$ that forms part of a color dipole with parton $l$.  This angle dependence arises because soft gluon emission from a color dipole depends on the angles among the partons. The angle $\theta_{kl}$ is small when partons $k$ and $l$ are the daughter partons of a previous splitting that was nearly collinear. When this previous splitting was a final state splitting, ordering in $\Lambda^2$ for the new splitting of parton $l$ implies $a_l  y < 1 - \cos\theta_{kl}$. When a previous splitting that produced partons $k$ and $l$ was an initial state splitting with a small momentum fraction $z_{kl}$, one can also have $a_l  y > 1 - \cos\theta_{kl}$. This happens in the case of multi-regge kinematics, as discussed in section 5.4 of ref.~\cite{ShowerTime}. This is the opposite kinematic regime from that of threshold logarithms, so we ignore this possibility in this paper. Thus we assume $a_l y \ll 1 - \cos\theta_{kl}$ in evaluating $\Vreal(\tau)$.

The terms in $\Vreal(\tau)$ contain operators like $[(\bm{T}_l\cdot \bm{T}_k) \otimes 1]$ that act on the color space. This means that, in the ket part of the color density matrix, color generator $T^a$ acts on the color of parton $l$, color generator $T^a$ acts on the color of parton $k$, and we sum over the octet color index $a$. In this example, a unit operator acts on the bra part of the color density matrix. In manipulating the color operators, we use the identity $\sum_k \bm T_k = 0$ and the identity $(\bm{T}_l\cdot \bm{T}_l) = C_{f_l}$, where $C_q = C_\LF$ for a quark or antiquark flavor $q$ and $C_\Lg = C_\LA$ for a gluon.

At any stage in the shower, there are initial state partons ``a'' and ``b'' as well as final state partons $l$ with $l \in \{1,\dots,m\}$. There is a term in $\Vreal(\tau)$ for each parton:
\begin{equation}
\label{eq:Vrealsum}
\Vreal(\tau)
= 
\Vreal_\La(\tau)
+ \Vreal_\Lb(\tau)
+ \sum_{l=1}^m \Vreal_l(\tau)
\;.
\end{equation}

The limiting form for $\Vreal_l(\tau)$ for a final state parton, from eq.~(\ref{eq:Rlk02}), is quite simple,
\begin{equation}
\begin{split}
\label{eq:calV2}
\Vreal_l(\tau)&\sket{\{p,f,c',c\}_{m}} 
\\={}& 
\frac{\as}{2\pi}\bigg\{
\left( 
2 C_{f_l} \log\!\left[\frac{2|\vec p_l|}{E_Q y}\right]
- \gamma_{f_l}
\right)[1\otimes 1]
\\&\qquad
- \sum_{k \ne l} 
\log\!
\left[
\frac{1 - \cos\theta_{lk}}{2}
\right]
[(\bm{T}_l\cdot \bm{T}_k) \otimes 1
+ 1\otimes(\bm{T}_l\cdot \bm{T}_k)]
\bigg\}
\\ & \times
\sket{\{p,f,c',c\}_{m}}
\;.
\end{split}
\end{equation}

The limiting form for $\Vreal_\La(\tau)$ for initial state parton ``a'' is not quite so simple because it involves parton distribution functions. From eq.~(\ref{eq:calRa2}), we have
\begin{equation}
\begin{split}
\label{eq:calRa2encore}
\Vreal_{\La}(t)&\sket{\{p,f,c',c\}_{m}}
\\
={}& \bigg[
\frac{\as}{2\pi}\,
\int_0^{1/(1+y)}\!dz 
\\&\times \Bigg\{
\sum_{\hat a}
\left(\frac{
f_{\hat a/A}(\eta_{\La}/z,y Q^2)}
{zf_{a/A}(\eta_{\La},y Q^2)}
P_{a\hat a}(z)
- \delta_{a\hat a}\frac{2C_a}{1-z}
\right)
[1\otimes 1]
\\&\quad +
\sum_{k\ne \La,\Lb}
\left[\frac{
f_{a/A}(\eta_{\La}/z,y Q^2)}
{f_{a/A}(\eta_{\La},y Q^2)} - 1\right]
\Delta_{\La k}(z,y)
\big([(\bm T_\La\cdot \bm T_k)\otimes 1] + [1 \otimes (\bm T_\La\cdot \bm T_k)]\big)
\Bigg\}
\\& -
\sum_{k\ne \La,\Lb}
\frac{\as}{2\pi}\,
\log\!
\left[
\frac
{1 - \cos\theta_{\La k}}{2}
\right]\,
\big([(\bm T_\La\cdot \bm T_k)\otimes 1] + [1 \otimes (\bm T_\La\cdot \bm T_k)]\big)
\\&
+\frac{\as}{2\pi}\,2 C_a
\log\left[\frac{1}{y}\right]
[1\otimes 1]
\bigg]
\\&\times
\sket{\{p,f,c',c\}_{m}}
\;.
\end{split}
\end{equation}
The first two terms involve ratios of parton distribution functions together with subtractions. The ratios of parton distribution functions arise naturally in the Sudakov factors for initial state parton shower evolution from hard emissions to soft emissions \cite{EarlyPythia}. Notice that there is an upper limit for the momentum fraction variable in the splitting: $z < 1/(1+y)$. With the definition of $z$ used in \textsc{Deductor}, $z \to 1$ implies that the emitted parton has zero momentum, so that the virtuality variable $y$ must vanish. Thus at finite $y$ there must be a limit on $z$. The precise limit depends on the splitting kinematics used in \textsc{Deductor}. See eq.~(\ref{eq:zlimit}).

In the second term, there is a non-trivial color factor and a function defined in eq.~(\ref{eq:wakdef}),
\begin{equation}
\begin{split}
\label{eq:wakdefencore}
\Delta_{\La k}(z,y)
=
\frac{1}{1-z}
-
\frac{1}
{\sqrt{(1-z)^2 + y^2 z^2/\psi_{\La k}^2}} 
\;,
\end{split}
\end{equation}
where $\psi_{\La k} = (1 - \cos\theta_{\La k})/\sqrt{8(1 + \cos\theta_{\La k})}$, as in eq.~(\ref{eq:psikl}). For $(1-z) \ll y/\psi_{\La k}$, $\Delta_{\La k}$ is approximately $1/(1-z)$. However, for $(1-z) \gg y/\psi_{\La k}$, $\Delta_{\La k}$ tends to zero faster than $1/(1-z)$. Thus, when we integrate over $z$, the important integration region is $y/(1+y) < (1-z) \lesssim y/\psi_{\La k}$ provided that $\psi_{\La k} \lesssim 1$. If $\psi_{\La k} \gg 1$, there is no important integration region.

In the final two terms, there is no dependence on parton distribution functions and the $z$-integration has been performed. One term depends on $\theta_{\La k}$, while the other term, with a trivial color factor, depends only on $y$.

\section{The Sudakov integrand $\Vvirt(t)$}
\label{sec:Vvirt}

In this section, we study the integrand $\Vvirt(t)$ for the Sudakov exponent. As in our discussion of $\Vreal(t)$, we calculate without an infrared cutoff associated with the end of the shower at $k_\perp^2 = m_\perp^2({\rm start}) \sim 1 \GeV^2$. 

Our first task is to identify the parts of $\Vvirt(t)$ that come from parton evolution and from virtual graphs. 

\subsection{Decomposition of $\Vvirt(t)$}
\label{sec:VvirtDecomposition}

The operator $\Vvirt(t)$ affects the evolution of $\sket{\rho(t)}$ through eq.~(\ref{eq:evolution}),
\begin{equation}
\label{eq:evolutionbis}
\frac{d}{dt}\,\sket{\rho(t)} = [{\cal H}_I(t) - \Vvirt(t)]\sket{\rho(t)}
\;.
\end{equation}
The color density matrix $\sket{\rho(t)}$ is defined to include the proper factors of parton distributions so that the color trace of $\rho$ is a differential cross section \cite{NSI}. It is related to the density matrix without parton distribution functions by
\begin{equation}
\label{eq:rhopertdef}
\sket{\rho(t)} = {\cal F}(t) \sket{\rho_{\rm pert}(t)}
\;,
\end{equation}
where
\begin{equation}
\begin{split}
\label{eq:Fdef}
{\cal F}(t)&\sket{\{p,f,c',c\}_{m}}\\ &= 
\frac{f_{a/A}(\eta_{\La},\mu_\La^2(t))
f_{b/B}(\eta_{\Lb}, \mu_\Lb^2(t))}
{4n_\Lc(a) n_\Lc(b)\,4\eta_{\La}\eta_{\Lb}p_\LA\!\cdot\!p_\LB}\
\sket{\{p,f,c',c\}_{m}} 
\;.
\end{split}
\end{equation}
Here the virtuality scale for the parton distributions is a function of the shower time $t$ as defined in eq.~(\ref{eq:musqamusqb}) or eq.~(\ref{eq:musqamusqbbis}), ({\it Cf.}\ section 2.3 of ref.~\cite{PartonDistFctns}.) Then $\sket{\rho_{\rm pert}(t)}$ obeys an evolution equation of the form
\begin{equation}
\label{eq:rhopertevolution}
\frac{d}{dt}\sket{\rho_{\rm pert}(t)} =
[{\cal H}^{\rm pert}_I(t) - \Vvirt^{\rm pert}(t)]
\sket{\rho_{\rm pert}(t)}
\;.
\end{equation}

In eq.~(\ref{eq:rhopertevolution}), 
\begin{equation}
\begin{split}
\label{eq:HpertfromF}
{\cal H}^{\rm pert}_I(t)
 ={}& 
{\cal F}(t)^{-1}
{\cal H}_I(t)
{\cal F}(t)
\;.
\end{split}
\end{equation}
is the perturbative splitting function, with no factors representing parton distribution functions. That is, the definition of ${\cal H}_I(t)$ based on eq.~(8.26) of ref.~\cite{NSI} contains explicit factors of ratios of parton distributions; to define ${\cal H}^{\rm pert}_I(t)$, we remove these factors. We then interpret $-\Vvirt^{\rm pert}(t)$ in eq.~(\ref{eq:rhopertevolution}), as giving the part of the evolution of $\sket{\rho_{\rm pert}(t)}$ that does not involve parton splitting. Thus we will calculate $-\Vvirt^{\rm pert}(t)$ using suitable approximations to virtual one loop Feynman graphs. We do that in appendix \ref{sec:CalculateS}.

Using eqs.~(\ref{eq:evolutionbis}), (\ref{eq:rhopertdef}), (\ref{eq:rhopertevolution}), and (\ref{eq:HpertfromF}), we see that the operator $\Vvirt(t)$ is related to $\Vvirt^{\rm pert}(t)$ by 
\begin{equation}
\begin{split}
\label{eq:VfromVpertandF}
\Vvirt(t)
 ={}& 
\Vvirt^{\rm pert}(t)
- {\cal F}(t)^{-1}\left[\frac{d}{dt}\,{\cal F}(t)\right]
\;.
\end{split}
\end{equation}
Acting on a statistical basis state, ${\cal F}(t)^{-1}\left[d{\cal F}(t)/dt\right]$ gives
\begin{equation}
\begin{split}
\label{eq:Vperteigenvalues1}
{\cal F}(t)^{-1}\left[\frac{d}{dt}\,{\cal F}(t)\right]&
\sket{\{p,f,c',c\}_{m}}
\\ ={}& 
[\lambda^{\cal F}_{\La}(a,\eta_\La,t)
+ \lambda^{\cal F}_{\Lb}(b,\eta_\Lb,t)]
\sket{\{p,f,c',c\}_{m}} 
\;,
\end{split}
\end{equation}
where
\begin{equation}
\lambda^{\cal F}_{\La}(a,\eta_\La,t)
=
\frac{\frac{d}{dt}\,f_{a/A}(\eta_\La,\mu_\La^2(t))}
{f_{a/A}(\eta_\La,\mu_\La^2(t))}
\end{equation}
with a corresponding expression for $\lambda^{\cal F}_{\Lb}$. Using the parton evolution equation, this is
\begin{equation}
\begin{split}
\label{eq:Vperteigenvalues2}
\lambda^{\cal F}_{\La}(a,\eta_\La,t)
={}&
-\sum_{\hat a} 
\int_0^{1}\!dz\
\bigg\{
\frac{\as}{2\pi}\,\frac{1}{z}
P_{a\hat a}\!\left(z\right)
\,
\frac{f_{\hat a/A}(\eta_{\La}/z, \mu_\La^2(t))}
{f_{a/A}(\eta_{\La}, \mu_\La^2(t))}
\\&-
\delta_{a\hat a}\,
\frac{\as}{2\pi}
\left[
\frac{2 C_a}{1-z}-\gamma_a
\right]
\bigg\}
+ {\cal O}(\as^2)
\;.
\end{split}
\end{equation}
To be more precise, we should use the evolution equation (\ref{eq:DGLAPLambda}) appropriate for $\Lambda^2$ as the evolution variable. For the moment, it suffices to use this simple form: we ignore the theta function that enforces an infrared cutoff $\mu_\perp^2 > m_\perp^2(a)$ and we do not specify the argument of $\as$.

\subsection{Approach to calculating $\Vvirt^{\rm pert}(t)$}
\label{sec:VvirtPertCalculation}

We have argued that $-\Vvirt^{\rm pert}(t)$ should be calculated from virtual one loop Feynman diagrams. There are some choices for how to do that.

First, we should choose a gauge. Ultimately, we expect that Feynman gauge provides the most powerful method, especially if one wants a method that can be extended to higher orders of perturbation theory. However, in Feynman gauge, factorization of softer interactions from harder interactions does not work graph by graph. Gluons that are nearly collinear to external legs of a diagram or are very soft but that carry unphysical longitudinal polarization can couple to the interior of ultraviolet dominated subgraphs. These attachments can be treated after a sum over graphs by the use of the gauge invariance of the theory. However, if we use a physical gauge, only transversely polarized gluons can propagate over long distances. Then we do not need to sum over graphs. For this reason, in appendix \ref{sec:CalculateS}, we use Coulomb gauge. 

Second, virtual graphs do not come with a definition of the shower time $t$, so to define $-\Vvirt^{\rm pert}(t)$ from virtual graphs, we need to provide a relation to $t$. Ultimately, we expect that the most powerful method is to consider
\begin{equation}
{\cal G}(t) = - \int_{t}^\infty\!d\tau\ \Vvirt^{\rm pert}(\tau)
\;.
\end{equation}
Here we integrate over the shower time with a lower cutoff at $t$. This integral is infrared divergent and we understand it to be regulated with dimensional regularization. Then we can obtain $\Vvirt^{\rm pert}(t)$ by differentiating with respect to $t$. This formulation gives us the possibility of incorporating the virtual graphs into a general theory that includes the real graphs and a systematic treatment of factorization and renormalization.

In this paper, we choose a method that is computationally a little involved, but relates the definition of the shower time in virtual graphs quite directly to that in the shower splitting functions. We recognize that the splitting functions arise simply from cut Feynman diagrams for real parton emissions in an approximation in which one is close to the infrared singular soft or collinear limits. There is a three dimensional integral over the momentum of the emitted parton, with a delta function that fixes this momentum as a function of the shower time. 

For virtual diagrams, one can imitate the structure of the real emission diagrams. We begin with an integration over four components of the loop momentum $q$. Working in a reference frame in which the total momentum $Q$ of the final state partons has only a time component, we can perform the integration over the energy $q^0$ in the loop. This leaves us with an integration $\int\!d\vec q$ over the three-momentum in the loop. Then we have an integrand with a similar structure to that of the real emission diagrams, but with $\mi\epsilon$ terms in the denominators modified. 

For both real and virtual diagrams, we obtain integrands corresponding to the exchange of a soft gluon between partons $l$ and $k$ that form a color dipole. We partition each of these contributions into two, one with a collinear singularity when the gluon momentum $q$ is collinear to the momentum of parton $l$ and one with a collinear singularity when $q$ is collinear to the momentum of parton $k$.

In the splitting function corresponding to a real emission diagram, when a parton with momentum $\vec q$ is emitted from a parton with velocity $\vec v_l$, there is an infrared singularity when $|\vec q\,| - \vec q \cdot \vec v_l \to 0$. This happens in the soft limit $|\vec q\,| \to 0$ or in the collinear limit that $\vec q$ becomes collinear with $\vec v_l$. Note that $(|\vec q\,| - \vec q \cdot \vec v_l)$ is proportional to the virtuality associated with the emission:
\begin{equation}
|\vec q\,| - \vec q \cdot \vec v_l = 2 p_l\cdot q\
\frac{\sqrt{Q^2}}{2 p_l \cdot Q}
\;.
\end{equation}
The shower time $t$ associated with this splitting is defined by
\begin{equation}
\label{eq:denomtoShowerTime}
\sqrt{Q^2}\  \frac{2 p_l\cdot Q_0}{2 p_l\cdot Q}\,e^{-t} =  |\vec q\,| - \vec q \cdot \vec v_l
\;.
\end{equation}
Here $Q_0$ is the total momentum of the final state partons at the start of the shower, as in section \ref{sec:ordering}.

In the virtual diagrams, we find that the infrared singularities are controlled by factors of the form $1/(|\vec q\,| - \vec q \cdot \vec v_l)$. We simply insert a delta function that identifies this denominator with the shower time according to eq.~(\ref{eq:denomtoShowerTime}). This leaves an integration over two dimensions, which we can arrange to have the form $\int \!dz\,d\phi$ to match the integrations in splitting functions.

With these manipulations, there is a term in $\Vvirt^{\rm pert}(t)$ for each parton in the shower at shower time $t$:
\begin{equation}
\Vvirt^{\rm pert}(t)
= 
\Vvirt_\La^{\rm pert}(t)
+ \Vvirt_\Lb^{\rm pert}(t)
+ \sum_{l=1}^m \Vvirt_l^{\rm pert}(t)
\;.
\end{equation}
In the following two subsections, we state our results for these terms. In each case, the results apply in the limit that the dimensionless virtuality variable $y$ is small: $y \ll 1$ and $y \ll (1 - \cos\theta_{k l})$.

\subsection{Virtual contributions for final state partons}
\label{sec:FinalStateVirtual}

We sketch the calculation for $\Vvirt_{l}^{\rm pert}(t)$ for final state partons in appendix \ref{sec:CalculateS}.  After combining the definitions (\ref{eq:VirtualTotal}), (\ref{eq:Virtuall}), and (\ref{eq:Slkcolordef}) with the results in eqs.~(\ref{eq:Glktotal}), (\ref{eq:Gkatotal}), and (\ref{eq:Gllnet}), we obtain a simple result,
\begin{equation}
\begin{split}
\label{eq:calGl2}
\Vvirt_{l}^{\rm pert}(t)&\sket{\{p,f,c',c\}_{m}} 
\\={}& %%%%%% 1 %%%%%%
\bigg\{
-\frac{\as}{2\pi}\,
\sum_{k \ne l}
\log\!
\left[
\frac
{1 - \cos\theta_{k l}}{2}
\right]
\big(
[(\bm{T}_l\cdot \bm{T}_k)\otimes 1]
+ [1 \otimes (\bm{T}_l\cdot \bm{T}_k)]
\big)
\\&\ \; %%%%%% 2 %%%%%%
+ \sum_{k \ne \La,\Lb,l}
\frac{\as}{2\pi}\,
\mi \pi\,
\big(
[(\bm{T}_l\cdot \bm{T}_k)\otimes 1]
- [1 \otimes (\bm{T}_l\cdot \bm{T}_k)]
\big)
\\&\ \; %%%%%% 3 %%%%%%
+\frac{\as}{2\pi}\,
\left(
2 C_{f_l}\log\!
\left[
\frac
{2 |\vec p_l|}
{E_Q y}
\right]
-\gamma_{f_l}
\right)
[1\otimes 1]
\bigg\}
\\&\times
\sket{\{p,f,c',c\}_{m}}
\;.
\end{split}
\end{equation}
The constants $C_f$ and $\gamma_f$ are defined in appendix \ref{sec:notation}. In assembling this result, we have used the color identities $T_l\cdot T_l = C_{f_l}$ and $\sum_k T_l\cdot T_k = 0$.

The most notable feature of eq.~(\ref{eq:calGl2}) is that $\Vvirt_{l}^{\rm pert}(t)$ has contributions proportional to $\mi\pi$. There is one such contribution for each final state parton $k$ that can form a color dipole with parton $l$, but there is no contribution for $k = \La$ or $k = \Lb$. These contributions persist no matter how small $e^{-t}$ is.

\subsection{Virtual contributions for initial state partons}
\label{sec:InitialStateVirtual}

We sketch the calculation for $\Vvirt_{l}^{\rm pert}(t)$ for initial state partons in appendix \ref{sec:CalculateS}.  After combining the definitions (\ref{eq:VirtualTotal}), (\ref{eq:Virtuall}), and (\ref{eq:Slkcolordef}) with the results in eqs.~(\ref{eq:Gabtotal}), (\ref{eq:Gaktotal}), and (\ref{eq:Gllnet}), we obtain
\begin{equation}
\begin{split}
\label{eq:calGa2}
\Vvirt_{\La}^{\rm pert}(t)&\sket{\{p,f,c',c\}_{m}} 
\\={}& %%%%%% 1 %%%%%%
\bigg\{
-\frac{\as}{2\pi}\,
\sum_{k \ne \La,\Lb}
\log\!
\left[
\frac
{1 - \cos\theta_{\La k}}{2}
\right]
\big(
[(\bm{T}_\La\cdot \bm{T}_k)\otimes 1]
+ [1 \otimes (\bm{T}_\La\cdot \bm{T}_k)]
\big)
\\&\ \; %%%%%% 2 %%%%%%
+ \frac{\as}{2\pi}\,
\mi \pi\,
\big(
[(\bm{T}_\La\cdot \bm{T}_\Lb)\otimes 1]
- [1 \otimes (\bm{T}_\La\cdot \bm{T}_\Lb)]
\big)
\\&\ \; %%%%%% 3 %%%%%%
+\frac{\as}{2\pi}\,
\left(
2 C_{a}\log\!
\left[
\frac
{1}
{y}
\right]
-\gamma_{a}
\right)
[1\otimes 1]
\bigg\}
\\&\times
\sket{\{p,f,c',c\}_{m}}
\;.
\end{split}
\end{equation}
In assembling this result, we have used the color identities $T_\La\cdot T_\La = C_{a}$ and $\sum_k T_\La\cdot T_k = 0$ and the kinematic identity $E_Q = 2 |\vec p_\La|$.

As in the final state $\Vvirt_{l}^{\rm pert}(t)$, there is an $\mi \pi$ contribution, this time associated with the color dipole formed by parton ``a'' and parton ``b.''

\subsection{The complete $\Vvirt(t)$}

The complete operator $\Vvirt(t)$ is a sum of pieces associated with the individual partons,
\begin{equation}
\Vvirt(t)
= 
\Vvirt_\La(t)
+ \Vvirt_\Lb(t)
+ \sum_{l=1}^m \Vvirt_l(t)
\;.
\end{equation}
For the final state partons, we use simply $\Vvirt_{l}^{\rm pert}(t)$ from eq.~(\ref{eq:calGl2}). For  initial state parton ``a'', we use $\Vvirt_{\La}^{\rm pert}(t)$ from eq.~(\ref{eq:calGa2}) and add the contribution from parton evolution according to eq.~(\ref{eq:VfromVpertandF}), (\ref{eq:Vperteigenvalues1}) and (\ref{eq:Vperteigenvalues2}). For final state partons, this gives
\begin{equation}
\begin{split}
\label{eq:Vvirtl}
\Vvirt_{l}(t)&\sket{\{p,f,c',c\}_{m}} 
\\={}& %%%%%% 1 %%%%%%
\bigg\{
-\frac{\as}{2\pi}\,
\sum_{k \ne l}
\log\!
\left[
\frac
{1 - \cos\theta_{k l}}{2}
\right]
\big(
[(\bm{T}_l\cdot \bm{T}_k)\otimes 1]
+ [1 \otimes (\bm{T}_l\cdot \bm{T}_k)]
\big)
\\&\ \; %%%%%% 2 %%%%%%
+ \sum_{k \ne \La,\Lb,l}
\frac{\as}{2\pi}\,
\mi \pi\,
\big(
[(\bm{T}_l\cdot \bm{T}_k)\otimes 1]
- [1 \otimes (\bm{T}_l\cdot \bm{T}_k)]
\big)
\\&\ \; %%%%%% 3 %%%%%%
+\frac{\as}{2\pi}\,
\left(
2 C_{f_l}\log\!
\left[
\frac
{2 |\vec p_l|}
{E_Q y}
\right]
-\gamma_{f_l}
\right)
[1\otimes 1]
\bigg\}
\\&\times
\sket{\{p,f,c',c\}_{m}}
\;.
\end{split}
\end{equation}
For initial state partons, there is a small simplification because terms proportional to $\gamma_a$ cancel, giving
\begin{equation}
\begin{split}
\label{eq:calAa}
\Vvirt_\La(t)&\sket{\{p,f,c',c\}_{m}} 
\\={}& %%%%%% 1 %%%%%%
\bigg\{
-\frac{\as}{2\pi}\,
\sum_{k \ne \La,\Lb}
\log\!
\left[
\frac
{1 - \cos\theta_{\La k}}{2}
\right]
\big(
[(\bm{T}_\La\cdot \bm{T}_k)\otimes 1]
+ [1 \otimes (\bm{T}_\La\cdot \bm{T}_k)]
\big)
\\&\ \; %%%%%% 2 %%%%%%
+ \frac{\as}{2\pi}\,
\mi \pi\,
\big(
[(\bm{T}_\La\cdot \bm{T}_\Lb)\otimes 1]
- [1 \otimes (\bm{T}_\La\cdot \bm{T}_\Lb)]
\big)
\\&\ \; %%%%%% 3 %%%%%%
+\frac{\as}{2\pi}\,
2 C_{a}\log\!
\left[
\frac
{1}
{y}
\right]
[1\otimes 1]
\\&\ \; %%%%%% 4 %%%%%%
+
\sum_{\hat a} 
\int_0^{1}\!dz\
\frac{\as}{2\pi}
\bigg(
\frac{1}{z}
P_{a\hat a}\!\left(z\right)
\,
\frac{f_{\hat a/A}(\eta_{\La}/z, \mu_\La^2(t))}
{f_{a/A}(\eta_{\La}, \mu_\La^2(t))}
-
\delta_{a\hat a}
\frac{2 C_a}{1-z}
\bigg)
[1\otimes 1]
\bigg\}
\\&\times
\sket{\{p,f,c',c\}_{m}}
\;.
\end{split}
\end{equation}
We will use these results in the following section.

\section{The cross section changing exponent}
\label{sec:enhancementintegrand}

The operator
\begin{equation}
{\cal N}_{\Vreal}(t_2,t_1) = \mathbb T \exp\left[
-\int_{t_1}^{t_2} d\tau\ \Vreal(\tau)\right]
\;,
\end{equation}
inserted between parton splittings, will preserve the Born cross section throughout the shower. This statement is exact in color as long as we use the exact $\Vreal(\tau)$ and the exact splitting operator ${\cal H}_\LI(t)$. In the simplest application, we use the leading color approximation or the LC+ approximation for both ${\cal H}_\LI(t)$ and $\Vreal(\tau)$. Then the Born cross section is also preserved in the shower.

In a leading order shower, it would be better to use
\begin{equation}
{\cal N}_{\Vvirt}(t_2,t_1) = \mathbb T \exp\!\left[
-\int_{t_1}^{t_2} d\tau\ \Vvirt(\tau)\right]
\;,
\end{equation}
where $\Vvirt(t)$ is obtained from virtual one loop graphs and parton evolution as explained above. Then we have
\begin{equation}
{\cal N}_{\Vvirt}(t_2,t_1) = \mathbb T \exp\!\left[
\int_{t_1}^{t_2} d\tau\ [-\Vreal(\tau) + (\Vreal(\tau) - \Vvirt(\tau))]\right]
\;.
\end{equation}
The first term creates a cross section preserving shower, while the second term sums effects that change the cross section. We thus need the cross section changing integrand $(\Vreal(\tau) - \Vvirt(\tau))$ and its integral over $\tau$. We assemble this from our previous results.

\subsection{Final state partons}

For the contribution from final state partons, we subtract eq.~(\ref{eq:Vvirtl}) for $\Vvirt_l(\tau)$ from eq.~(\ref{eq:calV2}) for $\Vreal_l(\tau)$. Almost everything cancels and we are left with
\begin{equation}
\begin{split}
\label{eq:VminusSl}
(\Vreal_{l}(t) - \Vvirt_{l}(t))&\sket{\{p,f,c',c\}_{m}} 
\\={}&
\bigg\{
- \sum_{k \ne \La,\Lb,l}
\frac{\as(\mu^2_l(t))}{2\pi}\,
\mi \pi\,
\big(
[(\bm{T}_l\cdot \bm{T}_k)\otimes 1]
- [1 \otimes (\bm{T}_l\cdot \bm{T}_k)]
\big)
\bigg\}
\\&\times
\sket{\{p,f,c',c\}_{m}}
\;.
\end{split}
\end{equation}
That is, only the ``$\mi \pi$'' terms remain. We have supplied the virtuality of the splitting as the argument of $\as$, using the definition (\ref{eq:musql}).

Integrating over $t$ gives
\begin{equation}
\begin{split}
\label{eq:VminusSlintegrated}
\int_{t_1}^{t_2}\!d\tau\
(\Vreal_{l}(t)& - \Vvirt_{l}(t))\sket{\{p,f,c',c\}_{m}} 
\\={}&
\bigg\{
- 
\int_{\mu^2_l(t_2)}^{\mu^2_l(t_1)} \frac{d\mu^2}{\mu^2}
\sum_{k \ne \La,\Lb,l}
\frac{\as(\mu^2)}{2\pi}\,
\mi \pi\,
\big(
[(\bm{T}_l\cdot \bm{T}_k)\otimes 1]
- [1 \otimes (\bm{T}_l\cdot \bm{T}_k)]
\big)
\bigg\}
\\&\times
\sket{\{p,f,c',c\}_{m}}
\;.
\end{split}
\end{equation}

\subsection{Initial state partons}

For the contribution from initial state parton ``a,'' we subtract eq.~(\ref{eq:calAa}) for $\Vvirt_\La(\tau)$ from eq.~(\ref{eq:calRa2encore}) for $\Vreal_\La(\tau)$. Quite a lot cancels and we are left with
\begin{equation}
\begin{split}
\label{eq:VminusSa0}
(\Vreal_{\La}(t) - \Vvirt_{\La}(t))&\sket{\{p,f,c',c\}_{m}}
\\
={}& \bigg[
-
\int_{1/(1+y)}^1\!dz\
\frac{\as}{2\pi}
\sum_{\hat a}\
\left(\frac{
f_{\hat a/A}(\eta_{\La}/z,\mu_\La^2(t))}
{zf_{a/A}(\eta_{\La},\mu_\La^2(t))}
P_{a\hat a}(z)
- \delta_{a\hat a}\frac{2C_a}{1-z}
\right)
[1\otimes 1]
\\& -
\int_0^{1/(1+y)}\!dz\
\frac{\as}{2\pi}
\left[
1 -
\frac{
f_{a/A}(\eta_{\La}/z,\mu_\La^2(t))}
{f_{a/A}(\eta_{\La},\mu_\La^2(t))}
\right]
\sum_{k\ne \La,\Lb}
\Delta_{\La k}(z,y)
\\&\qquad\quad \times
\big([(\bm T_\La\cdot \bm T_k)\otimes 1] + [1 \otimes (\bm T_\La\cdot \bm T_k)]\big)
\\&
- \frac{\as}{2\pi}\,
\mi \pi\,
\big(
[(\bm{T}_\La\cdot \bm{T}_\Lb)\otimes 1]
- [1 \otimes (\bm{T}_\La\cdot \bm{T}_\Lb)]
\big)
\bigg]
\\&\times
\sket{\{p,f,c',c\}_{m}}
\;.
\end{split}
\end{equation}

There is an ``$\mi \pi$'' term that does not cancel. There is a term with a non-trivial color factor that is proportional to the function $\Delta_{\La k}(z,y)$ defined in eq.~(\ref{eq:wakdefencore}). Finally, there is a term with a trivial color factor $[1 \otimes 1]$. This term arises from using a contribution from $\Vreal_{\La}(t)$ that is integrated over $0 < z < 1/(1+y)$ and subtracting a term with the same integrand from $\Vvirt_{\La}(t)$ that is integrated over $0 < z < 1$. In the difference, we integrate over $1/(1+y) < z < 1$. For small $y$, this is a small range near $z = 1$.

We can integrate this over a range of shower times, $t_1 < \tau < t_2$. This gives the probability changing Sudakov exponent associated with going from a splitting at shower time $t_1$ to a later splitting at shower time $t_2$. We use $y = \mu_\La^2(\tau)/Q^2$, where $Q^2$ is the square of the total momentum of the final state particles at shower time $t_1$. Additionally, in the $[1 \otimes 1]$ term, we decompose the splitting function $P_{a\hat a}(z)$ into $P_{a\hat a}(z) = \delta_{a \hat a}\,{2 z C_a}/(1-z) + P_{a\hat a}^{\rm reg}(z)$ as in eq.~(\ref{Pregdef}). We keep the nonsingular $P_{a\hat a}^{\rm reg}(z)$ term because it can give a large contribution for certain flavor choices. This gives
\begin{equation}
\begin{split}
\label{eq:VminusSa1}
\int_{t_1}^{t_2}\!d\tau\
(\Vreal_{\La}(\tau)& - \Vvirt_{\La}(\tau))\sket{\{p,f,c',c\}_{m}}
\\
={}&
\int_{\mu^2_\La(t_2)}^{\mu^2_\La(t_1)} \frac{d\mu^2}{\mu^2}\
\Bigg\{ 
\bigg[
\int_{1/(1+\mu^2/Q^2)}^1\!dz\
\frac{\as}{2\pi}
\left[1
- \frac{
f_{a/A}(\eta_{\La}/z,\mu^2)}
{f_{a/A}(\eta_{\La},\mu^2)}
\right]
\frac{2C_a}{1-z}\,
[1\otimes 1]
\\&
-
\int_{1/(1+\mu^2/Q^2)}^1\!dz\
\frac{\as}{2\pi}
\sum_{\hat a}\,
\frac{
f_{\hat a/A}(\eta_{\La}/z,\mu^2)}
{zf_{a/A}(\eta_{\La},\mu^2)}\,
P_{a\hat a}^{\rm reg}(z)\,
[1\otimes 1]
\\& -
\int_0^{1/(1+\mu^2/Q^2)}\!dz\
\frac{\as}{2\pi}
\left[
1 -
\frac{
f_{a/A}(\eta_{\La}/z,\mu^2)}
{f_{a/A}(\eta_{\La},\mu^2)}
\right]
\sum_{k\ne \La,\Lb}
\Delta_{\La k}(z,\mu^2/Q^2)
\\&\qquad\quad \times
\big([(\bm T_\La\cdot \bm T_k)\otimes 1] + [1 \otimes (\bm T_\La\cdot \bm T_k)]\big)
\\&
- \frac{\as}{2\pi}\,
\mi \pi\,
\big(
[(\bm{T}_\La\cdot \bm{T}_\Lb)\otimes 1]
- [1 \otimes (\bm{T}_\La\cdot \bm{T}_\Lb)]
\big)
\bigg]
\Bigg\}
\\&\times
\sket{\{p,f,c',c\}_{m}}
\;.
\end{split}
\end{equation}
In the remainder of this section, we discuss the physics of the four terms in eq.~(\ref{eq:VminusSa1}), revise the $\Delta_{\La k}$ term to better reflect the physics, implement an infrared cutoff, and specify the argument of $\as$ in each term. 

\subsubsection{Integration ranges}

We first consider the effective ranges of $\mu^2$ and $z$ that appear in each term in eq.~(\ref{eq:VminusSa1}). All but the last term in eq.~(\ref{eq:VminusSa1}) involve the ratio $y = \mu^2/Q^2$ for a potential splitting that might occur in hadron A between the previous splitting at shower time $t_1$ and the next shower time $t_2$ at which a splitting occurs. Suppose that $\tilde y = \tilde \mu^2/\tilde Q^2$ is the corresponding hardness variable for a previous splitting in hadron A, perhaps the one at $t_1$ if it was in hadron A or perhaps an earlier splitting. How is $\mu^2/Q^2$ related to $\tilde \mu^2/\tilde Q^2$? The relation between $\mu^2/Q^2$ and the shower time for initial state splittings is given in eq.~(\ref{eq:yfromt}),
\begin{equation}
\begin{split}
\label{eq:yfromt}
\frac{\mu^2}{Q^2}
\equiv y ={}&  \frac{2 p_\La\cdot Q_0}{2 p_\La\cdot p_\Lb}\,e^{-t}
.
\end{split}
\end{equation}
As the shower progresses with increasing $t$, the momentum $p_\Lb$ can stay the same or increase. Thus $\mu^2/Q^2$ decreases:
\begin{equation}
\label{eq:decreasingy}
\frac{\mu^2}{Q^2} < \frac{\tilde \mu^2}{\tilde Q^2}
\;.
\end{equation}

We now examine the terms in eq.~(\ref{eq:VminusSa1}), beginning with the simplest term, the one proportional to $\mi \pi$. This term can have important effects, but it does not change the inclusive probability $\sbrax{1}\sket{\rho(t)}$ associated with the statistical state. Its effect continues for all $\mu^2$ from the beginning of the shower to the end of the shower.

Consider next the second term, proportional to $P_{a\hat a}^{\rm reg}(z)$. The integrand here has no $z \to 1$ singularity. The integration over $z$ covers the range $0 < 1-z < \mu^2/Q^2/(1+\mu^2/Q^2)$. Thus this term contributes only near the start of the shower, when $\mu^2/Q^2 \sim 1$. It does not contribute when $\mu^2/Q^2 \ll 1$.

Now consider the first term. Here there is a factor $2 C_a/(1-z)$, so we are sensitive to small $(1-z)$. We can understand the integration region, at least qualitatively, by approximating\footnote{We will explore this approximation further in section \ref{sec:comparison}.} the parton distribution functions as $f_{a/A}(\eta_{\La},\mu^2) \propto \eta_{\La}^{-N}$. We consider the case that $N$ is large, which happens when $\eta_\La$ is bigger than about 0.1. Here ``$N$ is large'' means something like $N \sim 3$ in realistic cases. With this approximation, we have
\begin{equation}
\label{eq:pdfratio}
\left[1
- \frac{
f_{a/A}(\eta_{\La}/z,\mu^2)}
{f_{a/A}(\eta_{\La},\mu^2)}
\right]
\sim [1 - z^N] \sim [1 - \exp(-N(1-z))]
\;.
\end{equation}
 This factor in eq.~(\ref{eq:pdfratio}) is 1 for $(1-z) \gg 1/N$ and tends to zero for $(1-z) \ll 1/N$. Thus we approximate
\begin{equation}
\label{eq:pdfapproximation}
\left[1
- \frac{
f_{a/A}(\eta_{\La}/z,\mu^2)}
{f_{a/A}(\eta_{\La},\mu^2)}
\right]
\sim 
\theta\!\left(
\frac{1}{N} < (1-z)
\right)
\;.
\end{equation}
We also approximate $1 - 1/(1+\mu^2/Q^2) \sim \mu^2/Q^2$. Thus the integral in the first term is roughly
\begin{equation}
\begin{split}
\int_{\mu^2(t_2)}^{\mu^2(t_1)}& \frac{d\mu^2}{\mu^2}\
\int_{1/(1+\mu^2/Q^2)}^1\!\frac{dz}{1-z}\ 
\left[1
- \frac{
f_{a/A}(\eta_{\La}/z,\mu^2)}
{f_{a/A}(\eta_{\La},\mu^2)}
\right]
\\
\sim{}&
\int_{\mu^2(t_2)}^{\mu^2(t_1)} \frac{d\mu^2}{\mu^2}\
\theta\!\left(
\frac{1}{N} < \frac{\mu^2}{Q^2}
\right)
\int_{0}^1\!\frac{dz}{1-z}\
\theta\!\left(
\frac{1}{N} < (1-z) < \frac{\mu^2}{Q^2}
\right)
\;.
\end{split}
\end{equation}
Evidently, this is a rather crude approximation, but it is instructive. The integral would be a small perturbative correction that we could simply ignore except that $N$ is large. This gives a contribution to the exponent proportional to $\log^2 N$. However, the range of $\mu^2/Q^2$ does not extend down to infinitesimal values. As soon as $\mu^2/Q^2 < 1/N$, there is no more contribution. That is, the threshold factor associated with this term in eq.~(\ref{eq:VminusSa1}) comes from the first few steps in shower evolution.

The remaining term in eq.~(\ref{eq:VminusSa1}) has more complicated structure. It contains a sum over final state partons $k$ and  factor $\Delta_{\La k}$, defined in eq.~(\ref{eq:wakdefencore}). This factor depends on the angle $\theta_{\La k}$ the initial state parton ``a'' and parton $k$. The angle $\theta_{\La k}$ appears in the combination $\psi_{\La k} = (1 - \cos\theta_{\La k})/\sqrt{8(1 + \cos\theta_{\La k})}$. From the definition, we see that $\Delta_{\La k} \sim 1/(1-z)$ for small $(1-z)$ with $(1-z) \ll (\mu^2/Q^2)/ \psi_{\La k}$. However, when  $(\mu^2/Q^2)/\psi_{\La k} \ll (1-z)$, $\Delta_{\La k}$ is small compared to $1/(1-z)$ Thus we can roughly approximate $\Delta_{\La k}$ by
\begin{equation}
\Delta_{\La k} \sim \frac{1}{1-z}\ \theta\!\left(
(1-z) <\frac{\mu^2}{2Q^2\psi_{\La k}}
\right)
\;.
\end{equation}
(The factor 2 here is rather arbitrary.)  For the parton factor in this term, we can use the rough approximation (\ref{eq:pdfapproximation}). We can also approximate the upper limit of the $z$-integration as $(1-z) < \mu^2/Q^2$. This gives us the integral
\begin{equation}
\begin{split}
\label{eq:DeltaIntegral}
\int_{\mu^2(t_2)}^{\mu^2(t_1)}& \frac{d\mu^2}{\mu^2}\
\int_0^{1/(1+\mu^2/Q^2)}\!dz\ 
\left[
1 -
\frac{
f_{a/A}(\eta_{\La}/z,\mu^2)}
{f_{a/A}(\eta_{\La},\mu^2)}
\right]
\Delta_{\La k}(z,\mu^2/Q^2)
\\
\sim{}&
\int_0^{1}\! 
\frac{dz}{1-z}\
\theta\!\left(
\frac{1}{N} < (1-z)
\right)
\int_{\mu^2(t_2)}^{\mu^2(t_1)} \frac{d\mu^2}{\mu^2}\
\theta\!\left(
2 \psi_{\La k} (1-z) < \frac{\mu^2}{Q^2}
< (1-z)
\right)
\;.
\end{split}
\end{equation}
We will consider the behavior of this integral both for $\psi_{\La k} \sim 1$  and for $\psi_{\La k} \ll 1$.

Let us first consider the case that $\psi_{\La k} \sim 1$. Then we integrate $(1-z)$ over a range that is large when $N$ is large, giving a $\log N$. On the other hand, $\mu^2/Q^2$ is integrated over a finite range. Thus we have only a single power of $\log N$. We note also that $\mu^2/Q^2$ is never smaller than a number of order $1/N$.

Now consider the case that $\psi_{\La k} \ll 1$, supposing that $\mu^2(t_2) \to 0$. Then the lower bound $2 \psi_{\La k} (1-z)$ on $\mu^2/Q^2$ in eq.~(\ref{eq:DeltaIntegral}) is very small and one may wonder if this leads to a large integral. When $\psi_{\La k} \ll 1$, we have
\begin{equation}
2\psi_{\La k} \approx \theta^2_{\La k}/4
\;.
\end{equation}
Then parton $k$ makes a very small angle with respect to the beam axis. This can happen with a large probability when, late in the shower, parton $k$ was emitted from the initial state parton from hadron A. The most important case to consider is that the latest real emission before the shower time interval under consideration was the emission of parton $k$. Then the upper endpoint of the $\mu^2$ integration, $\mu^2 < \mu^2(t_1)$ corresponds to the bound (\ref{eq:decreasingy}), in which $\tilde \mu^2$ and $\tilde Q^2$ refer to the splitting at which parton $k$ was created. We have $\tilde \mu^2 = 2 p_\La\cdot p_k \approx E_a E_k \theta^2_{\La k}$ and $\tilde Q^2 = 2 \tilde p_\La\cdot p_\Lb \approx 4\tilde E_\La E_\Lb$. Here $\tilde E_\La = \tilde z E_\La$ is the momentum of the mother parton of the previous splitting, while the energy of the of the emitted parton is approximately $E_k \approx (1-\tilde z) E_\La$. Thus
\begin{equation}
\frac{\tilde \mu^2}{\tilde Q^2} \approx
\frac{E_\La [(1-\tilde z) E_\La] \theta^2_{\La k}}{4\tilde z E_\La E_\Lb}
\;.
\end{equation}
Since we define energies and angles in the $\vec Q = 0$ frame, we have $E_\La = E_\Lb$. Thus
\begin{equation}
\frac{\tilde \mu^2}{\tilde Q^2} \approx
\frac{(1-\tilde z) \theta^2_{\La k}}{4\tilde z }
\;.
\end{equation}
The splitting function for the previous emission contained a ratio $f_{a/A}(\tilde \eta/\tilde z, \tilde \mu^2)/ f_{a/A}(\tilde \eta, \tilde \mu^2)$ of parton distributions. Applying the approximation (\ref{eq:pdfapproximation}) to this ratio, we have $(1 - \tilde z) < 1/N$. Thus also, $\tilde z$ is close to 1. This gives
\begin{equation}
\frac{\tilde \mu^2}{\tilde Q^2} \lesssim 
\frac{\theta^2_{\La k}}{4 N }
\;.
\end{equation}
Since $\mu^2/Q^2 < {\tilde \mu^2}/{\tilde Q^2}$, we have
\begin{equation}
\label{eq:upperbound}
\frac{\mu^2}{Q^2} \lesssim 
\frac{\theta^2_{\La k}}{4 N }
\;.
\end{equation}
In the integrand of eq.~(\ref{eq:DeltaIntegral}), we have $(1-z)\theta^2_{\La k}/4 < \mu^2/Q^2$ and $1/N < (1-z)$. This gives
\begin{equation}
\label{eq:lowerbound}
\frac{\theta^2_{\La k}}{4 N }
< \frac{\mu^2}{Q^2}\;.
\end{equation}
When we combine the upper bound (\ref{eq:upperbound}) with the lower bound (\ref{eq:lowerbound}), we see that ${\mu^2}/{Q^2}$ can vary only over a small range around ${\theta^2_{\La k}}/({4 N })$. Similarly, $(1-z)$ varies only over a small range around $1/N$.

\subsubsection{Revised $\Delta_{\La k}$ term}
\label{sec:Deltacut}

We conclude that there are no $\log N$ factors associated with the integration in 
the $\Delta_{\La k}$ term eq.~(\ref{eq:VminusSa1}) in the case that $\psi_{\La k} \ll 1$. The integral does have a finite contribution proportional to $\as$ with no $\log N$ factors. However, a first order parton shower is not adequate to calculate this contribution accurately: the definition of the parton shower splitting functions incorporates the strong ordering condition $\mu^2/Q^2 \ll {\tilde \mu^2}/{\tilde Q^2}$ and that condition is violated here. Ordinarily, the inclusion of an inaccurately calculated small perturbative correction to the cross section would be of little consequence. However, this correction can occur many times as the shower progresses, leading to a large, inaccurately calculated, correction. Thus, we eliminate this contribution from the $\Delta_{\La k}$ term at small $\psi_{\La k}$ by multiplying this term by $\theta(\psi_{\La k} > \psi_{\rm min})$, where the default value of $\psi_{\rm min}$ is $10^{-2}$.

\subsubsection{Infrared cutoff}

Recall now from section \ref{sec:partonevolution} that the shower algorithm has an infrared cutoff that vetoes initial state splittings unless the transverse momentum in the splitting is above a minimum: $(1-z) \mu^2 > m_\perp^2(a)$. Here $m_\perp(a)$ is the shower cutoff scale of order 1 GeV or the heavy quark mass in the case of a charm or bottom quark. Eq.~(\ref{eq:VminusSa1}) was derived with no infrared cutoff, but we can insert the cutoff by inserting a factor $\theta((1-z) \mu^2 > m_\perp^2(a))$ into the integrations over $z$. This cutoff has negligible effect as long as $\mu^2 \gg m_\perp(a) Q$. Since the whole integral is negligible unless $\mu^2 > Q^2/N$, we see that inserting the infrared cutoff has negligible effect as long as $m_\perp(a) < Q/N$. This is the case in situations of phenomenological interest, in which $Q$ is of order 1 TeV, or at least 100 GeV for Tevatron studies, and $m_\perp(a)$ is at most 5 GeV. Thus we make eq.~(\ref{eq:VminusSa1}) consistent with the rest of the shower algorithm and with our treatment of parton evolution by inserting a factor $\theta((1-z) \mu^2 > m_\perp^2(a))$ into the $z$-integrations in Eq.~(\ref{eq:VminusSa1}).\footnote{In a future publication, we hope to derive the real and virtual shower splitting functions in a more general framework in which the infrared cutoff is included from the start.}

\subsubsection{Running coupling}

Now we need to specify the argument of the running coupling $\as$. The simplest choice would be $\as(\mu^2)$. Instead, in the two soft-sensitive terms (proportional to $1/(1-z)$ and $\Delta_{\La k}$) we use $\lambda_\LR \mu_\perp^2 = \lambda_\LR (1-z) \mu^2$ as the argument of $\as$. Here $\lambda_\LR$ \cite{lambdaR} is the constant defined in eq.~(\ref{eq:lambdaR}). The factors $(1-z)$ and  $\lambda_\LR$ in the argument of $\as$ are, strictly speaking, beyond the order of perturbation theory that we control in a leading order shower, but it is helpful in generating next-to-leading logarithms for at least some inclusive observables \cite{lambdaR, NSpT}. The use of $\as(\lambda_\LR (1-z) \mu^2)$ inside the $z$ integrations would create an artificial problem if we integrated down to $(1-z) = 0$. However, with the cutoff $(1-z) \mu^2 > m_\perp^2(a)$ to keep us out of the nonperturbative region, we do not encounter this problem. For the remaining two terms in eq.~(\ref{eq:VminusSa1}), we use simply $\as(\lambda_\LR \mu^2)$.

\subsubsection{Result}

With these substitutions, we have
\begin{equation}
\begin{split}
\label{eq:VminusSaintegrated}
\int_{t_1}^{t_2}\!d\tau\ &
(\Vreal_{\La}(\tau) - \Vvirt_{\La}(\tau))\sket{\{p,f,c',c\}_{m}}
\\
={}& 
\int_{\mu^2_\La(t_2)}^{\mu^2_\La(t_1)} \frac{d\mu^2}{\mu^2}\
\Bigg\{
\int_{1/(1 + \mu^2/Q^2)}^1\!dz\
\frac{\as(\lambda_\LR (1-z) \mu^2)}{2\pi}\,
\theta((1-z) \mu^2 > m_\perp^2(a))
\\&\quad\times
\left[1 -
\frac{
f_{a/A}(\eta_{\La}/z,\mu^2)}
{f_{a/A}(\eta_{\La},\mu^2)}
\right]\frac{2C_a}{1-z}\
[1\otimes 1]
\\&-
\int_{1/(1 + \mu^2/Q^2)}^1\!dz\
\frac{\as(\lambda_\LR \mu^2)}{2\pi}\,
\theta((1-z) \mu^2 > m_\perp^2(a))
\\&\quad\times
\sum_{\hat a}
\frac{
f_{\hat a/A}(\eta_{\La}/z,\mu_\La^2(t))}
{zf_{a/A}(\eta_{\La},\mu_\La^2(t))}\,
P_{a\hat a}^{\rm reg}(z)\
[1\otimes 1]
\\& -
\int_0^{1/(1 + \mu^2/Q^2)}\!dz\
\frac{\as(\lambda_\LR (1-z) \mu^2)}{2\pi}\,
\theta((1-z) \mu^2 > m_\perp^2(a))
\\&\quad\times
\left[
1 -
\frac{
f_{a/A}(\eta_{\La}/z,\mu^2)}
{f_{a/A}(\eta_{\La},\mu^2)}
\right]
\sum_{k\ne \La,\Lb}
\Delta_{\La k}(z,\mu^2/Q^2)\,
\theta(\psi_{\La k} > \psi_{\rm min})
\\&\quad \times
\big([(\bm T_\La\cdot \bm T_k)\otimes 1] 
+ [1 \otimes (\bm T_\La\cdot \bm T_k)]\big)
\\&
- \frac{\as(\lambda_\LR \mu^2)}{2\pi}\,
\mi \pi\,
\big(
[(\bm{T}_\La\cdot \bm{T}_\Lb)\otimes 1]
- [1 \otimes (\bm{T}_\La\cdot \bm{T}_\Lb)]
\big)
\Bigg\}
\\&\times
\sket{\{p,f,c',c\}_{m}}
\;.
\end{split}
\end{equation}
Some discussion of this result may be useful.

In the first term here, there is a singular factor $1/(1-z)$. As discussed above, the singularity is cancelled because the ratio of parton distribution functions approaches 1 as $(1-z) \to 0$. This constant will be large if the derivative of $f_{a/A}(\eta_{\La},\mu_\La^2(t))$ with respect to $\eta_\La$ is large. Then there is a ``threshold enhancement'' of the cross section. Since the integration range in $z$ disappears when $\mu^2/Q^2 \to 0$, the important contribution comes from the region in which $\mu^2/Q^2$ is not too small.

In the second term, there is no singular factor $1/(1-z)$. In standard treatments, there is no ``threshold log.''  However, the ratio of parton distributions can be large, for instance for $a = \Lg$ and $\hat a = \Lu$. For this reason, we retain this contribution.

The term in eq.~(\ref{eq:VminusSaintegrated}) proportional to $\Delta_{\La k}$ appears when there is a final state parton that can form a color dipole with the initial state parton. This does not happen in the starting configuration of the Drell-Yan process, before any final state partons have been emitted. However, the $\Delta_{\La k}$ term does appear in the starting parton configuration for jet production. The $\Delta_{\La k}$ term also appears for any hard process once one or more final state partons have been emitted by initial state radiation. The color factor for this term is non-trivial. If we use the LC+ approximation instead of full color, the $\Delta_{\La k}$ term contributes when parton $k$ is color connected to parton ``a.'' For the reasons given above in this section, we turn this term off when $\psi_{\La k}$ is too small. Here $\Delta_{ak}(z,y)$ and $\psi_{\La k}$ were defined in eqs.~(\ref{eq:wakdef}) and (\ref{eq:Psiak}). 

The last term in eq.~(\ref{eq:VminusSaintegrated}) is proportional to $\mi \pi$. This term is not associated with $1/(1-z)$ singularities, but it is potentially important.

\section{The cross section changing exponent in the LC+ approximation}
\label{sec:enhancementLPplus}

The color structure of $\Vreal(\tau)$ and of $(\Vreal(\tau) - \Vvirt(\tau))$ is non-trivial. However, the current version of \textsc{Deductor} uses the LC+ approximation. In this approximation, the operators are diagonal in color and thus commute with each other. Then
\begin{equation}
{\cal N}_{\Vvirt}(t_2,t_1) = 
K(t_2,t_1)\,
\exp\!\left[
-\int_{t_1}^{t_2} d\tau\ \Vreal(\tau)\right]
\hskip 1 cm
{\rm for\ LC+}
\;,
\end{equation}
where
\begin{equation}
K(t_2,t_1) = \exp\!\left[\int_{t_1}^{t_2} d\tau\ (\Vreal(\tau) - \Vvirt(\tau))
\right]
\hskip 1 cm
{\rm for\ LC+}
\;.
\end{equation}
Thus, within the LC+ approximation, we can generate the shower using $\Vreal(\tau)$ and then, at each splitting, multiply by a numerical factor $K(t_2,t_1)$.

The exponent in $K$ is a sum over partons
\begin{equation}
\begin{split}
K(t_2,&t_1) = 
\\& \exp\!\left[\int_{t_1}^{t_2} d\tau 
\left( (\Vreal_\La(\tau) - \Vvirt_\La(\tau))
+ (\Vreal_\Lb(\tau) - \Vvirt_\Lb(\tau))
+
\sum_{l = 1}^m(\Vreal_l(\tau) - \Vvirt_l(\tau))
\right)
\right]_{\rm LC+}
.
\end{split}
\end{equation}
The contributions for final state partons $l$ are given in eq.~(\ref{eq:VminusSlintegrated}). The contribution for initial state parton ``a'' is given in eq.~(\ref{eq:VminusSaintegrated}) and for parton ``b'' we simply have to substitute $\La \leftrightarrow \Lb$. With exact color, the operators in the exponent change the color state. However, in the LC+ approximation these operators are color diagonal. For a splitting of parton $l$ with helper parton $k$, $[\bm{T}_l\cdot \bm{T}_k \otimes 1]$ vanishes unless $k$ is color connected to $l$ in the ket state and, if $k$ is color connected to $l$, equals $C_\LA/2$ if parton $l$ is a gluon and equals $C_\LF$ if parton $l$ is a quark or antiquark. For $[1 \otimes \bm{T}_l\cdot \bm{T}_k]$, we have the same factors if $k$ is color connected to $l$ in the bra state. This rule applies for $l$ and $k$ being either initial state or final state parton indices.

\section{Comparison to the standard summation}
\label{sec:comparison}

In this section, we consider the cross section $d\sigma/(dQ^2 dY)$ to produce a muon pair with squared momentum $Q^2$ and rapidity $Y$ and compare our result to standard results. To do that, we need two manipulations, which are interesting in their own right.

\subsection{The single power approximation for the Mellin transform}

A parton distribution function can be expressed as an integral over its Mellin transform,
\begin{equation}
\label{eq:MellinTransform}
f_{a/A}(\eta_\La,\mu^2)
= \frac{1}{2\pi}\int_{-\infty}^\infty\!d\omega \
\tilde f_{a/A}(N + \mi\,\omega,\mu^2)\, \eta_\La^{- (N + \mi\,\omega)}
\;.
\end{equation}
Here $\tilde f$ is the Mellin transform of $f$ and is a function of the Mellin variable $n = N + \mi\,\omega$. The integration contour runs from $N - \mi\, \infty$ to $N + \mi\, \infty$, where $N$ is chosen such that the contour runs to the right of any singularities of $\tilde f_{a/A}(n,\mu^2)$.

In the standard method, it is not the parton distribution function that appears in an exponent, but rather the Mellin moment variable $n$. However, there is a simple method that allows us to compare the results of this paper to the standard results. We note that, with a reasonable model of the behavior of $f_{a/A}(\eta_\La,\mu^2)$, its Mellin transform $\tilde f_{a/A}(n,\mu^2)$ has a saddle point at some point $n = N$ along the real axis. If we choose to let the integration contour in eq.~(\ref{eq:MellinTransform}) run through the saddle point, the integration will be dominated\footnote{The large $N$ dependence of $\tilde f_{a/A}(N,\mu^2)$ is determined by how fast $f_{a/A}(\eta_\La,\mu^2)$ decreases as $\eta_\La \to 1$. If we suppose that in this limit, $f_{a/A}(\eta_\La,\mu^2) \sim (1 - \eta_\La)^{\beta - 1}$, then large $\beta$ corresponds to a fast decrease. Then one can easily show that the saddle point approximation is valid for $\beta \to \infty$, $\eta_\La \to 1$. However, the saddle point approximation is {\em not} valid in the limit $\eta_\La \to 1$ at fixed $\beta$, even though $N \to \infty$ in this limit.} by $n \approx N$. ({\it Cf.} ref.~\cite{BFRSaddlePt} for a related use of the saddle point approximation.)

Now, when addressing threshold summation, we encounter $f_{a/A}(\eta_\La/z,\mu^2)$, where $\eta_\La$ is the momentum fraction that appears in the Born cross section and we integrate over $z$. Thus what enters our calculation is 
\begin{equation}
f_{a/A}(\eta_\La/z,\mu^2)
=\frac{1}{2\pi}\int_{-\infty}^\infty\!d\omega\
\tilde f_{a/A}(N + \mi\,\omega,\mu^2)\, (\eta_\La/z)^{- (N + \mi\, \omega)}
\;.
\end{equation}
The integration over $z$ is dominated by $z$ near 1. Using the saddle point approximation, with the location $N$ of the saddle point determined for $z = 1$, we have
\begin{equation}
f_{a/A}(\eta_\La/z,\mu^2)
= \tilde f_{a/A}(N,\mu^2)\,
(\eta_\La/z)^{- N}\,
I(N,\mu^2,z)
\;.
\end{equation}
Here
\begin{equation}
\begin{split}
I(N,\mu^2,z)
={}&
\frac{1}{2\pi}
\int_{-\infty}^\infty\!d\omega \
\exp\left(\mi\,\omega\log z - \frac{1}{2}\omega^2 B(N,\mu^2) + \cdots\right)
\;,
\end{split}
\end{equation}
where 
\begin{equation}
B(N,\mu^2) = \left[\frac{d^2}{dn^2}\,\log(\tilde f_{a/A}(n,\mu^2))
\right]_{n = N}
\;.
\end{equation}
Keeping the order $\omega$ and $\omega^2$ terms indicated, we have
\begin{equation}
\begin{split}
I(N,\mu^2,z)
\approx{}& 
\frac{1}{\sqrt{2\pi B(N,\mu^2)}}\
\exp\left(
-\frac{\log^2(z)}{2 B(N,\mu^2)}
\right)
\;.
\end{split}
\end{equation}
We are interested in the behavior of $f_{a/A}(\eta_\La/z,\mu^2)$ for $(1-z) \ll 1$, so we neglect the $z$ dependence of $I(\eta_\La,\mu^2,z)$, which starts at order $(1-z)^2$. Then
\begin{equation}
f_{a/A}(\eta_\La/z,\mu^2)
\approx 
\left[
\tilde f_{a/A}(N,\mu^2)
I(N,\mu^2,1)
\right]
(\eta_\La/z)^{- N}
\;.
\end{equation}
This gives us what Sterman and Zeng \cite{StermanZeng} call the ``single power approximation.'' Sterman and Zeng argue that the single power approximation is numerically quite accurate in practice.  

With the single power approximation, the ratios of parton distributions in eq.~(\ref{eq:VminusSaintegrated}) becomes
\begin{equation}
\label{eq:singlepower}
\frac{f_{a/A}(\eta_\La/z,\mu^2)}{f_{a/A}(\eta_\La,\mu^2)} = z^{N}
\;.
\end{equation}
Then the Mellin moment variable appears in the exponent of our expressions, so that we can compare to standard results that are written in this form.

\subsection{Rapidity dependence}
\label{subsec:rapidity}

We now consider the cross section to produce a muon pair with squared momentum $Q^2$ and rapidity $Y$:
\begin{equation}
\label{eq:DY0}
\frac{d\sigma}{dQ^2\,dY}
= \sum_{a,b}\int_{x_{\La}}^1 d\eta_\La \int_{x_{\Lb}}^1 d\eta_\Lb\
f_{a/A}^{\MSbar}(\eta_\La,Q^2)\,f_{b/B}^{\MSbar}(\eta_\Lb,Q^2)\,
\frac{d\hat\sigma(a,b)}{dQ^2\,dY}
\;.
\end{equation}
Here the lower limits on $\eta_\La$ and $\eta_b$ are the momentum fractions at the Born level:
\begin{equation}
\begin{split}
x_\La ={}& \sqrt{Q^2/s}\ e^Y
\;,
\\
x_\Lb ={}& \sqrt{Q^2/s}\ e^{-Y}
\;.
\end{split}
\end{equation}
We are interested in the threshold region for $Q^2$ and $Y$, by which we mean that $x_\La$ and $x_\Lb$ are close to 1 and $f^{\MSbar}_{a/A}(\eta_\La,\mu^2)$ and $f^{\MSbar}_{b/B}(\eta_\Lb,\mu^2)$ are fast decreasing functions for $\eta_\La > x_\La$ and $\eta_\Lb > x_\Lb$, respectively. In the threshold region, the flavor structure simplifies. Parton $a$ must be a quark and $b$ an antiquark, or vice versa. The flavor structure is carried by a function $\sigma_0$ that appears in the Born cross section:
\begin{equation}
\frac{d\sigma_{\rm Born}}{dQ^2\,dY}
= \sum_{a,b}
f_{a/A}^{\MSbar}(x_\La,Q^2)\,f_{b/B}^{\MSbar}(x_\Lb,Q^2)\,
\sigma_0(a,b,Q^2,s)
\;.
\end{equation}
In the threshold region, we can write the parton level cross section in eq.~(\ref{eq:DY0}) in terms of a dimensionless and flavor independent coefficient function $C$ as
\begin{equation}
\frac{d\hat\sigma(a,b)}{dQ^2\,dY}
\approx
\frac{\sigma_0(a,b,Q^2,s)}{\eta_\La \eta_\Lb}\,
C(\as(Q^2),z,y)
\;,
\end{equation}
where
\begin{equation}
\begin{split}
\label{eq:zydef}
z = {}& \frac{Q^2/s}{\eta_\La \eta_\Lb}
\;,
\\
y ={}& Y - \frac{1}{2} \log(\eta_\La/\eta_\Lb)
\;.
\end{split}
\end{equation}
This gives
\begin{equation}
\begin{split}
\frac{d\sigma}{dQ^2\,dY}
\approx{}&  \sum_{a,b}\frac{d\sigma(a,b)}{dQ^2\,dY}
\;,
\end{split}
\end{equation}
where
\begin{equation}
\begin{split}
\frac{d\sigma(a,b)}{dQ^2\,dY}
={}&  \sigma_0(a,b,Q^2,s)
\int_{x_{\La}}^1 \frac{d\eta_\La}{\eta_\La} 
\int_{x_{\Lb}}^1 \frac{d\eta_\Lb}{\eta_\Lb}\
f^{\MSbar}_{a/A}(\eta_\La,Q^2)\,f^{\MSbar}_{b/B}(\eta_\La,Q^2)\,
C(\as(\mu^2),z,y)
\;.
\end{split}
\end{equation}

Now we change integration variables to $z$ and $y$ defined in eq.~(\ref{eq:zydef}):
\begin{equation}
\begin{split}
\frac{d\sigma(a,b)}{dQ^2\,dY}
={}& \sigma_0(a,b,Q^2,s)
\int_{0}^1 \frac{dz}{z} \int_{-\frac{1}{2}\log(1/z)}^{\frac{1}{2}\log(1/z)} dy
\\&\times
f^{\MSbar}_{a/A}(x_\La\,e^{-y}/\sqrt z,Q^2)\,f^{\MSbar}_{b/B}(x_\Lb\,e^{y}/\sqrt z,Q^2)\
C(\as(Q^2),z,y)
\;.
\end{split}
\end{equation}
The limits $-\frac{1}{2}\log(1/z) < y < \frac{1}{2}\log(1/z)$ come from the requirement that real emissions have positive components along the directions of $p_\La$ and $p_\Lb$, so that $\eta_\La > x_\La$ and $\eta_\Lb > x_\Lb$. Separately, the arguments $x_\La\,e^{-y}/\sqrt z$ and $x_\Lb\,e^{y}/\sqrt z$ of the parton distribution functions must be less than 1 or else the parton distribution functions will vanish. The requirements on the arguments of the parton distribution functions also implies that $z > Q^2/s$.

We can now apply the single power approximation for the parton distributions, giving
\begin{equation}
\begin{split}
\label{eq:zyintegral}
\frac{d\sigma(a,b)}{dQ^2\,dY}
={}& \sigma_0(a,b,Q^2)\,
f^{\MSbar}_{a/A}(x_\La,Q^2)\,f^{\MSbar}_{b/B}(x_\Lb,Q^2)
\int_{0}^1 \frac{dz}{z}\
z^{(N_\La + N_\Lb)/2}
\\&\times
\int_{-\frac{1}{2}\log(1/z)}^{\frac{1}{2}\log(1/z)} dy\
e^{y(N_\La - N_\Lb)}\,
C(\as(Q^2),z,y)
\;.
\end{split}
\end{equation}
Here $N_\La$ and $N_\Lb$ are the saddle point Mellin powers for hadrons A and B, respectively. We can simplify this for the purpose of comparing to standard results. For a given choice of quark-antiquark flavors $a,b$, the saddle point Mellin powers $N_\La$ and $N_\Lb$ depend on the rapidity $Y$. If we imagine replacing $Y \to Y + \delta Y$ while keeping the definitions of $x_\La$ and $x_\Lb$ unchanged, then the parton distribution function factor in eq.~(\ref{eq:zyintegral}) becomes $f_{a/A}(x_\La e^{\delta Y},\mu^2)\,f_{b/B}(x_\Lb e^{-\delta Y},\mu^2)$. According to the single power approximation, this factor is
\begin{equation}
f^{\MSbar}_{a/A}(x_\La e^{\delta Y},\mu^2)\,f^{\MSbar}_{b/B}(x_\Lb e^{-\delta Y},\mu^2)
\approx 
f^{\MSbar}_{a/A}(x_\La,\mu^2)\,f^{\MSbar}_{b/B}(x_\Lb,\mu^2)\,
e^{(N_\Lb - N_\La)\delta Y}
\;.
\end{equation}
This implies that the parton distribution factor (and thus also the Born cross section) is maximum at that value of $Y$ such that $(N_\Lb - N_\La) = 0$. 

For our purpose of comparing to standard results, let us choose $Y$ such that the parton factor is close to its maximum. Then $N_\Lb \approx N_\La$. Now look at the integration in eq.~(\ref{eq:zyintegral}). For the kinematic regime in which threshold summation is needed, $(N_\La + N_\Lb)/2$ is large. Then in the $z$-integration, $z$ near 1 dominates. That means that the range of $y$ is small. Since $(N_\Lb - N_\La)$ is small (compared to $(N_\La + N_\Lb)/2$), the factor $\exp({y(N_\La - N_\Lb)})$ can be approximated by 1, as argued in in refs.~\cite{MukherjeeVogelsang, Bolzoni2006}. This gives
\begin{equation}
\begin{split}
\label{eq:zintegral}
\frac{d\sigma(a,b)}{dQ^2\,dY}
\approx{}& \sigma_0(a,b,Q^2)\,
f^{\MSbar}_{a/A}(x_\La,Q^2)\,f^{\MSbar}_{b/B}(x_\Lb,Q^2)
\int_{0}^1 \frac{dz}{z}\
z^{N}\,
C(\as(Q^2),z)
\;,
\end{split}
\end{equation}
where $N = (N_\La + N_\Lb)/2$ and
\begin{equation}
C(\as(Q^2),z) 
= 
\int_{-\frac{1}{2}\log(1/z)}^{\frac{1}{2}\log(1/z)} dy\
C(\as(Q^2),z,y)
\;.
\end{equation}

These manipulations have given us a function of one variable to work with instead of a function of two variables. The function of one variable, $C(\as(Q^2),z)$ is the function that appears in the rapidity-integrated cross section $d\sigma/dQ^2$ and is well studied. Eq.~(\ref{eq:zintegral}) gives us
\begin{equation}
\begin{split}
\label{eq:mellinC}
\frac{d\sigma(a,b)}{dQ^2\,dY}
\approx{}& \sigma_0(a,b,Q^2)\,
f^{\MSbar}_{a/A}(x_\La,\mu^2)\,f^{\MSbar}_{b/B}(x_\Lb,\mu^2)\,
\widetilde C(\as(Q^2),N)
\;,
\end{split}
\end{equation}
where $\widetilde C$ is the Mellin transform of $C$:
\begin{equation}
\begin{split}
\label{eq:ztoNforC}
\widetilde C(\as(\mu^2),N)
=
\int_{0}^1 \frac{dz}{z}\
z^{N}\,
C(\as(Q^2),z)
\;.
\end{split}
\end{equation}

\subsection{The standard result}
\label{subsec:thestandard}

Now we need $\widetilde C$. We use the standard result \cite{Sterman1987, Catani32} as given in eq.~(3.1) of ref.~\cite{StermanZeng} with factorization scale $\mu_\Lf^2 = Q^2$. We use the first term only in the cusp anomalous dimension, set the hard scattering function to 1, and omit the function $D$, thus dropping terms that contribute non-leading logarithms of $N$:
\begin{equation}
\begin{split}
\label{eq:mellinCStermanZeng}
\widetilde C(& \as(Q^2),N) 
\\={}& 
\exp\!\Bigg( 
\int_{0}^1\!dz\ 
\frac{4 C_\LF}{1-z} 
\left[
1 - z^{N-1}
\right]
\int_{(1-z)^2 Q^2}^{Q^2} \frac{d\mu_\perp^2}{\mu_\perp^2}\,
\frac{\as(\mu_\perp^2)}{2\pi}\,
\theta(\mu_\perp^2 > m_\perp^2)
\Bigg)
\;.
\end{split}
\end{equation}

\subsection{The comparison}
\label{subsec:thecomparison}

How does this compare to our results? In the parton shower approach, there are Sudakov factors $K$ for the shower interval between the hard scattering that produces the muon pair and the first real parton splitting, then for the interval between the first real splitting and the second, and so forth. The integrands in the exponent of $K$ decrease with decreasing splitting scale. Therefore we assume for the purposes of making a comparison that the first splitting occurs at a sufficiently small scale that we can just set that scale to zero and ignore the Sudakov factors for further splittings. This gives
\begin{equation}
\begin{split}
\label{eq:showerDY}
\frac{d\sigma(a,b)}{dQ^2\,dY}
\approx{}& \sigma_0(a,b,Q^2)\,
f_{a/A}(x_\La,Q^2)\, f_{b/B}(x_\Lb,Q^2)\,
K_\La K_\Lb
\;.
\end{split}
\end{equation}
The factor $K_\La$ comes from the exponential of $\Vreal_{\La} - \Vvirt_{\La}$ for initial state radiation from parton ``a.'' The factor $K_\Lb$ is the same expression applied to parton ``b.'' The parton distribution functions in eq.~(\ref{eq:showerDY}) are those appropriate for a $\Lambda^2$-ordered shower. They are related to the $\MSbar$ parton distribution functions by factors $Z_\La$ and $Z_\Lb$ defined in eq.~(\ref{eq:Zadef}), so that
\begin{equation}
\begin{split}
\label{eq:showerDYZZ}
\frac{d\sigma(a,b)}{dQ^2\,dY}
\approx{}& \sigma_0(a,b,Q^2)\,
f^{\MSbar}_{a/A}(x_\La,Q^2)\, f^{\MSbar}_{b/B}(x_\Lb,Q^2)\,
Z_\La K_\La Z_\Lb K_\Lb
\;.
\end{split}
\end{equation}
This matches the form of eq.~(\ref{eq:mellinC}). We need to check whether the factor $\widetilde C$ given by eq.~(\ref{eq:mellinCStermanZeng}) is the same as the leading approximation to $Z_\La K_\La Z_\Lb K_\Lb$.

We worked out the leading approximation to $Z_\La$ in eq.~(\ref{eq:Za1}):
\begin{equation}
\begin{split}
\label{eq:Za1encore}
Z_a ={}& 
\exp\!\Bigg( 
\int_0^1\!dz\
\frac{2 C_\LF}{1-z} \left\{
1 - 
\frac{f_{a/A}(\eta_\La/z,Q^2)}{f_{a/A}(\eta_\La,Q^2)}
\right\}
\\&\qquad\times
\int_{(1-z)\,Q^2}^{Q^2} \frac{d\mu_\perp^2}{\mu_\perp^2}\,
\frac{\as(\lambda_\LR \mu_\perp^2)}{2\pi}\,\theta(\mu_\perp^2 > m_\perp^2(a))
\Bigg)
\;.
\end{split}
\end{equation}

The factor $K_\La$ comes from the exponential of $\Vreal_{\La} - \Vvirt_{\La}$ for initial state radiation from parton ``a,'' integrated from an upper scale $\mu_\La^2(t_0) = Q^2$, to a lower scale $\mu_\La^2(\infty) = 0$. We use eq.~(\ref{eq:VminusSaintegrated}) for $K_\La$. The term in eq.~(\ref{eq:VminusSaintegrated}) proportional to $\Delta_{\La k}$ is absent from the first factor $K_\La$ for the Drell-Yan process and the term proportional to $P_{a\hat a}^{\rm reg}(z)$ can be omitted because it is not large. We also omit the $\mi\,\pi$ term. Thus in eq.~(\ref{eq:VminusSaintegrated}) we include only the main term, proportional to $1/(1-z)$, we take the argument of the parton distributions to by fixed at $Q^2$, and we approximate the lower endpoint of the $z$-integration as $1 - \mu^2/Q^2$ instead of $1/(1+ \mu^2/Q^2)$. This gives
\begin{equation}
\begin{split}
\label{eq:Ka1}
K_\La
={}& 
\exp\!\Bigg(
\int_{0}^{Q^2} \frac{d\mu^2}{\mu^2}\
\int_{1 - \mu^2/Q^2}^1\!dz\
\frac{\as(\lambda_\LR(1-z) \mu^2)}{2\pi}\,
\theta((1-z) \mu^2 > m_\perp^2(a))
\\&\qquad\times
\left[1 -
\frac{
f_{a/A}(\eta_{\La}/z,Q^2)}
{f_{a/A}(\eta_{\La},Q^2)}
\right]
\frac{2C_a}{1-z}\
\Bigg)
\;.
\end{split}
\end{equation}
In eq.~(\ref{eq:Ka1}) it is useful to interchange the order of integrations and change variables from $\mu^2$ to $\mu_\perp^2 = (1-z) \mu^2$. This gives
\begin{equation}
\begin{split}
\label{eq:Ka2}
K_\La
={}& 
\exp\!\Bigg(
\int_{0}^1\!dz\
\frac{2C_\LF}{1-z}
\left[1 -
\frac{
f_{a/A}(\eta_{\La}/z,Q^2)}
{f_{a/A}(\eta_{\La},Q^2)}
\right]
\\&\qquad\times
\int_{(1-z)^2 Q^2}^{(1-z)Q^2} \frac{d\mu_\perp^2}{\mu_\perp^2}\
\frac{\as(\lambda_\LR\mu_\perp^2)}{2\pi}\,
\theta(\mu_\perp^2 > m_\perp^2(a))
\Bigg)
\;.
\end{split}
\end{equation}

In the product $Z_\La K_\La$, the integrands combine in a nice way to give
\begin{equation}
\begin{split}
\label{eq:ZK1}
Z_\La K_\La
={}& 
\exp\!\Bigg(
\int_{0}^1\!dz\
\frac{2C_\LF}{1-z}
\left[1 -
\frac{
f_{a/A}(\eta_{\La}/z,Q^2)}
{f_{a/A}(\eta_{\La},Q^2)}
\right]
\\&\qquad\times
\int_{(1-z)^2 Q^2}^{Q^2} \frac{d\mu_\perp^2}{\mu_\perp^2}\
\frac{\as(\lambda_\LR\mu_\perp^2)}{2\pi}\,
\theta(\mu_\perp^2 > m_\perp^2(a))
\Bigg)
\;.
\end{split}
\end{equation}
Although an analysis of other shower ordering schemes is beyond the scope of this paper, it is worth noting that if we had used $k_\perp^2$ ordering or angular ordering for the shower, then $Z_\La$ and $K_\La$ would be different from what we have with $\Lambda^2$ ordering, but $Z_\La K_\La$ would be the same.

Now, apply the single power approximation (\ref{eq:singlepower}). This gives
\begin{equation}
\begin{split}
\label{eq:ZK2}
Z_\La K_\La
={}& 
\exp\!\left(
\int_{0}^1\!dz\
\frac{2C_\LF}{1-z}
\left(1 -
z^{N}
\right)
\int_{(1-z)^2 Q^2}^{Q^2} \frac{d\mu_\perp^2}{\mu_\perp^2}\
\frac{\as(\lambda_\LR\mu_\perp^2)}{2\pi}\,
\theta(\mu_\perp^2 > m_\perp^2)\,
\right)
\;.
\end{split}
\end{equation}

We get the same factor for parton ``b'', so
\begin{equation}
\begin{split}
\label{eq:ZaKaZbKb}
Z_\La K_\La& Z_\Lb K_\Lb 
\\={}& 
\exp\!\Bigg( 
\int_{0}^1\!dz\ 
\frac{4 C_\LF}{1-z} 
\left[
1 - z^{N}
\right]
\int_{(1-z)^2 Q^2}^{Q^2} \frac{d\mu_\perp^2}{\mu_\perp^2}\,
\frac{\as(\lambda_\LR \mu_\perp^2)}{2\pi}\,
\theta(\mu_\perp^2 > m_\perp^2)
\Bigg)
\;.
\end{split}
\end{equation}
This matches the standard result (\ref{eq:mellinCStermanZeng}) with two small changes. First, we have used a factor $\lambda_\LR$ in the argument of $\as$. This does not affect the leading logarithms. Second, we have $\left[1 - z^{N}\right]$ instead of $\left[1 - z^{N-1}\right]$. These forms are equivalent for large $N$.

We conclude that the form of threshold logarithm summation that arises naturally in a parton shower is equivalent to the traditional forms that one gets in a direct-QCD analysis \cite{SudakovFactorization, KidonakisSterman, StermanZeng} or in soft-collinear-effective-theory as in ref.~\cite{BecherNeubertXu}. The shower form is less precise in that it does not allow an analysis beyond the leading approximation. On the other hand, it applies immediately to many processes with no further analysis.

\section{Numerical comparisons}
\label{sec:numerical}

In this section, we exhibit two numerical tests of the threshold summation presented in this paper. In the first test, we look at the inclusive Drell-Yan cross section, $p + p \to e^+ + e^- + X$ at 13 TeV. In the second test we look at the one jet inclusive cross section in proton-proton collisions at 13 TeV. We compare cross sections $d\sigma$, differential in whatever variables we choose, calculated with threshold summation to the corresponding cross sections without threshold summation.

The full cross section including threshold factors is $d\sigma({\rm full})$. This includes all of the terms in eq.~(\ref{eq:VminusSaintegrated}) for $(\Vreal_{\La}(\tau) - \Vvirt_{\La}(\tau))$  except the $\mi \pi$ term, with the color matrices calculated using the leading color approximation.\footnote{We could use the LC+ approximation, but we find that this makes very little difference.} It also includes a factor $Z_a(\eta_\La, \mu_\Lf^2) Z_b(\eta_\Lb, \mu_\Lf^2)$, as defined in eq.~(\ref{eq:Zadef}). This factor relates the $\Lambda^2$-ordered parton distributions to the $\MSbar$ parton distributions. Here $\mu_\Lf^2$ is the factorization scale, characteristic of the hard scattering. For the jet cross section, $\as$ at the hard interaction is evaluated at $\mu_\Lr = \mu_\Lf$. The parton shower then starts at scale $\mu^2(t) = \mu_\Lf^2$ as given in eqs.~(\ref{eq:musqamusqb}) and (\ref{eq:musql}). 

We will sometimes find it of interest to exhibit a cross section $d\sigma({\rm no\ }\Delta)$ in which we omit the term proportional to $\Delta_{ak}$ in eq.~(\ref{eq:VminusSaintegrated}). 

We can turn off all of the threshold effects to obtain a standard parton shower cross section $d\sigma({\rm std.})$. 

In all of these cross sections the default calculation in \textsc{Deductor} begins with a factor $f^{\MSbar}_{a/A}(\eta_\La,\mu_\Lf^2) f^{\MSbar}_{b/B}(\eta_\Lb,\mu_\Lf^2)$, where these functions obey the first order evolution equation (\ref{eq:DGLAP0}). In the calculations presented in this section, we modify the initial parton factor to use NLO parton distributions by multiplying by a factor
\begin{equation}
R_{\rm pdf} =
\frac{f^{\MSbar,{\rm NLO}}_{a/A}(\eta_\La,\mu_\Lf^2)}
{f^{\MSbar}_{a/A}(\eta_\La,\mu_\Lf^2)}
\,
\frac{f^{\MSbar,{\rm NLO}}_{b/B}(\eta_\Lb,\mu_\Lf^2)}
{f^{\MSbar}_{b/B}(\eta_\Lb,\mu_\Lf^2)}
\;.
\end{equation}
The NLO parton distribution functions used are from the central CT14 NLO fit \cite{CT14}. The parton distributions that obey the first order evolution equation (\ref{eq:DGLAP0}) are simply obtained by using eq.~(\ref{eq:DGLAP0}) with the same starting distributions at the starting scale $\mu_{\rm start}^2$. For the Drell-Yan cross section, we find that $R_{\rm pdf}$ is within about 5\% of 1.

The \textsc{Deductor}  (full) results depend on the parameter $\psi_{\rm min}$ introduced in section \ref{sec:Deltacut}. We use $\psi_{\rm min} = 0.01$. We have checked that varying $\psi_{\rm min}$ by a factor 2 or 1/2 affects the cross sections examined by $\pm 2\%$ or somewhat less, depending on the observable.

\subsection{Drell-Yan}

\begin{figure}
%-------------------- Figure -----------------------------
\begin{center}
\begin{tikzpicture}
\begin{semilogyaxis}[title = {Drell-Yan cross section},
   xlabel={$Q\,\mathrm{[TeV]}$}, ylabel={$d\sigma/dQ\,\mathrm{[nb/GeV]}$},
   xmin=0, xmax=7000,
   ymin=1e-15, ymax=1e-5,
  legend cell align=left,
  width=14cm,	
  height=11cm,
  x coord trafo/.code={
    \pgflibraryfpuifactive{
    \pgfmathparse{(#1)/(1000)}
  }{
     \pgfkeys{/pgf/fpu=true}
     \pgfmathparse{(#1)/(1000)}
     \pgfkeys{/pgf/fpu=false}
   }
 },
 xminorgrids=false,
 yminorgrids=false,
 minor x tick num=4
]

\pgfplotstableread{  
%  Q         Full     Std          MCFM
400.	2.240439e-6	2.189078e-6	2.651039609125948e-6
500.	8.129607e-7	7.773716e-7	9.399731113884471e-7
600.	3.506541e-7	3.292006e-7	4.0238373210865825e-7
700.	1.708194e-7	1.577547e-7	1.940820228586128e-7
800.	9.067667e-8	8.249447e-8	1.0087394646903929e-7
900.	5.100633e-8	4.577134e-8	5.617894395295582e-8
1000.	3.031512e-8	2.685946e-8	3.344249734537043e-8
1100.	1.876477e-8	1.643271e-8	2.0421278259789148e-8
1200.	1.197963e-8	1.036476e-8	1.2969266173775543e-8
1300.	7.856658e-9	6.725094e-9	8.336951452794e-9
1400.	5.218321e-9	4.421908e-9	5.584453752141672e-9
1500.	3.550311e-9	2.979312e-9	3.768486046603487e-9
1600.	2.47915e-9	2.06071e-9	2.6080020438335587e-9
1700.	1.732646e-9	1.426385e-9	1.8294502439487958e-9
1800.	1.227988e-9	1.001499e-9	1.2950108380576446e-9
1900.	8.941797e-10	7.231346e-10	9.255280621616675e-10
2000.	6.492504e-10	5.208344e-10	6.73867978110656e-10
2100.	4.78222e-10	3.805756e-10	4.937828758819516e-10
2200.	3.526745e-10	2.782754e-10	3.643695727578509e-10
2300.	2.653895e-10	2.078505e-10	2.702696204820565e-10
2400.	1.965457e-10	1.527065e-10	2.014180183964401e-10
2500.	1.486498e-10	1.145983e-10	1.5061663140475808e-10
2600.	1.133493e-10	8.671663e-11	1.1417394495883553e-10
2700.	8.687528e-11	6.599987e-11	8.689783933633633e-11
2800.	6.626865e-11	4.992879e-11	6.577303306837398e-11
2900.	5.060466e-11	3.7877e-11	5.086163906770884e-11
3000.	3.950178e-11	2.934495e-11	3.88557989637459e-11
3100.	2.969871e-11	2.191335e-11	3.0071169687820366e-11
3200.	2.332995e-11	1.70841e-11	2.3166466966760353e-11
3300.	1.82516e-11	1.328505e-11	1.799761909797025e-11
3400.	1.419874e-11	1.02481e-11	1.3974419373557255e-11
3500.	1.110441e-11	7.964686e-12	1.0788590525282332e-11
3600.	8.659983e-12	6.16704e-12	8.452453402768767e-12
3700.	6.811683e-12	4.81832e-12	6.5974951076291014e-12
3800.	5.315855e-12	3.734303e-12	5.149964940897568e-12
3900.	4.132891e-12	2.878492e-12	4.037044128750859e-12
4000.	3.3181e-12	2.298798e-12	3.1812364356880425e-12
4100.	2.609094e-12	1.791773e-12	2.504473307764408e-12
4200.	2.078552e-12	1.417965e-12	1.964115121316236e-12
4300.	1.611786e-12	1.091348e-12	1.54875277258037e-12
4400.	1.260665e-12	8.475063e-13	1.2222525621817522e-12
4500.	1.016275e-12	6.788656e-13	9.539010466948343e-13
4600.	7.948615e-13	5.261433e-13	7.522734079285966e-13
4700.	6.213409e-13	4.075485e-13	5.936558008014305e-13
4800.	5.085005e-13	3.311695e-13	4.703384667791365e-13
4900.	3.977662e-13	2.57067e-13	3.6583201997754635e-13
5000.	3.111847e-13	1.996331e-13	2.9098130800107293e-13
5100.	2.507014e-13	1.596265e-13	2.297338831219369e-13
5200.	1.9743e-13	1.246532e-13	1.8047362935168596e-13
5300.	1.552282e-13	9.714536e-14	1.4259882120526996e-13
5400.	1.219306e-13	7.553737e-14	1.122377668639666e-13
5500.	9.686108e-14	5.956177e-14	8.750460280357539e-14
5600.	7.682895e-14	4.680331e-14	6.878703693340874e-14
5700.	6.124469e-14	3.696854e-14	5.4224209878397364e-14
5800.	4.684884e-14	2.795215e-14	4.245290007220492e-14
5900.	3.796507e-14	2.247779e-14	3.300055009290906e-14
6000.	2.910671e-14	1.710573e-14	2.601174361565691e-14
6100.	2.326592e-14	1.353689e-14	2.041729237811613e-14
6200.	1.7993e-14	1.036269e-14	1.58743613512427e-14
6300.	1.407761e-14	8.03189e-15	1.2413198168061396e-14
6400.	1.123045e-14	6.344625e-15	9.641367739797055e-15
6500.	8.549849e-15	4.778302e-15	7.42231743770954e-15
6600.	6.898109e-15	3.810637e-15	5.78831230385481e-15
6700.	5.334363e-15	2.914271e-15	4.4714234340200476e-15
6800.	4.054122e-15	2.19061e-15	3.423520232917807e-15
}\xsec

\addplot [red,thick] table [x={0},y={1}]{\xsec};
\addplot [blue,thick] table [x={0},y={2}]{\xsec};
\addplot [black,semithick, dashed] table [x={0},y ={3}]{\xsec};

\node[anchor=west, fill=white] (source) at (axis cs:750,1e-11)
{{\sc Deductor} (std.)};
\node(destination) at (axis cs:2500,1.5e-10){};
\draw[-stealth](source)--(destination);

\node[anchor=west, fill=white] (source) at (axis cs:3500,1e-10)
{{\sc Deductor} (full) \& NLO};
\node(destination) at (axis cs:3600,1.0e-11){};
\draw[-stealth](source)--(destination);

\end{semilogyaxis}
\end{tikzpicture}
\end{center}

\caption{
The cross section $d\sigma/dQ$ for $\Lp + \Lp \to \Le^+ + \Le^- + X$ versus the mass $Q$ of the $\Le^+ \Le^-$ pair for $\sqrt s = 13 \TeV$. The lower, blue curve is $d\sigma({\rm std.})/dQ$, with no threshold effects. The middle, black dashed curve is a perturbative NLO calculation using \textsc{MCFM} \cite{MCFM}. The upper, red curve the shower cross section is $d\sigma({\rm full})/dQ$ including threshold effects. In each case, the factorization scale is $\mu_\Lf = Q$.
}
\label{fig:DYQ}
%-------------------- Figure -----------------------------
\end{figure}
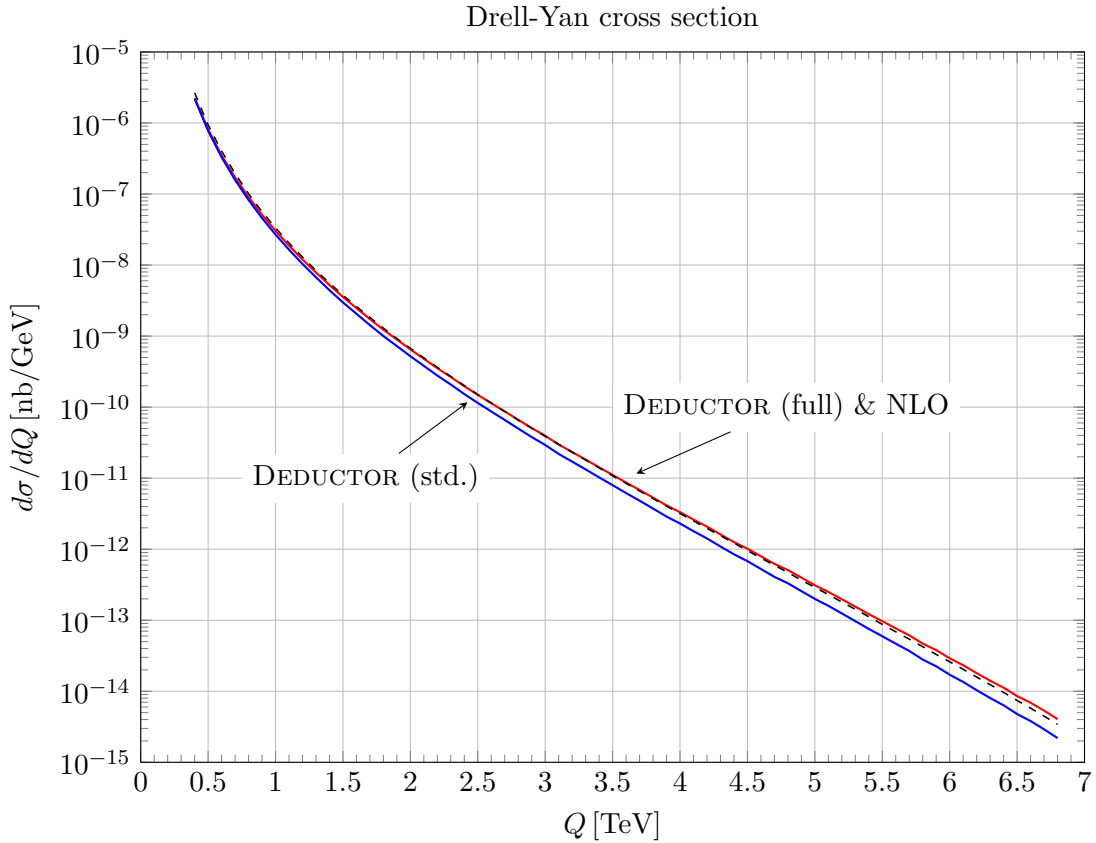

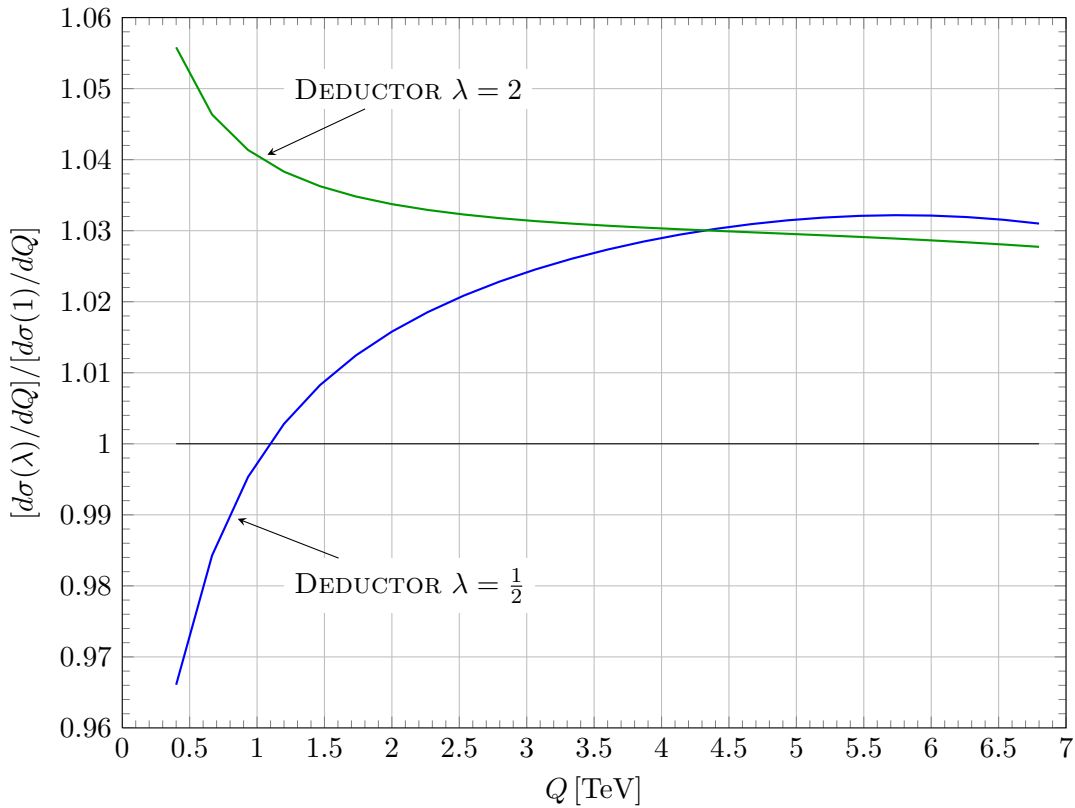
\begin{figure}
%-------------------- Figure -----------------------------
\begin{center}
\begin{tikzpicture}
\begin{axis}[title = {Scale dependence of Drell-Yan cross section},
   xlabel={$Q\,\mathrm{[TeV]}$}, ylabel={$[d\sigma(\lambda)/dQ]/[d\sigma(1)/dQ]$},
   xmin=0, xmax=7000,
   ymin=0.96, ymax=1.06,
   legend style={at={(0.95,0.3)}},
   legend cell align=left,
   width=14cm,	
   height=11cm,
   x coord trafo/.code={
     \pgflibraryfpuifactive{
     \pgfmathparse{(#1)/(1000)}
    }{
     \pgfkeys{/pgf/fpu=true}
     \pgfmathparse{(#1)/(1000)}
     \pgfkeys{/pgf/fpu=false}
    }
  },
 xminorgrids=false,
 yminorgrids=false,
 minor x tick num=4,
 minor y tick num=4
]

\addplot[black,forget plot,domain=400:6800]{1};
\addplot[blue,thick,domain=400:6800]{1.0383 + 5.57062e-10*(1.0-x/7000.0)*x*x - 57.8451/(400.0+x)};
\addplot[green!60!black,thick,domain=400:6800]{1.02488 + 1.49107e-10*(1.0-x/7000.0)*x*x + 18.5487/(200.0+x)};
%\legend{{\sc Deductor} $\lambda=\frac12$,{\sc Deductor} $\lambda=2$}

\node[anchor=west, fill=white] (source) at (axis cs:1200,0.98){{\sc Deductor} $\lambda=\frac12$};
\node(destination) at (axis cs:780,0.99){};
\draw[-stealth](source)--(destination);

\node[anchor=west, fill=white] (source) at (axis cs:1200,1.05){{\sc Deductor} $\lambda=2$};
\node(destination) at (axis cs:1000,1.04){};
\draw[-stealth](source)--(destination);

\end{axis}
\end{tikzpicture}
\end{center}

\caption{
Dependence of the Drell-Yan cross section $d\sigma({\rm full},\lambda)/dQ$ on the factorization scale $\mu_\Lf = \lambda Q$. The blue curve is $[d\sigma({\rm full},1/2)/dQ]/[d\sigma({\rm full},1)/dQ]$. The green curve is $[d\sigma({\rm full},2)/dQ]/[d\sigma({\rm full},1)/dQ]$.
}
\label{fig:DYscales}
%-------------------- Figure -----------------------------
\end{figure}

In figure \ref{fig:DYQ} we look at the inclusive Drell-Yan cross section, $p + p \to e^+ + e^- + X$ at $\sqrt s = 13 \TeV$, as a function of the mass $Q$ of the $e^+ e^-$ pair. We take $\mu_\Lf = Q$. We show two curves from \textsc{Deductor}, one, $d\sigma({\rm full})/dQ$, with the threshold effects turned on, the other, $d\sigma({\rm std.})/dQ$, with the threshold effects turned off. Since the parton shower does not change $Q$, $d\sigma({\rm std.})/dQ$ equals the leading order (LO) perturbative cross section. For comparison, we show a perturbative next-to-leading order (NLO) calculation obtained from \textsc{MCFM} \cite{MCFM} with $\mu_\Lf = Q$ and CT14 NLO parton distributions.\footnote{The MCFM results were adjusted to use the same running $\alpha_{\rm em}(Q)$ as \textsc{Deductor}.}

We note that $d\sigma/dQ$ decreases approximately exponentially as $Q$ increases in the threshold region $Q > 1 \TeV$. This reflects the fast decrease of the parton distribution functions as the momentum fraction increases. All three computed cross sections display the same approximately exponential behavior. However, the threshold correction has an effect that is large enough to notice even in this multi-decade semilog plot.

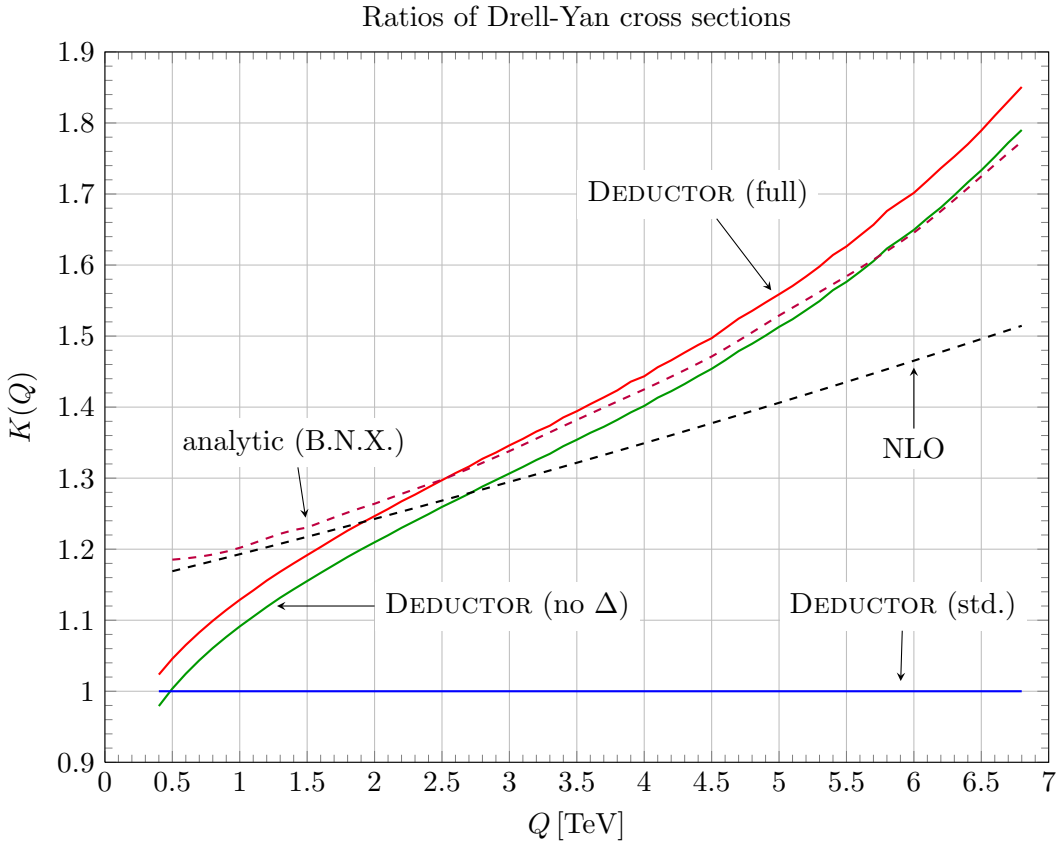
\begin{figure}
%-------------------- Figure -----------------------------
\begin{center}
\begin{tikzpicture}
\begin{axis}[title = {Ratios of Drell-Yan cross sections},
   xlabel={$Q\,\mathrm{[TeV]}$}, ylabel={$K(Q)$},
   xmin=0, xmax=7000,
   ymin=0.9, ymax=1.9,
  legend style={at={(0.4,0.95)}},
  legend cell align=left,
  width=14cm,	
  height=11cm,
  x coord trafo/.code={
    \pgflibraryfpuifactive{
    \pgfmathparse{(#1)/(1000)}
  }{
     \pgfkeys{/pgf/fpu=true}
     \pgfmathparse{(#1)/(1000)}
     \pgfkeys{/pgf/fpu=false}
   }
 },
 xminorgrids=false,
 yminorgrids=false,
 minor x tick num=4,
 minor y tick num=4
]

\pgfplotstableread{  
% Q       Full/LO           no Delta / LO  
400.	1.0234623891885075	0.979159719297348
500.	1.045781322600414	1.0036893552581545
600.	1.065168471746406	1.024681607506183
700.	1.0828165499981934	1.0435549622293345
800.	1.0991848302074068	1.0607806802080189
900.	1.1143726620195082	1.0764723077803708
1000.	1.128657091393498	1.091258349944489
1100.	1.1419157278379526	1.1049759899614853
1200.	1.1558038970511617	1.1187041475152342
1300.	1.1682599529463829	1.1316848507991117
1400.	1.1801061894548688	1.14358620758279
1500.	1.1916546504696386	1.1551385017749065
1600.	1.2030562281931956	1.1666173309199257
1700.	1.2147113156686309	1.177848897737988
1800.	1.2261500011482787	1.1890935487703933
1900.	1.2365328667719675	1.1995887072752431
2000.	1.2465582150487755	1.2097031225280048
2100.	1.2565755660636155	1.2195421882012405
2200.	1.2673578045346445	1.2300656112613622
2300.	1.276828778376766	1.2396871790060644
2400.	1.2870814274441493	1.2493436756130223
2500.	1.297137915658435	1.2595317731589388
2600.	1.3071229820623793	1.2686493928557878
2700.	1.3162947139138306	1.27802175973983
2800.	1.3272632883753042	1.2882214850389926
2900.	1.336026084431185	1.297344826675819
3000.	1.346118497390522	1.3065532570340042
3100.	1.3552793160333771	1.3157842137327245
3200.	1.365594324547386	1.325565291703982
3300.	1.3738450363378385	1.3342576806259667
3400.	1.3854997511733882	1.3450366409383203
3500.	1.3942056221676538	1.3540571467600857
3600.	1.4042365543275217	1.3637777604815273
3700.	1.4137049843098841	1.3722832854604927
3800.	1.4235199982433135	1.3820490731469837
3900.	1.435783389358039	1.3925663159737807
4000.	1.4434065107068998	1.4016803564297515
4100.	1.4561520906945242	1.4132900763657001
4200.	1.4658697499585671	1.4223679710006947
4300.	1.4768763034339183	1.432588871743935
4400.	1.4874992669671012	1.443678943743545
4500.	1.4970194394884642	1.4538886636765807
4600.	1.5107319621859672	1.465920025970111
4700.	1.5245814915280023	1.4788323352926094
4800.	1.5354689969939865	1.4892032629816454
4900.	1.5473250164354038	1.500659750181859
5000.	1.5587830875741546	1.5129379847329927
5100.	1.5705500026624652	1.5236724478704977
5200.	1.5838341895755583	1.5364041998119584
5300.	1.5978961836159753	1.5492433194956508
5400.	1.6141758708305571	1.5646811637736395
5500.	1.6262290391974583	1.5763393196676325
5600.	1.6415281312368721	1.590588144300051
5700.	1.6566705095738161	1.60547941574106
5800.	1.6760370848038522	1.6232239738267003
5900.	1.6890036787424387	1.6360950075607967
6000.	1.7015766062015476	1.649528549790041
6100.	1.718704961036102	1.6658900234839762
6200.	1.7363252205749666	1.6809776226057136
6300.	1.7527144918568356	1.6982901907272134
6400.	1.7700730933664321	1.7164087712039722
6500.	1.789306954646232	1.7333626045402737
6600.	1.810224642231732	1.7522852478470137
6700.	1.8304279183370384	1.7721780163890042
6800.	1.850681773569919	1.7901274074344589
}\xsec

\pgfplotstableread{  
500     1.18525
600     1.18698
700     1.18933
800     1.19248
900     1.19664
1000    1.20195
1100    1.20828
1200    1.21515
1300    1.22166
1400    1.22691
1500    1.23053
1600    1.2376
1700    1.24471
1800    1.25159
1900    1.25803
2000    1.26395
2100    1.27095
2200    1.27785
2300    1.2846
2400    1.29118
2500    1.29766
2600    1.30542
2700    1.31341
2800    1.32154
2900    1.32977
3000    1.33807
3100    1.34681
3200    1.35563
3300    1.36447
3400    1.37329
3500    1.38204
3600    1.39071
3700    1.39932
3800    1.40788
3900    1.41641
4000    1.42496
4100    1.43383
4200    1.44285
4300    1.45205
4400    1.46148
4500    1.47123
4600    1.4822
4700    1.49356
4800    1.50521
4900    1.51703
5000    1.52891
5100    1.53999
5200    1.551
5300    1.56199
5400    1.57303
5500    1.58419
5600    1.5957
5700    1.60751
5800    1.61968
5900    1.63231
6000    1.64552
6100    1.66037
6200    1.67586
6300    1.69184
6400    1.70817
6500    1.72471
6600    1.74127
6700    1.75767
6800    1.77374
%6900    1.78925
%7000    1.804
}\bnx

\addplot[red,thick] table [x={0},y={1}]{\xsec};
\addplot[green!60!black,thick] table [x={0},y={2}]{\xsec};
\addplot[blue,thick, domain=400:6800]{1};
\addplot[purple,dashed,thick] table [x={0},y={1}]{\bnx};
\addplot[black,thick, dashed, domain=500:6800]{1.14563 + 0.0000460973*x + 1.198e-9*x*x};
%\legend{{\sc Deductor} (full), {\sc Deductor} (no $\Delta$),  {\sc Deductor} (std.), analytic (B.N.X.), NLO}

\node[anchor=west, fill=white] (source) at (axis cs:5000,1.12)
{{\sc Deductor} (std.)};
\node(destination) at (axis cs:5900,1){};
\draw[-stealth](source)--(destination);

\node[anchor=west, fill=white] (source) at (axis cs:2000,1.12)
{{\sc Deductor} (no $\Delta$)};
\node(destination) at (axis cs:1200,1.12){};
\draw[-stealth](source)--(destination);

\node[anchor=north, fill=white] (source) at (axis cs:6000,1.37)
{NLO};
\node(destination) at (axis cs:6000,1.47){};
\draw[-stealth](source)--(destination);

\node[anchor=west, fill=white] (source) at (axis cs:500,1.35)
{analytic (B.N.X.)};
\node(destination) at (axis cs:1500,1.23){};
\draw[-stealth](source)--(destination);

\node[anchor=west, fill=white] (source) at (axis cs:3500,1.7)
{{\sc Deductor} (full)};
\node(destination) at (axis cs:5000,1.55){};
\draw[-stealth](source)--(destination);

\end{axis}
\end{tikzpicture}
\end{center}

\caption{
Ratios, $K$, of Drell-Yan cross sections $d\sigma/dQ$, as in figure \ref{fig:DYQ}, to the Born cross section $d\sigma({\rm std.})/dQ = d\sigma({\rm LO})/dQ$ calculated with a factorization scale $\mu_\Lf = Q$. The solid green curve is $K({\rm no\ }\Delta)$ corresponding to $d\sigma({\rm no\ }\Delta)/dQ$. The solid red curve is $K({\rm full})$. In each case, we take $\mu_\Lf = Q$. The dashed, black curve is $K({\rm NLO})$ obtained from a perturbative calculation using \textsc{MCFM} \cite{MCFM} with $\mu_\Lf = Q$. The purple, dashed curve is the analytic result of Becher, Neubert, and Xu, comparable to the NNLO curve of figure 8 of ref.~\cite{BecherNeubertXu}.
}
\label{fig:DYK}
%-------------------- Figure -----------------------------
\end{figure}

There is a theoretical uncertainty associated with the parton shower calculation, which we can estimate by changing the factorization scale $\mu_\Lf$ at which the initial parton distributions are evaluated and at which shower evolution starts. It is rather standard for the Drell-Yan cross section to choose the factorization scale to be $\mu_\Lf = Q$. However, the maximum value of the transverse momentum of the $e^-$ or $e^+$ is $Q/2$, so, by analogy with jet production, for which $\mu_\Lf = P_\LT({\rm jet})$ is a widely used choice, $\mu_\Lf = Q/2$ might seem a sensible choice here. On the other hand, one could choose $\mu_\Lf = 2 Q$. In figure \ref{fig:DYscales}, we plot the ratios of $d\sigma({\rm full})$ with these two scale choices to $d\sigma({\rm full})$ with $\mu_\Lf = Q$. Based on this result, one might estimate a $\pm 5\%$ uncertainty. The precision of the \textsc{Deductor} calculation could be improved by matching to a NLO calculation of the Drell-Yan cross section, but we have not done this.

It is difficult to see small effects in semilog plots like figure \ref{fig:DYQ}, so, in figure \ref{fig:DYK}, we show ratios $K$ of cross sections to the Born cross section, $d\sigma({\rm std.})/dQ = d\sigma({\rm LO})/dQ$. In each case shown, we use $\mu_\Lf = Q$. 

We show first, in red, $K({\rm full})$, corresponding to the cross section with the threshold correction, $d\sigma({\rm full})/dQ$. We see that the threshold correction is quite substantial and increases with $Q$.

We next show, in green, $K({\rm no}\ \Delta)$ for the cross section including just the part of the threshold correction in which we omit the term proportional to $\Delta_{ak}$ in eq.~(\ref{eq:VminusSaintegrated}). The ratio $K_\Delta \equiv [d\sigma({\rm full})/dQ]/[\sigma({\rm no}\ \Delta)/dQ]$ is of some interest. For the Drell-Yan process, the $\Delta_{ak}$ term does not occur in the Sudakov factor between the hard interaction and the first real parton emission. After the first emission, there is color flow transverse to the beam so that the pattern of virtual gluon exchange is changed and $\Delta_{ak}$ can be non-zero. Since $\Delta_{ak}$ is itself proportional to $\as$, the perturbative expansion of $K_\Delta - 1$ begins at order $\as^2$. Thus we expect $K_\Delta$ to be close to 1. In fact, we find that $K_\Delta \approx 1.03$ for $Q > 1 \TeV$. The parameter $\psi_{\rm min}$ described in section \ref{sec:Deltacut} controls the integration range over which $\Delta_{ak}$ operates. As noted earlier, the cross section is sensitive to a factor 2 or 1/2 change in $\psi_{\rm min}$ at a level of about $\pm 2 \%$, so the numerical value of $K_\Delta$ is not highly significant. What is significant is that $K_\Delta$ is indeed close to 1.

We display next, as a dashed black curve, the ratio $K({\rm NLO})$ of the NLO cross section to the Born cross section, as given by \textsc{MCFM} \cite{MCFM}. We note that $K({\rm NLO})$ is an increasing function of $Q$, as we might have expected since it includes some of the effect of threshold logs. We note also that the slope of the NLO curve remains rather constant as $Q$ increases, in contrast to $K({\rm full})$, which has an increasing slope as the threshold logs build up.

Finally, we show as a purple, dashed curve, $K$ obtained with the analytic threshold summation of Becher, Neubert, and Xu \cite{BecherNeubertXu}. This curve is obtained by adapting the code for figure 8 of ref.~\cite{BecherNeubertXu} to CT14 parton distributions and $\sqrt s = 13\ \TeV$. We regard the analytic B.N.X.\ curve as more precise than the \textsc{Deductor} (full) curve since the analytic result contains a high order of approximation in the summation of logarithms, while \textsc{Deductor} (full) is based on only a leading order parton shower. We note that, for $Q > 2 \TeV$, the BNX curve agrees within about 3\% with the \textsc{Deductor} result for $d\sigma({\rm full})/dQ$.

\begin{figure}
%-------------------- Figure -----------------------------
\begin{center}
\begin{tikzpicture}
\begin{axis}[title = {Drell-Yan cross section},
   xlabel={$Q_\LT\,\mathrm{[GeV]}$}, ylabel={$dN/dQ_\LT\,\mathrm{[GeV^{-1}]}$},
   xmin=0, xmax=100,
%   ymin=1e-15, ymax=1e-5,
  legend cell align=left,
  width=14cm,	
  height=11cm,
  xminorgrids=false,
  yminorgrids=false,
  minor x tick num=4,
  minor y tick num=4
]

\pgfplotstableread{  
0.5	0.005650502	0.005821711
1.5	0.01425114	0.01435135
2.5	0.0194214	0.01955953
3.5	0.02280124	0.02297272
4.5	0.02440241	0.02458711
5.5	0.02522246	0.02541276
6.5	0.02529579	0.02548216
7.5	0.02506059	0.02524575
8.5	0.02468628	0.02486812
9.5	0.0240963	0.02426689
10.5	0.02350895	0.02366956
11.5	0.02266288	0.02280833
12.5	0.02184813	0.02198396
13.5	0.02121858	0.02134926
14.5	0.02054524	0.02066427
15.5	0.01978989	0.01989951
16.5	0.01919543	0.01929347
17.5	0.01870806	0.0187966
18.5	0.01789857	0.0179812
19.5	0.01730661	0.0173814
20.5	0.0167358	0.01680303
21.5	0.01617651	0.01624109
22.5	0.01569053	0.01574469
23.5	0.0153385	0.015386
24.5	0.01472049	0.01476413
25.5	0.01426088	0.0142961
26.5	0.0139305	0.0139635
27.5	0.01332661	0.01334951
28.5	0.0131325	0.01315694
29.5	0.01274387	0.0127604
30.5	0.01230374	0.01231339
31.5	0.01206434	0.01207368
32.5	0.01169342	0.01170158
33.5	0.01135032	0.01135047
34.5	0.01103892	0.01103674
35.5	0.01070796	0.01069717
36.5	0.0105458	0.01053788
37.5	0.01028911	0.01027412
38.5	0.009907184	0.009890602
39.5	0.009689421	0.009670715
40.5	0.009519422	0.009496337
41.5	0.009267068	0.009243929
42.5	0.009052757	0.009022261
43.5	0.008844779	0.008819091
44.5	0.00853171	0.008505144
45.5	0.008459141	0.008426547
46.5	0.008205055	0.008175143
47.5	0.007995724	0.007956996
48.5	0.007869108	0.007830409
49.5	0.007670794	0.007628977
50.5	0.007454833	0.007412836
51.5	0.007325545	0.007282986
52.5	0.007205377	0.007163422
53.5	0.00703862	0.006993163
54.5	0.006924306	0.006879698
55.5	0.006789442	0.006744556
56.5	0.006641341	0.006596112
57.5	0.006553377	0.006505318
58.5	0.00637089	0.006319819
59.5	0.006237813	0.006186012
60.5	0.006113665	0.00605816
61.5	0.005981015	0.005928851
62.5	0.005893285	0.005837229
63.5	0.005759016	0.005700101
64.5	0.005671162	0.005614767
65.5	0.005533169	0.005476666
66.5	0.005473938	0.005417077
67.5	0.005359979	0.005303552
68.5	0.005187693	0.005130212
69.5	0.00515369	0.005099729
70.5	0.005026259	0.004970223
71.5	0.004960437	0.004905214
72.5	0.004898096	0.004836951
73.5	0.004806342	0.004749849
74.5	0.004703591	0.004644987
75.5	0.004618914	0.00455341
76.5	0.004536291	0.004479621
77.5	0.004490879	0.004429578
78.5	0.004417616	0.004354085
79.5	0.004321693	0.004260806
80.5	0.004238368	0.004176738
81.5	0.004202509	0.004142467
82.5	0.004156209	0.004088975
83.5	0.004036633	0.003974485
84.5	0.004029204	0.003964998
85.5	0.003930171	0.003868219
86.5	0.003880354	0.003816138
87.5	0.003827527	0.003765166
88.5	0.003760164	0.003693664
89.5	0.003766153	0.003700447
90.5	0.003639929	0.003575604
91.5	0.003554372	0.003490881
92.5	0.003524887	0.003462691
93.5	0.003510254	0.003447273
94.5	0.00343548	0.003372772
95.5	0.003369838	0.003309517
96.5	0.003360226	0.003295948
97.5	0.003281165	0.003216216
98.5	0.003245917	0.003183636
99.5	0.003169857	0.003106888
}\xsec

\pgfplotstableread{  
0.5      0.00270587
1.5      0.00883755
2.5      0.0135983
3.5      0.0171181
4.5      0.0196568
5.5      0.020964
6.5      0.0216509
7.5      0.0219991
8.5      0.0219673
9.5      0.0217916
10.5      0.0215672
11.5      0.0212597
12.5      0.0207951
13.5      0.0203295
14.5      0.019868
15.5      0.0193716
16.5      0.0188878
17.5      0.0184465
18.5      0.0180028
19.5      0.0174863
20.5      0.0170324
21.5      0.0165773
22.5      0.0161545
23.5      0.0157219
24.5      0.0152628
25.5      0.0149284
26.5      0.014526
27.5      0.0140715
28.5      0.0137186
29.5      0.0133133
30.5      0.01303
31.5      0.0127002
32.5      0.0124343
33.5      0.0120487
34.5      0.01176
35.5      0.0115668
36.5      0.0111793
37.5      0.0109402
38.5      0.0106992
39.5      0.010531
40.5      0.0102152
41.5      0.0100404
42.5      0.00976311
43.5      0.00954145
44.5      0.00933789
45.5      0.00912951
46.5      0.00898488
47.5      0.00876441
48.5      0.00861545
49.5      0.00843629
50.5      0.0082177
51.5      0.00804609
52.5      0.00791573
53.5      0.00778268
54.5      0.00757076
55.5      0.00743518
56.5      0.00724082
57.5      0.00718896
58.5      0.0069986
59.5      0.00684893
60.5      0.006699
61.5      0.00662059
62.5      0.00652552
63.5      0.00637303
64.5      0.0062582
65.5      0.00615157
66.5      0.00601693
67.5      0.00594694
68.5      0.00588354
69.5      0.005734
70.5      0.00561756
71.5      0.00555492
72.5      0.00549924
73.5      0.00539484
74.5      0.0052409
75.5      0.00515962
76.5      0.00514174
77.5      0.00503421
78.5      0.00491944
79.5      0.00487378
80.5      0.0047481
81.5      0.00469404
82.5      0.00465083
83.5      0.00454712
84.5      0.00450064
85.5      0.00444006
86.5      0.00432044
87.5      0.00427321
88.5      0.00418244
89.5      0.00419085
90.5      0.00401486
91.5      0.0040562
92.5      0.0039918
93.5      0.00386719
94.5      0.00386745
95.5      0.0039057
96.5      0.00367034
97.5      0.00361572
98.5      0.00359091
99.5      0.0036522
}\resbos

\addplot [blue,thick] table [x={0},y={2}]{\xsec};
\addplot [red,thick] table [x={0},y={1}]{\xsec};
\addplot [black,semithick, dashed] table [x={0},y={1}]{\resbos};

\node[anchor=west, fill=white] (source) at (axis cs:18,0.023)
{ {\sc Deductor} (full) \& {\sc Deductor} (std.)};
\node(destination) at (axis cs:11,0.023){};
\draw[-stealth](source)--(destination);

\node[anchor=west, fill=white] (source) at (axis cs:60,0.008)
{{\sc ResBos}};
\node(destination) at (axis cs:65,0.006){};
\draw[-stealth](source)--(destination);

%\legend{{\sc Deductor} (full),  {\sc Deductor} (std.), {\sc ResBos}}
\end{axis}
\end{tikzpicture}
\end{center}

\caption{
The normalized Drell-Yan transverse momentum distribution, $dN/dQ_\LT = (1/\sigma)\, d\sigma/dQ_\LT$ for the LHC at 13 TeV. Here $Q_T$ is the transverse momentum of the $e^+$$e^-$ pair, $d\sigma/dQ_\LT$ is $d\sigma/(dQ_\LT\,dQ)$ integrated over $2 \TeV < Q < 2.1 \TeV$ and $\sigma$ is this cross section integrated over $0 < Q_\LT < 100 \GeV$, so that the area under the curve is 1. The red curve that is lowest at small $Q_\LT$ is the full result with threshold effects, $dN({\rm full})/dQ_\LT$. The blue curve that is very slightly higher at small $Q_\LT$ is the result with no threshold effects, $dN({\rm std.})/dQ_\LT$. In these calculations, we have chosen $\mu_\Lf = Q$. We also show the corresponding result obtained using \textsc{ResBos} \cite{ResBos1, ResBos} as a dashed black curve.
}
\label{fig:DYkT}
%-------------------- Figure -----------------------------
\end{figure}
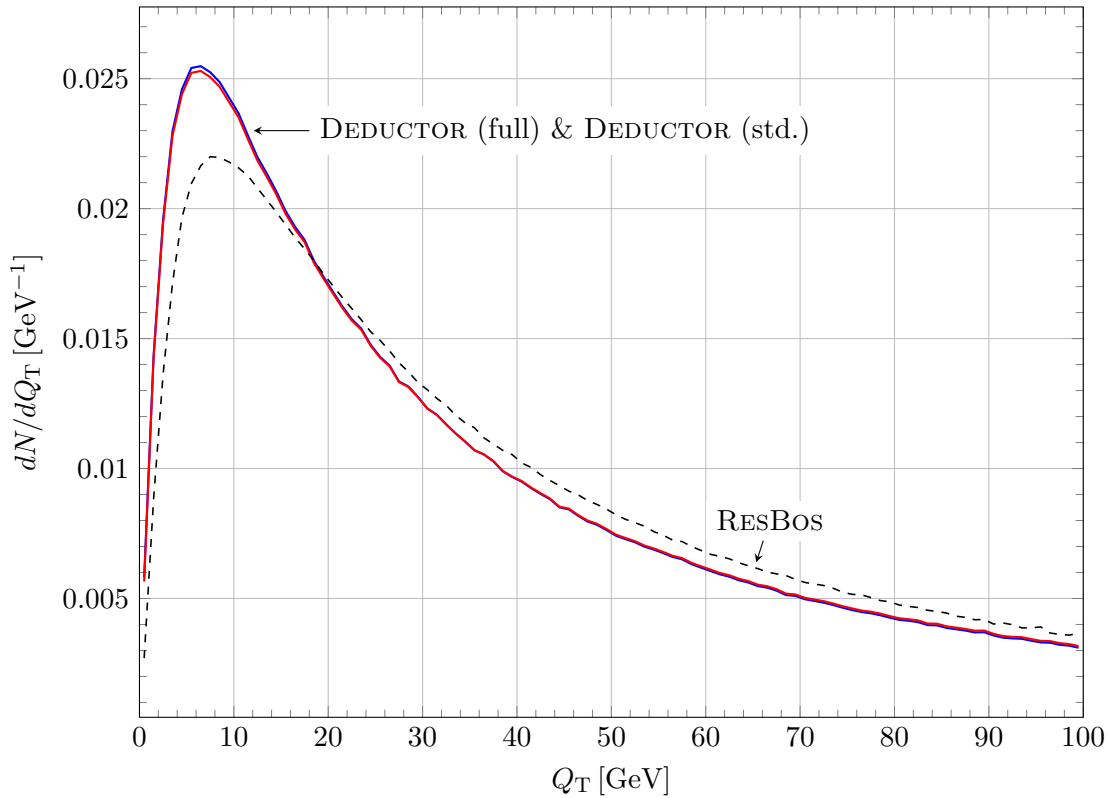

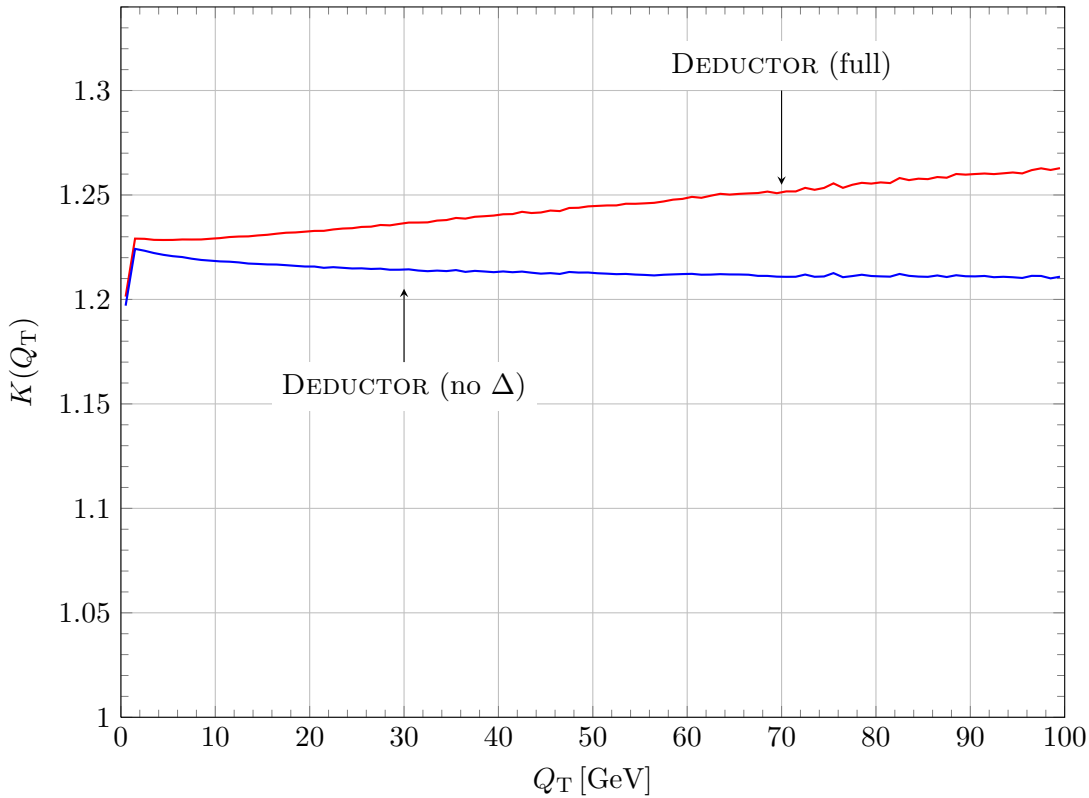
\begin{figure}
%-------------------- Figure -----------------------------
\begin{center}
\begin{tikzpicture}
\begin{axis}[title = {Drell-Yan cross section},
   xlabel={$Q_\LT\,\mathrm{[GeV]}$}, ylabel={$K(Q_\LT)$},
   xmin=0, xmax=100,
   ymin=1, ymax=1.34,
  legend cell align=left,
  legend style={at={(0.95,0.7)}},
  width=14cm,	
  height=11cm,
  xminorgrids=false,
  yminorgrids=false,
  minor x tick num=4,
  minor y tick num=4
]

\pgfplotstableread{  
 0.5	1.201354	1.197016
1.5	1.229112	1.224194
2.5	1.229013	1.223314
3.5	1.228516	1.222205
4.5	1.228456	1.221348
5.5	1.228486	1.220699
6.5	1.228702	1.220282
7.5	1.228676	1.219495
8.5	1.228704	1.218904
9.5	1.229053	1.218561
10.5	1.229356	1.218206
11.5	1.229861	1.218063
12.5	1.230107	1.217721
13.5	1.230178	1.217179
14.5	1.230625	1.216996
15.5	1.230936	1.216758
16.5	1.231464	1.216704
17.5	1.231924	1.216399
18.5	1.232067	1.216071
19.5	1.232429	1.215759
20.5	1.232801	1.215747
21.5	1.232832	1.215133
22.5	1.233497	1.215443
23.5	1.233933	1.215138
24.5	1.234096	1.214819
25.5	1.234705	1.214887
26.5	1.23483	       1.214591
27.5	1.235632	1.214725
28.5	1.235455	1.214211
29.5	1.236151	1.214227
30.5	1.236785	1.214403
31.5	1.236798	1.213833
32.5	1.236891	1.213523
33.5	1.237738	1.213801
34.5	1.237999	1.213564
35.5	1.239004	1.214052
36.5	1.238685	1.213188
37.5	1.23956 	1.213671
38.5	1.23983 	1.213379
39.5	1.240149	1.213047
40.5	1.240763	1.213398
41.5	1.240853	1.213025
42.5	1.241938	1.213329
43.5	1.24136	        1.2128
44.5	1.241621	1.212342
45.5	1.242542	1.212598
46.5	1.242283	1.212206
47.5	1.243779	1.213143
48.5	1.243872	1.212863
49.5	1.244539	1.212893
50.5	1.244767	1.212527
51.5	1.244987	1.212338
52.5	1.245004	1.212096
53.5	1.2458	        1.212209
54.5	1.24578 	1.211903
55.5	1.245992	1.211738
56.5	1.246242	1.211489
57.5	1.246899	1.211812
58.5	1.247757	1.211984
59.5	1.248119	1.212143
60.5	1.249095	1.212244
61.5	1.248645	1.21183
62.5	1.249641	1.21188
63.5	1.250548	1.212101
64.5	1.250186	1.21196
65.5	1.250524	1.211938
66.5	1.250747	1.21183
67.5	1.250924	1.211278
68.5	1.251623	1.211263
69.5	1.250851	1.210903
70.5	1.251709	1.210822
71.5	1.251689	1.210857
72.5	1.253401	1.211912
73.5	1.252476	1.210844
74.5	1.253371	1.210999
75.5	1.25556 	1.212607
76.5	1.253413	1.210644
77.5	1.254884	1.211183
78.5	1.255815	1.211823
79.5	1.255442	1.211216
80.5	1.256018	1.211032
81.5	1.255695	1.210906
82.5	1.258106	1.212196
83.5	1.257109	1.211294
84.5	1.257798	1.210942
85.5	1.257578	1.210817
86.5	1.258583	1.211471
87.5	1.258255	1.210725
88.5	1.260039	1.211583
89.5	1.259733	1.211098
90.5	1.260022	1.211032
91.5	1.260267	1.211241
92.5	1.259987	1.210642
93.5	1.260368	1.21084
94.5	1.260767	1.21062
95.5	1.260315	1.210273
96.5	1.261893	1.211286
97.5	1.26275 	1.211237
98.5	1.261968	1.21009
99.5	1.262841	1.210866
}\xsec

\addplot [red,thick] table [x={0},y={1}]{\xsec};
\addplot [blue,thick] table [x={0},y={2}]{\xsec};
%\legend{{\sc Deductor} (full),  {\sc Deductor} (std.), NLO}

\node[anchor=south, fill=white] (source) at (axis cs:70,1.3)
{{\sc Deductor} (full)};
\node(destination) at (axis cs:70,1.25){};
\draw[-stealth](source)--(destination);

\node[anchor=north, fill=white] (source) at (axis cs:30,1.17)
{{\sc Deductor} (no $\Delta$)};
\node(destination) at (axis cs:30,1.21){};
\draw[-stealth](source)--(destination);

\end{axis}
\end{tikzpicture}
\end{center}

\caption{
Ratios, $K$, of the Drell-Yan cross section $d\sigma/dQ_\LT$ obtained by integrating $d\sigma/(dQ_\LT\,dQ)$ over $2 \TeV < Q < 2.1 \TeV$ for the LHC at 13 TeV. The numerator in the upper curve is the full result with threshold factors, $d\sigma({\rm full})/dQ_\LT$. The numerator in the lower curve is the cross section $d\sigma({\rm no}\ \Delta)/dQ_\LT$ obtained by omitting the $\Delta$ term in the threshold factor. In each case the denominator is the cross section $d\sigma({\rm std.})/dQ_\LT$ with no threshold factors. The factorization scale in all of these cross sections is chosen to be  $\mu_\Lf = Q$.
}
\label{fig:DYkTratios}
%-------------------- Figure -----------------------------
\end{figure}

\subsection{Drell-Yan transverse momentum}

In the previous subsection, we examined the $Q$ dependence of the Drell-Yan cross section $d\sigma/dQ$, looking for the effects of steeply falling parton distribution functions when $Q$ is large. Parton shower event generators can also predict the distribution of the transverse momentum $Q_\LT$ of the $e^+ e^-$ pair in the region of small $Q_\LT/Q$, where logarithms of $Q_\LT/Q$ need to be summed. We have found \cite{NSpT} that a parton shower with virtuality based ordering, like \textsc{Deductor}, gives the same result at the next-to-leading-log level for logarithms of $Q_\LT/Q$ (without threshold logs) as the analytical treatment of ref.~\cite{CSSpT}. Now, with threshold effects included in a parton shower, we can examine both the logs of $Q_\LT/Q$ and the threshold effect at the same time, as in the analytical treatments of refs.~\cite{LSVJointR, KSVJointR, Li:2016axz, Lustermans:2016nvk}. We do not, however, have analytical knowledge of the level of accuracy of the parton shower treatment. 

To study the $Q_\LT$ distribution at large $Q$, we examine
\begin{equation}
\label{eq:dsigmadQT}
\frac{d\sigma}{d Q_\LT}
=
\int_{2.0 \TeV}^{2.1 \TeV}\!dQ\ \frac{d\sigma}{dQ\,dQ_\LT}
\;.
\end{equation}
We we divide by the integral of this over the $Q_\LT$ range $0 < Q_\LT < 100 \GeV$ to produce a distribution $dN/dQ_\LT$ normalized to 
\begin{equation}
\label{eq:QTnorm}
\int_0^{100 \GeV}\!dQ_\LT\ \frac{dN}{dQ_\LT} = 1
\;.
\end{equation}
In figure \ref{fig:DYkT}, we show the result with the threshold correction, $dN({\rm full})/dQ_\LT$. We choose the factorization scale to be $\mu_\Lf = Q$. For comparison, we show the result $dN({\rm std.})/dQ_\LT$ with the threshold correction omitted. We see that the threshold correction has only a very small effect. We also show the result $dN({\rm ResBos})/dQ_\LT$ obtained with the analytical summation of logs of $Q_\LT/Q$ contained in \textsc{ResBos} \cite{ResBos1, ResBos}. The \textsc{ResBos} result does not contain a summation of threshold logs and so should be compared to \textsc{Deductor} (std.).
The \textsc{ResBos} calculation contains smearing with non-perturbative functions that are fit to data. This smearing has not been included in \textsc{Deductor}. Thus it is not surprising that the \textsc{Deductor} distributions are somewhat narrower than the distribution from \textsc{ResBos}.

In figure \ref{fig:DYkTratios}, we examine directly $d\sigma/dQ_\LT$ defined in eq.~(\ref{eq:dsigmadQT}) so that we can see the effect of the threshold factors on the normalization of the cross section. We take $\mu_\Lf = Q$. We examine ratios $K$ obtained by dividing $d\sigma/dQ_\LT$ by the cross section obtained with no threshold corrections, $d\sigma({\rm std.})/dQ_\LT$. We show two curves. In the upper curve, the numerator of $K$ is the result with the full threshold correction, $d\sigma({\rm full})/dQ_\LT$. We see that there is a substantial, about 25\%, threshold enhancement. This enhancement is weakly $Q_\LT$ dependent,\footnote{There is a strong $Q_\LT$ dependence at about $Q_\LT = 1 \GeV$. This arises from the minimum $p_\LT$ allowed for emissions in the shower and is not really physical.} increasing from 23\% to 26\% over the range $0 < Q_\LT < 100 \GeV$. We examine where this $Q_\LT$ dependence comes from by plotting also the ratio $K$ obtained using $d\sigma({\rm no}\ \Delta)/dQ_\LT$ in the numerator. Without the $\Delta$ contribution in the Sudakov exponent, the threshold enhancement is flat as a function of $Q_\LT$. Thus the small $Q_\LT$ dependence seen in $d\sigma({\rm full})/dQ_\LT$ comes mainly from the $\Delta$ term in the Sudakov exponent. This is easy to understand. The $\Delta$ term appears only after we have an initial state emission. Having an initial state emission gives a transverse momentum recoil to the $e^+ e^-$ pair, so larger $Q_\LT$ should have a positive correlation with a larger threshold factor from the $\Delta$ term.

We can offer two observations. First, the \textsc{Deductor} (std.) curve for ${dN}/(dQ_\LT)$ agrees nicely with the \textsc{ResBos} curve, considering that there should be differences from non-perturbative smearing. Second, the effect of threshold logs reflected in the \textsc{Deductor} (full) curves in figures \ref{fig:DYkT} and \ref{fig:DYkTratios} is small and its sign appears to us to be quite sensible.

\subsection{Jets}

\begin{figure}
%-------------------- Figure -----------------------------
\begin{center}
\begin{tikzpicture}
\begin{semilogyaxis}[title = {One jet inclusive cross section},
   xlabel={$P_\LT\,\mathrm{[TeV]}$}, ylabel={$d\sigma/dP_\LT\,\mathrm{[nb/GeV]}$},
   xmin=0, xmax=4000,
   ymin=1e-11, ymax=1,
  legend cell align=left,
  width=14cm,	
  height=11cm,
  x coord trafo/.code={
    \pgflibraryfpuifactive{
    \pgfmathparse{(#1)/(1000)}
  }{
     \pgfkeys{/pgf/fpu=true}
     \pgfmathparse{(#1)/(1000)}
     \pgfkeys{/pgf/fpu=false}
   }
 },
 xminorgrids=false,
 yminorgrids=false,
 minor x tick num=4
 %ytickten={-11,-9,...,-1},
]

\pgfplotstableread{  
300.	0.1653886	0.1343367	0.2225
400.	0.03706564	0.029705	0.0434
500.	0.009656414	0.007630301	0.01145
600.	0.003097295	0.002415791	0.003697
700.	0.001146672	0.00088015	0.001372
800.	0.0004624991	0.0003501973	0.0005667
900.	0.0002074315	0.000155166	0.0002502
1000.	0.00009766637	0.00007219534	0.0001175
1100.	0.00004842463	0.00003530886	0.0000588
1200.	0.00002514599	0.00001811828	0.00003058
1300.	0.00001352146	9.620043e-6	0.00001646
1400.	7.394166e-6	5.192994e-6	9.108e-6
1500.	4.195198e-6	2.9091e-6	5.117e-6
1600.	2.44036e-6	1.67136e-6	2.949e-6
1700.	1.447869e-6	9.785723e-7	1.722e-6
1800.	8.566396e-7	5.710588e-7	1.027e-6
1900.	5.24151e-7	3.451373e-7	6.21e-7
2000.	3.171256e-7	2.060246e-7	3.782e-7
2100.	1.955349e-7	1.252569e-7	2.317e-7
2200.	1.213029e-7	7.653455e-8	1.444e-7
2300.	7.655961e-8	4.765446e-8	8.944e-8
2400.	4.793326e-8	2.935761e-8	5.584e-8
2500.	3.020853e-8	1.822934e-8	3.501e-8
2600.	1.888834e-8	1.122442e-8	2.223e-8
2700.	1.209238e-8	7.077261e-9	1.397e-8
2800.	7.499057e-9	4.30921e-9	8.777e-9
2900.	4.771755e-9	2.699815e-9	5.522e-9
3000.	2.997824e-9	1.666519e-9	3.435e-9
3100.	1.905721e-9	1.040968e-9	2.194e-9
3200.	1.191893e-9	6.392307e-10	1.362e-9
3300.	7.489374e-10	3.939758e-10	8.457e-10
3400.	4.600686e-10	2.372244e-10	5.202e-10
3500.	2.871026e-10	1.451444e-10	3.25e-10
3600.	1.768839e-10	8.760783e-11	1.974e-10
3700.	1.078654e-10	5.226937e-11	1.198e-10
3800.	6.504836e-11	3.077432e-11	7.302e-11
}\xsec

\addplot [red, semithick] table [x={0},y={1}]{\xsec};
\addplot [blue,thick] table [x={0},y={2}]{\xsec};
\addplot [black,thick,dashed] table [x={0},y={3}]{\xsec};
 
\node[anchor=west, fill=white] (source) at (axis cs:1800,1e-5)
{{\sc Deductor} (full) \& NLO};
\node(destination) at (axis cs:1400,1e-5){};
\draw[-stealth](source)--(destination);

\node[anchor=east, fill=white] (source) at (axis cs:2200,1e-8)
{{\sc Deductor} (std.)};
\node(destination) at (axis cs:2600,1e-8){};
\draw[-stealth](source)--(destination);

%\legend{ {\sc Deductor} (full), {\sc Deductor} (std.), NLO}
\end{semilogyaxis}
\end{tikzpicture}
\end{center}
\caption{
One jet inclusive cross section $d\sigma/dP_\LT$ for the production of a jet with transverse momentum $P_\LT$ and rapidity in the range $-2 < y < 2$. The cross section is for the LHC at 13 TeV. We use the anti-$k_\LT$ algorithm \cite{antikt} with $R = 0.4$. The lower, blue curve is $d\sigma({\rm std.})/dP_\LT$, obtained with no threshold effects. The red curve is $d\sigma({\rm full})/dP_\LT$, obtained with threshold effects. The dashed, black curve is an NLO calculation \cite{EKS}. In each case, we take the renormalization and factorization scales and the starting scale of the shower to be $\mu_\Lf = \mu_\Lr = P_\LT$.
}
\label{fig:jetcrosssection}
%-------------------- Figure -----------------------------
\end{figure}
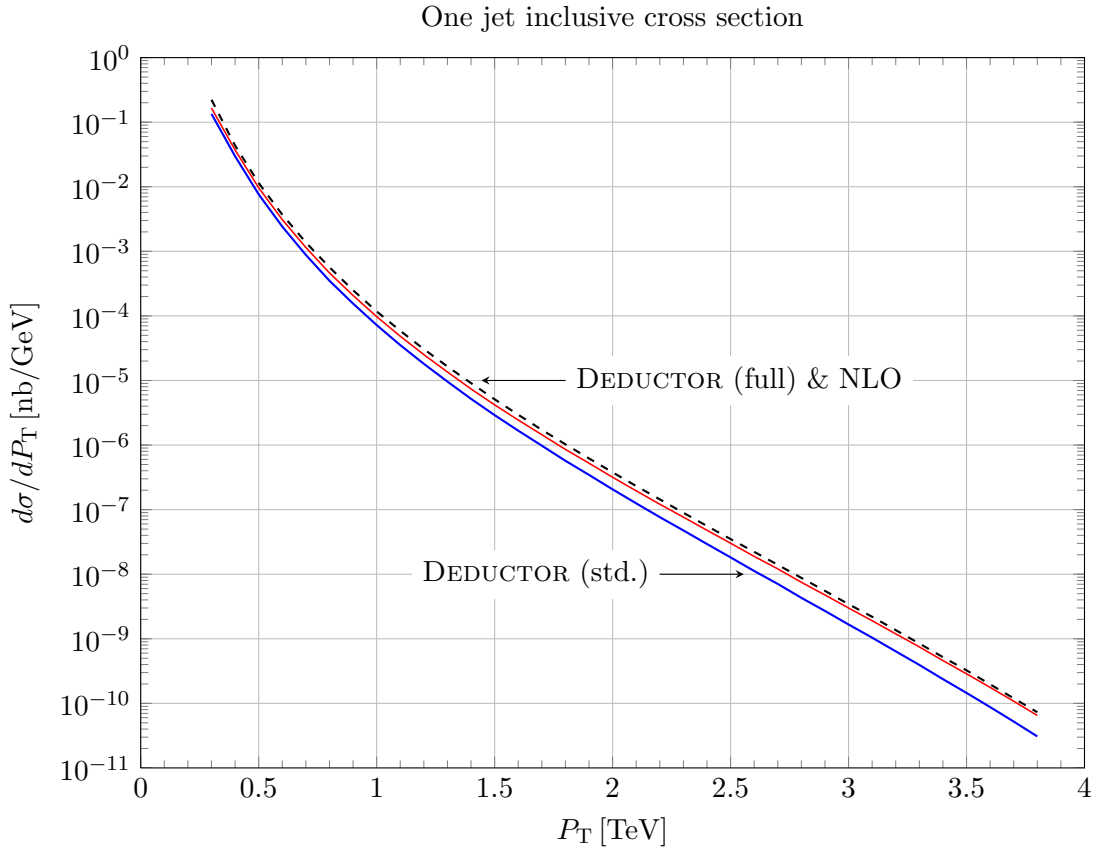

We now examine the one jet inclusive cross section $d\sigma/dP_\LT$ in proton-proton collisions at $\sqrt s = 13 \TeV$ as a function of the jet transverse momentum, $P_\LT$, integrated over the rapidity range $-2 < y < 2$. The jet is defined using the anti-$k_\LT$ algorithm \cite{antikt} with $R = 0.4$ with the aid of \textsc{FastJet} \cite{FastJet}. Notice that the cross section $d\sigma({\rm std.})/dP_\LT$ obtained with a standard shower with no threshold corrections is not the same as the Born cross section $d\sigma({\rm LO})/dP_\LT$ because partons generated in the shower from initial state splittings can become part of the jet, while partons generated as daughters of the starting final state partons can escape from the jet.

In figure \ref{fig:jetcrosssection}, we display three versions of $d\sigma/dP_\LT$ as functions of $P_\LT$. In each case, we take the renormalization and factorization scales and the starting scale of the shower to be $\mu_\Lf = \mu_\Lr = P_\LT$. The lower, blue curve is $d\sigma({\rm std.})/dP_\LT$, obtained with the parton shower with threshold effects omitted. The solid red curve is $d\sigma({\rm full})/dP_\LT$, obtained with threshold effects. We see that the threshold effect is large enough that it is evident even in this semilog plot. The black, dashed curve is the result of a purely perturbative next-to-leading order (NLO) calculation \cite{EKS}. We note that the parton shower calculation including the threshold effect is quite close to the NLO result.

There is a fairly substantial theoretical uncertainty associated with the parton shower calculation. To estimate this uncertainty, we examine the effect of changing the scale  $\mu_\Lf = \mu_\Lr$ at which the initial parton distributions and strong coupling are evaluated and at which shower evolution starts. In figure \ref{fig:jetcrosssection}, we used $\mu_\Lf = \mu_\Lr = P_\LT$. However, the minimum value of the dijet mass in the Born process is $Q = 2 P_\LT$. Thus $\mu_\Lf = \mu_\Lr = 2 P_\LT$ might seem a sensible choice. One the other hand, jet cross sections are sometimes evaluated with $\mu_\Lf = \mu_\Lr = P_\LT/2$, so $P_\LT/2$ might seem a sensible choice. In figure \ref{fig:jetscales}, we plot the ratios of $d\sigma({\rm full})/dP_\LT$ with $\mu_\Lf = \mu_\Lr = 2 P_\LT$ and with $\mu_\Lf = \mu_\Lr = P_\LT/2$ to $d\sigma({\rm full})/dP_\LT$ with $\mu_\Lf = \mu_\Lr = P_\LT$. Based on this result, we estimate a $\pm 30\%$ uncertainty in $d\sigma({\rm full})/dP_\LT$. This uncertainty could be reduced by performing a showered calculation matched to the NLO calculation.

\begin{figure}
%-------------------- Figure -----------------------------
\begin{center}
\begin{tikzpicture}
\begin{axis}[title = {Scale dependence of one jet inclusive cross section},
   xlabel={$P_\LT\,\mathrm{[TeV]}$}, ylabel={$[d\sigma(\lambda)/dP_\LT]/[d\sigma(1)/dP_\LT]$},
   xmin=0, xmax=4000,
   ymin=0.6, ymax=1.6,
  legend style={at={(0.4,0.97)}},
  legend cell align=left,
  width=14cm,	
  height=11cm,
  x coord trafo/.code={
    \pgflibraryfpuifactive{
    \pgfmathparse{(#1)/(1000)}
  }{
     \pgfkeys{/pgf/fpu=true}
     \pgfmathparse{(#1)/(1000)}
     \pgfkeys{/pgf/fpu=false}
   }
 },
 xminorgrids=false,
 yminorgrids=false,
 minor x tick num=4,
 minor y tick num=4
]

\addplot[blue, thick,name path=pluserror,no markers,domain=300:3800] {1.25432 + (0.0000835479 - 1.07399e-8*x)*x};
\addplot[green!60!black, thick,name path=minuserror,no markers,domain=300:3800]{0.782232 - (0.0000171292 - 5.83559e-9*x)*x};

%%%%%%% BAND IS OPTIONAL >>>>>>
%\addplot[forget plot,fill=lightgray!40!white, opacity=0.5] fill between[on layer={},of=pluserror and minuserror];
%%%%%%% BAND IS OPTIONAL <<<<<<

\addplot [black,semithick,forget plot,domain=300:3800]{1.0};
%\legend{{\sc Deductor} $\lambda=\frac12$,{\sc Deductor} $\lambda=2$}

\node[anchor=north, fill=white] (source) at (axis cs:3000,1.32)
{{\sc Deductor} $\lambda=\frac12$};
\node(destination) at (axis cs:3000,1.41){};
\draw[-stealth](source)--(destination);

\node[anchor=south, fill=white] (source) at (axis cs:1000,0.85)
{{\sc Deductor} $\lambda=2$};
\node(destination) at (axis cs:1000,0.76){};
\draw[-stealth](source)--(destination);

\end{axis}
\end{tikzpicture}
\end{center}

\caption{
Scale dependence of the one jet inclusive cross section after showering and threshold effects. We plot as functions of $P_\LT$ the ratios $[d\sigma({\rm full},\lambda)/dP_\LT]/[d\sigma({\rm full},1)/dP_\LT]$ with $\mu_\Lf = \mu_\Lr = \lambda P_\LT$ in the numerator and $\mu_\Lf = \mu_\Lr = P_\LT$ in the denominator. From top to bottom, the three curves are for $\lambda = 0.5$, $\lambda = 1$ and $\lambda = 2$. 
}
\label{fig:jetscales}
%-------------------- Figure -----------------------------
\end{figure}
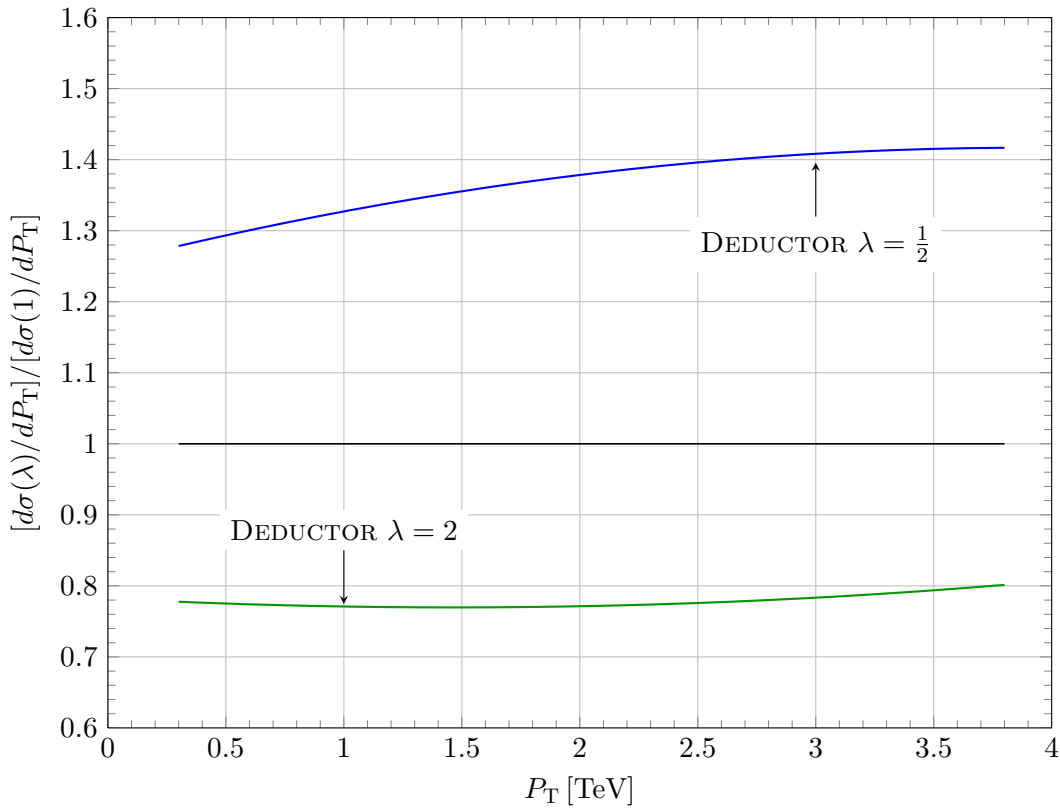

In figure \ref{fig:jetK}, we turn to several calculations of $d\sigma/dP_\LT$ presented as ratios $K$ to the perturbative Born cross section, $d\sigma({\rm LO})/dP_\LT$. In this figure, all cross sections are evaluated at $\mu_\Lf = \mu_\Lr = P_\LT$. 

The lowest, blue curve is $K({\rm std.})$, obtained using $d\sigma({\rm std.})/dP_\LT$, in which there is a standard shower but the threshold effects are turned off. We see that $d\sigma({\rm std.})/dP_\LT$ is only about 60\% of the Born cross section. Since the cross section is so steeply falling as a function of $P_\LT$, just a small amount of $P_\LT$ leakage out of the jet because of showering makes the cross section substantially smaller.

We now include the threshold correction, plotting the ratio $K({\rm full})$. This gives the red curve. We see that the threshold effect is very large and multiplies $d\sigma({\rm std.})/dP_\LT$ by a factor between 1.3 and 2 for $P_\LT > 1 \TeV$. This produces a result $d\sigma({\rm full})/dP_\LT$ that ranges from 90\% to 120\% of the Born cross section for $P_\LT > 1 \TeV$. We also show, as a solid green curve, the factor $K({\rm no}\ \Delta)$ obtained by omitting the term proportional to $\Delta_{\La k}$ in the Sudakov exponent. We see that, for the jet cross section, this term makes a contribution to the exponent that is not negligible.

We show also as a black, dashed curve, $K({\rm NLO})$ corresponding to the perturbative NLO calculation from figure \ref{fig:jetcrosssection}.

There are analytic summations of threshold logarithms \cite{SudakovFactorization, KidonakisOderdaSterman, KidonakisOwensJets, deFlorianVogelsang, deFlorianJets}. We have used the computer programs of Kidonakis and Owens \cite{KidonakisOwensJets} and of de Florian, Hinderer, Mukherjee, Ringer and Vogelsang \cite{deFlorianJets} to produce the cross sections $d\sigma({\rm K.O.})/dP_\LT$ and $d\sigma({\rm FHMRV})/dP_\LT$, respectively. In these programs, the all-order threshold effect is expanded to order $\as^2$ to produce the calculated cross sections.  In the Kidonakis-Owens formulation of threshold summation, there is no dependence on the algorithm used to define the jet. The FHMRV calculation is more sophisticated and includes the dependence on the jet algorithm (for which we use the anti-$k_\LT$ algorithm with $R = 0.4$). In figure \ref{fig:jetK}, we plot $K({\rm K.O.})$ as the green, dashed curve and $K({\rm FHMRV})$ as the purple, dashed curve.

We note that the \textsc{Deductor} (full) curve is roughly 30\% below the analytic FHMRV curve for $P_\LT > 2 \TeV$. We also note that the scale variation test in figure \ref{fig:jetscales} suggests that the {\textsc Deductor} (full) curve in figure \ref{fig:jetK} should be regarded as being uncertain to $\pm 30 \%$. In fact, we see in figure \ref{fig:jetscales} that changing the scales to $\mu_\Lf = \mu_\Lr = P_\LT/2$ makes the \textsc{Deductor} (full) cross section roughly 30\% bigger. Thus the level of agreement between the \textsc{Deductor} (full) and FHMRV curves seems not unreasonable. It would, of course, be desirable to improve the precision of the shower cross section. This can be achieved by matching the shower calculation to the perturbative NLO correction to the cross section, but we have not yet undertaken this task.

\begin{figure}
%-------------------- Figure -----------------------------
\begin{center}
\begin{tikzpicture}
\begin{axis}[title = {Ratios of one jet inclusive cross section},
   xlabel={$P_\LT\,\mathrm{[TeV]}$}, ylabel={$K(P_\LT)$},
   xmin=0, xmax=4000,
   ymin=0.4, ymax=1.6,
  legend style={at={(0.99,0.48)}},
  legend cell align=left,
  width=14cm,	
  height=11cm,
  x coord trafo/.code={
    \pgflibraryfpuifactive{
    \pgfmathparse{(#1)/(1000)}
  }{
     \pgfkeys{/pgf/fpu=true}
     \pgfmathparse{(#1)/(1000)}
     \pgfkeys{/pgf/fpu=false}
   }
 },
 xminorgrids=false,
 yminorgrids=false,
 minor x tick num=4,
 minor y tick num=4
]

\pgfplotstableread{  
  3.000000E+02    7.674850E+01  5.546433E+01  5.185896E+01
  4.000000E+02    1.421291E+01  1.069771E+01  9.968846E+00
  5.000000E+02    3.657718E+00  2.829493E+00  2.625802E+00
  6.000000E+02    1.157285E+00  9.128755E-01  8.439009E-01
  7.000000E+02    4.226511E-01  3.389261E-01  3.120649E-01
  8.000000E+02    1.717208E-01  1.395706E-01  1.279734E-01
  9.000000E+02    7.544097E-02  6.203423E-02  5.658198E-02
  1.000000E+03    3.530728E-02  2.934168E-02  2.663040E-02
  1.100000E+03    1.738327E-02  1.459335E-02  1.317032E-02
  1.200000E+03    8.927096E-03  7.560978E-03  6.784811E-03
  1.300000E+03    4.752090E-03  4.056798E-03  3.618204E-03
  1.400000E+03    2.612175E-03  2.245235E-03  1.990584E-03
  1.500000E+03    1.468547E-03  1.268283E-03  1.117336E-03
  1.600000E+03    8.430769E-04  7.328800E-04  6.413904E-04
  1.700000E+03    4.929805E-04  4.305788E-04  3.742689E-04
  1.800000E+03    2.922396E-04  2.564211E-04  2.212942E-04
  1.900000E+03    1.758695E-04  1.548979E-04  1.328090E-04
  2.000000E+03    1.067734E-04  9.431057E-05  8.030056E-05
  2.100000E+03    6.535941E-05  5.791791E-05  4.899699E-05
  2.200000E+03    4.050599E-05  3.593829E-05  3.018553E-05
  2.300000E+03    2.514176E-05  2.235903E-05  1.864498E-05
  2.400000E+03    1.568507E-05  1.398801E-05  1.158789E-05
  2.500000E+03    9.837599E-06  8.782475E-06  7.230093E-06
  2.600000E+03    6.177560E-06  5.517126E-06  4.507843E-06
  2.700000E+03    3.888066E-06  3.475283E-06  2.818828E-06
  2.800000E+03    2.447807E-06  2.189890E-06  1.764426E-06
  2.900000E+03    1.542020E-06  1.379038E-06  1.102261E-06
  3.000000E+03    9.681946E-07  8.666898E-07  6.883335E-07
  3.100000E+03    6.069212E-07  5.432059E-07  4.282223E-07
  3.200000E+03    3.805088E-07  3.402525E-07  2.663265E-07
  3.300000E+03    2.367336E-07  2.116890E-07  1.643406E-07
  3.400000E+03    1.468669E-07  1.312006E-07  1.011930E-07
  3.500000E+03    9.049092E-08  8.082744E-08  6.179179E-08
  3.600000E+03    5.543989E-08  4.946813E-08  3.754678E-08
  3.700000E+03    3.377952E-08  3.011089E-08  2.266200E-08
  3.800000E+03    2.034601E-08  1.811400E-08  1.352085E-08
%  3.900000E+03    1.216162E-08  1.081108E-08  8.007273E-09
%  4.000000E+03    7.197053E-09  6.383155E-09  4.687318E-09
%  4.100000E+03    4.189989E-09  3.710215E-09  2.700332E-09
%  4.200000E+03    2.418085E-09  2.135839E-09  1.539766E-09
%  4.300000E+03    1.374273E-09  1.210863E-09  8.647939E-10
%  4.400000E+03    7.656385E-10  6.727897E-10  4.759394E-10
%  4.500000E+03    4.194746E-10  3.676263E-10  2.574689E-10
%  4.600000E+03    2.242777E-10  1.956691E-10  1.355852E-10
%  4.700000E+03    1.173455E-10  1.021167E-10  7.000059E-11
%  4.800000E+03    5.980683E-11  5.184780E-11  3.515174E-11
%  4.900000E+03    2.951455E-11  2.546666E-11  1.705406E-11
%  5.000000E+03    1.409291E-11  1.211001E-11  8.003033E-12
}\vogelsang

\pgfplotstableread {
    300.00    0.7301E-01    0.5594E-01
    400.00    0.1371E-01    0.1068E-01
    500.00    0.3502E-02    0.2767E-02
    600.00    0.1094E-02    0.8729E-03
    700.00    0.3965E-03    0.3180E-03
    800.00    0.1604E-03    0.1291E-03
    900.00    0.7062E-04    0.5700E-04
   1000.00    0.3317E-04    0.2685E-04
   1100.00    0.1639E-04    0.1331E-04
   1200.00    0.8443E-05    0.6871E-05
   1300.00    0.4497E-05    0.3667E-05
   1400.00    0.2465E-05    0.2013E-05
   1500.00    0.1384E-05    0.1131E-05
   1600.00    0.7937E-06    0.6482E-06
   1700.00    0.4631E-06    0.3779E-06
   1800.00    0.2743E-06    0.2235E-06
   1900.00    0.1645E-06    0.1339E-06
   2000.00    0.9974E-07    0.8098E-07
   2100.00    0.6102E-07    0.4941E-07
   2200.00    0.3761E-07    0.3037E-07
   2300.00    0.2332E-07    0.1877E-07
   2400.00    0.1453E-07    0.1165E-07
   2500.00    0.9084E-08    0.7255E-08
   2600.00    0.5695E-08    0.4528E-08
   2700.00    0.3576E-08    0.2830E-08
   2800.00    0.2247E-08    0.1769E-08
   2900.00    0.1411E-08    0.1105E-08
   3000.00    0.8854E-09    0.6897E-09
   3100.00    0.5545E-09    0.4293E-09
   3200.00    0.3462E-09    0.2664E-09
   3300.00    0.2154E-09    0.1646E-09
   3400.00    0.1334E-09    0.1012E-09
   3500.00    0.8216E-10    0.6189E-10
   3600.00    0.5028E-10    0.3758E-10
   3700.00    0.3055E-10    0.2265E-10
   3800.00    0.1840E-10    0.1353E-10
%   3900.00    0.1098E-10    0.8000E-11
%   4000.00    0.6485E-11    0.4678E-11
%   4100.00    0.3783E-11    0.2702E-11
%   4200.00    0.2177E-11    0.1539E-11
%   4300.00    0.1235E-11    0.8628E-12
%   4400.00    0.6882E-12    0.4754E-12
%   4500.00    0.3765E-12    0.2569E-12
%   4600.00    0.2016E-12    0.1358E-12
%   4700.00    0.1054E-12    0.7001E-13
%   4800.00    0.5368E-13    0.3511E-13
%   4900.00    0.2650E-13    0.1706E-13
%   5000.00    0.1263E-13    0.7996E-14
}\kidow

\addplot [red,thick, domain=300:3800]{0.830114 + (0.0000324967 + 1.72707e-8*x)*x};
\addplot [blue,thick, domain=300:3800]{0.652178 + (9.38881e-6 - 8.90194e-9*x)*x};
\addplot [green!60!black,thick, domain=300:3800]{0.723559 + (0.000065642 + 6.58021e-9*x)*x};
\addplot [black, domain=300:3800]{1};
\addplot [black,thick, dashed, domain=300:3800]{1.03637 + (0.0000515808 + 7.16897e-9*x)*x};
\addplot [green!60!black,thick,dashed] table [x={0},y expr=\thisrow{1}/\thisrow{2}]{\kidow};
\addplot [purple,thick,dashed] table [x={0},y expr=\thisrow{1}/\thisrow{3}]{\vogelsang};
%\legend{{\sc Deductor} (full),{\sc Deductor} (std.), LO, NLO, analytic (K.O.), analytic (FHMRV)}

\node[anchor=north, fill=white] (source) at (axis cs:1000,0.55){{\sc Deductor} (std.)};
\node(destination) at (axis cs:1000,0.66){};
\draw[-stealth](source)--(destination);

\node[anchor=north, fill=white] (source) at (axis cs:2300,1.1){NLO};
\node(destination) at (axis cs:1750,1.15){};
\draw[-stealth](source)--(destination);

\node[anchor=south, fill=white] (source) at (axis cs:600,0.9){{\sc Deductor} (full)};
\node(destination) at (axis cs:1400,0.92){};
\draw[-stealth](source)--(destination);

\node[anchor=south, fill=white] (source) at (axis cs:2900,0.8){{\sc Deductor} (no $\Delta$)};
\node(destination) at (axis cs:2000,0.88){};
\draw[-stealth](source)--(destination);

\node[anchor=south, fill=white] (source) at (axis cs:2500,1.45){analytic (FHMRV)};
\node(destination) at (axis cs:2500,1.35){};
\draw[-stealth](source)--(destination);

\node[anchor=south, fill=white] (source) at (axis cs:1000,1.45){analytic (K.O.)};
\node(destination) at (axis cs:1000,1.23){};
\draw[-stealth](source)--(destination);

\end{axis}
\end{tikzpicture}
\end{center}

\caption{
Illustrations of the one jet inclusive cross section as in figure \ref{fig:jetcrosssection}. We show the results of calculations of $d\sigma/dP_\LT$ by plotting ratios $K$ of $d\sigma/dP_\LT$ to the perturbative Born cross section $d\sigma({\rm LO})/dP_\LT$. The scales in all cross sections are $\mu_\Lf = \mu_\Lr = P_\LT$. Going from the lowest to the highest curves at $P_\LT = 2 \TeV$, the lowest curve is $K({\rm std.})$ using in the numerator the showered cross section obtained with threshold effects turned off. The next is $K({\rm no}\ \Delta)$ obtained with the $\Delta_{\La k}$ contribution turned off.  The next curve is $K({\rm full})$ using the showered cross section with threshold effects. The next is $K({\rm LO}) = 1$. The next curve is $K({\rm NLO})$ using the perturbative NLO cross section.  The next highest curve is $K({\rm K.O.})$ using the cross section obtained using the Kidonakis-Owens threshold effects program \cite{KidonakisOwensJets}. The highest curve is $K({\rm FHMRV})$ using the cross section calculated with the de Florian, Hinderer, Mukherjee, Ringer and Vogelsang  algorithm \cite{deFlorianJets}.
}
\label{fig:jetK}
%-------------------- Figure -----------------------------
\end{figure}
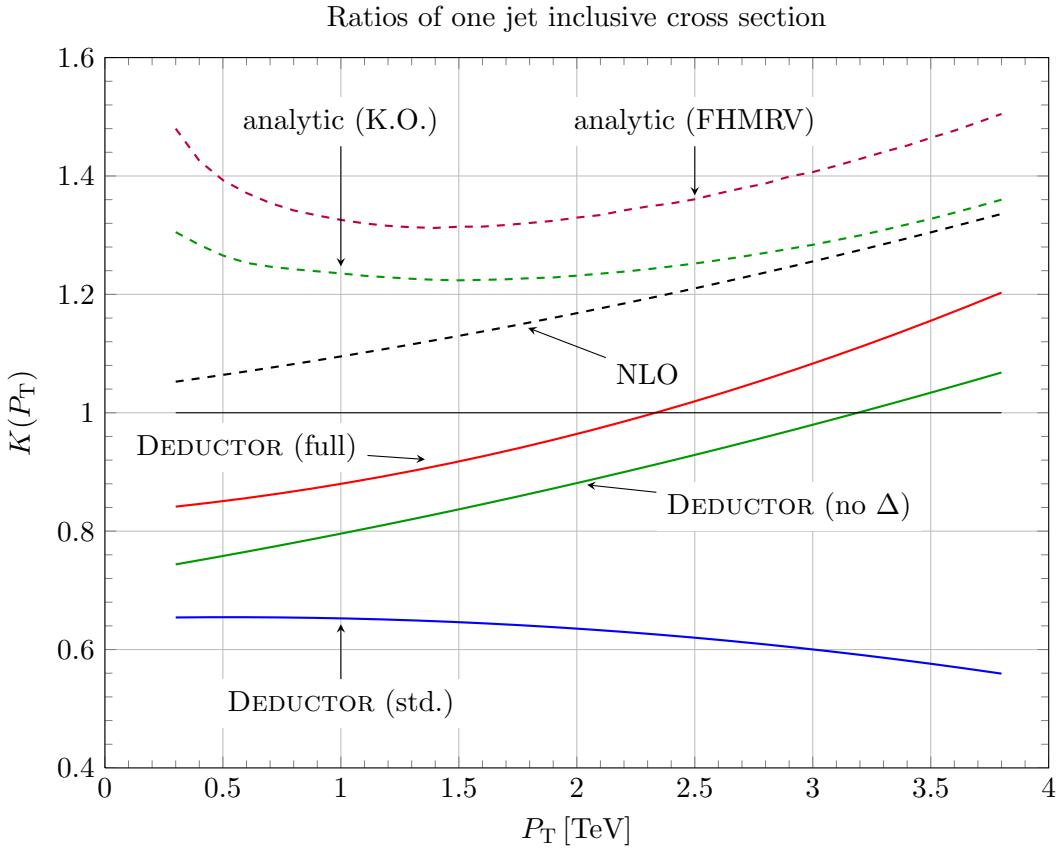

We can draw a further conclusion from these comparisons. The parton shower has a hard job to perform, since it needs to include two large effects that act in opposite directions: loss of $P_\LT$ from the jet from showering and also the threshold enhancement. It seems to us remarkable that the calculation works to within the expected uncertainty.

\section{Summary of the analysis}
\label{sec:summary}

We have viewed parton shower evolution as the solution of equation~(\ref{eq:evolution}),
\begin{equation}
\label{eq:evolutionencore}
\frac{d}{dt}\,\sket{\rho(t)} = [{\cal H}_I(t) - \Vvirt(t)]\sket{\rho(t)}
\;,
\end{equation}
where $\sket{\rho(t)}$ represents the probability distribution of parton flavors and momenta and the density matrix in the quantum color and spin space in a statistical ensemble of event generation trials \cite{NSI, NSII, NSspin, NScolor, Deductor, ShowerTime, PartonDistFctns, ColorEffects}. In this paper, we have ignored spin but still consider a full treatment of quantum color, even though in an actual implementation as computer code one has to make some approximations with respect to color. 

The shower time $t$ is the negative logarithm of the hardness scale $\mu^2$ considered. The shower starts at a hard interaction and evolves to softer scales. At shower time $t$, interactions that are softer than $\mu^2$ are regarded as unresolvable, so that it is not meaningful to measure properties of the states described by $\sket{\rho(t)}$ at a finer scale. In the evolution equation, ${\cal H}_I(t)$ represents real parton splittings, while $\Vvirt(t)$ leaves the number of partons, their momenta, and their flavors unchanged. Both operators are order $\as$; we do not examine higher order contributions. 

The operator $\Vvirt(t)$ gives us the Sudakov factor $\exp(-\int_{t_1}^{t_2}\!dt\, S(t))$ that appears between two parton splittings. We have argued that $\Vvirt(t)$ should consist of two parts, as in eq.~(\ref{eq:VfromVpertandF}),
\begin{equation}
\begin{split}
\label{eq:VfromVpertandFbis}
\Vvirt(t)
 ={}& 
\Vvirt^{\rm pert}(t)
- {\cal F}(t)^{-1}\left[\frac{d}{dt}\,{\cal F}(t)\right]
\;.
\end{split}
\end{equation}
Here ${\cal F}(t)$ represents the parton distribution factor in $\sket{\rho(t)}$, so that the second term in $\Vvirt(t)$ gives the effect of changing the scale parameter in the parton distributions. The first term, $\Vvirt^{\rm pert}(t)$, accounts for order $\as$ graphs that leave the number of partons unchanged. That is, $\Vvirt^{\rm pert}(t)$ represents one loop virtual graphs. We have (approximately) calculated $\Vvirt^{\rm pert}(t)$ in this paper.

In order to construct a parton shower based on eq.~(\ref{eq:evolutionencore}), one can use a trick. One can replace $\Vvirt(t)$ by $\Vreal(t)$, where $\Vreal(t)$ is constructed from the splitting operator ${\cal H}_I(t)$ in such a way that the Born level cross section contained in $\sket{\rho(0)}$ is exactly conserved by the shower evolution. In fact, parton shower algorithms are typically based on $\Vreal(t)$ instead of $\Vvirt(t)$. The difference $[\Vreal(t) - \Vvirt(t)]$ corrects $-\Vreal(t)$. If we approximate the color using the leading color (LC) approximation or the LC+ approximation \cite{NScolor}, then the color matrices are diagonal and $\exp(\int_{t_1}^{t_2}\!d t\,[\Vreal(t) - \Vvirt(t)])$ gives us a numerical weight factor that adjusts the cross section.

We found that $\Vvirt^{\rm pert}(t)$ contains factors $\pm \mi \pi$ times certain color matrices. These terms are very well known. (See, for example ref.~\cite{Manchester2005}.) As noted in ref.~\cite{NScolor}, these terms conserve the Born level cross section and could be included in $\Vreal(t)$ instead of $\Vvirt(t)$. Although the $\mi \pi$ terms are of considerable physical interest, they are of secondary interest in this paper and have not been included in our numerical results.

The integrand in $[\Vreal(t) - \Vvirt(t)]$ contains ratios of parton distribution functions. The most important term has the form 
\begin{equation}
\label{eq:thresholdterm}
\frac{\as}{2\pi}
\left[1 - \frac{f_{a/A}(\eta_\La/z,\mu^2)} {f_{a/A}(\eta_\La,\mu^2)}\right]
\frac{1}{1-z}
\end{equation}
integrated over $z$ in a small range near $z = 1$. This term is large if $f_{a/A}(\eta_\La/z,\mu^2)$ falls steeply as $1-z$ increases. Furthermore, $f_{a/A}(\eta_\La/z,\mu^2)$ {\em does} fall steeply as $1-z$ increases when $\eta_\La$ is large, say bigger than 0.1. Thus a straightforward analysis of parton shower evolution leads us to the conclusion that there can be large corrections to the Born level cross section for a hard process. These contributions are naturally summed in a parton shower algorithm that incorporates $\Vvirt(t)$ because $[\Vreal(t) - \Vvirt(t)]$ appears as part of the Sudakov exponent. 

Of course, these contributions are not summed if the Sudakov factor is $\exp[-\int \!dt\,\Vreal(t)]$, as is customary in parton showers (including ours, \textsc{Deductor 1.0}). In this paper, we have presented results from \textsc{Deductor 2.0} \cite{DeductorCode}, in which $[\Vreal(t) - \Vvirt(t)]$ is included within the LC+ approximation. (In our numerical results, we used the LC approximation.)

The effects that arise from the term in eq.~(\ref{eq:thresholdterm}) are clearly connected with the effects of what are usually called threshold logarithms, which have been extensively studied. It thus is puzzling that the formulation in this paper contains ratios of parton distribution functions in an exponent, whereas standard threshold summation results never, to our knowledge, contain such factors. How, then, can these formulations be connected? 

The answer can be understood in the treatment of the DGLAP parton evolution equation in the formulation of a parton shower. In a parton shower describing hadron-hadron collisions, at hardness scale $\mu^2$, one needs to include ``unresolvable'' initial state interactions as parton distribution factors, $f_{a/A}(\eta_\La,\mu^2)$ and $f_{b/B}(\eta_\Lb,\mu^2)$. When we come to a softer scale, we need to cancel the previous parton distribution functions and supply new ones. For this purpose, we can use
\begin{equation}
f_{a/A}(\eta_\La,\mu_2^2) = f_{a/A}(\eta_\La,\mu_1^2)\,
\exp\!\left[-
\int_{\mu_2^2}^{\mu_1^2} \!d\mu^2
\frac{d}{d\mu^2} \log\left(f_{a/A}(\eta_\La,\mu^2)\right)
\right]
\;.
\end{equation}
That is, making use of the first order evolution equation,
\begin{equation}
\begin{split}
\label{eq:DGLAP1}
f_{a/A}(\eta_\La,\mu_2^2) ={}& f_{a/A}(\eta_\La,\mu_1^2)\,
\exp\!\Bigg[-
\int_{\mu_2^2}^{\mu_1^2} \!\frac{d\mu^2}{\mu^2}
\int\!dz
\sum_{\hat a}
\frac{\as}{2 \pi}
\\&\times
\bigg\{
P_{a\hat a}(z)\,
\frac{f_{a/A}(\eta_\La/z,\mu^2)}{z f_{a/A}(\eta_\La,\mu^2)}
- \delta_{a, \hat a}\left(\frac{2 C_a}{1-z} - \gamma_a\right)
\bigg\}
\Bigg]
\;.
\end{split}
\end{equation}
This is what appears in shower evolution algorithms \cite{EarlyPythia}. On the other hand, if we use the Mellin transform of the parton distributions, eq.~(\ref{eq:MellinTransform}), we have
\begin{equation}
\label{eq:DGLAP2}
\tilde f_{a/A}(N,\mu_2^2)
= \tilde f_{a'/A}(N,\mu_1^2)
\exp\!\left[-
\int_{\mu_2^2}^{\mu_1^2} \!\frac{d\mu^2}{\mu^2}
\frac{\as}{2 \pi}\,
\gamma(N)
\right]_{a a'}
\;.
\end{equation}
Here $\gamma(N)$ it the matrix obtained by taking the Mellin transform of the evolution kernels. There are no parton distribution functions in the exponent. It is this formulation, or variations on it that do not work directly with the Mellin transformed evolution kernels, that typically appears in threshold summation calculations. For general purposes, eq.~(\ref{eq:DGLAP2}) is more powerful than eq.~(\ref{eq:DGLAP1}) because one does not need to have already solved the evolution equation to use it. However, eq.~(\ref{eq:DGLAP2}) is not more, or less, accurate than eq.~(\ref{eq:DGLAP1}).

In fact, we found in section \ref{sec:comparison} that a standard threshold summation in the case of the Drell-Yan process matches what the analysis of this paper gives when we use a leading saddle point approximation to connect the two results.

We now comment on the role of parton distributions in the formalism that we have presented. The $\sket{\rho(t)}$ in eq.~(\ref{eq:evolutionbis}) contains a factor representing the parton distribution functions at the resolution scale corresponding to $t$. In eq.~(\ref{eq:rhopertdef}), we defined an alternative statistical state $\sket{\rho_{\rm pert}(t)}$ in which this parton distribution factor has been removed. Then $\sket{\rho_{\rm pert}(t)}$ obeys the evolution equation (\ref{eq:rhopertevolution}) in which the parton distribution functions do not appear. At the end of the shower, we obtain the ordinary statistical state $\sket{\rho(t_\Lf)}$ by multiplying by parton distribution functions appropriate to a low $k_\perp^2$ scale, $m_\perp^2({\rm start})$, at which the shower turns off.

With this formulation, the whole hard scattering plus parton shower is a single perturbative process for which the resolution scale is $m_\perp^2({\rm start})$. There are no parton distribution functions except at the low scale. Now, the actual code works with $\sket{\rho(t)}$ and does use parton distribution functions at the hard scattering and at each shower stage. However, this use of parton distribution functions is only a trick \cite{EarlyPythia}. Actually, all of the parton distribution functions approximately cancel except for those at the final low scale.\footnote{The cancellation is approximate because the operator $[\Vreal(t) - \Vvirt(t)]$ is calculated approximately, using a limit in which successive splittings are strongly ordered.}

In order to make this cancellation work, \textsc{Deductor} 2.0 matches parton evolution to shower evolution by using parton distribution functions $f_{a/A}(\eta_\La,\mu_\Lambda^2)$ that evolve from the starting scale according a modified leading order evolution equation (\ref{eq:DGLAPLambda}). It would, of course, be better to use an appropriate next-to-leading order evolution equation for the parton distribution functions, but we do not have a next-to-leading order shower algorithm. Thus we are stuck with a parton shower summation of logarithms based on leading order perturbation theory.

\section{Choices in {\textsc Deductor}}
\label{sec:choices}

A user of \textsc{Deductor} 2.0 has some choices. 

The default choice is the calculation described above. With this choice, the parton shower sums threshold logarithms as described in this paper.

Another choice would be to eliminate the summation of threshold logarithms. This is easy to do, as described in section \ref{sec:numerical}. 

There is a third possibility. The user may want to retain the threshold logarithms but modify the way parton distribution functions appear in the calculated cross section. The default result for a cross section has the form
\begin{equation}
\label{eq:defaultdsigma}
d\sigma = d\hat \sigma\
Z_\La Z_\Lb\
f^{\MSbar}_{a/A}(\eta_\La,Q^2)f^{\MSbar}_{b/B}(\eta_\Lb,Q^2)
\;,
\end{equation}
where $Q^2$ is the scale of the hard scattering and some of the summation of threshold logarithms is contained in the factor
\begin{equation}
Z_\La Z_\Lb = 
\frac{f_{a/A}(\eta_\La,Q^2)} {f^{\MSbar}_{a/A}(\eta_\La,Q^2)}
\frac{f_{b/B}(\eta_\Lb,Q^2)}{f^{\MSbar}_{b/B}(\eta_\Lb,Q^2)}
\;.
\end{equation}
See section \ref{subsec:thecomparison}. Here the parton distribution functions $f^{\MSbar}_{a/A}(\eta_\La,Q^2)$ are obtained from the parton distribution functions at the scale $m_\perp^2({\rm start})$ using the first order evolution equation (\ref{eq:DGLAP0}).

Suppose that the user is only slightly interested in the details of the final state that a parton shower naturally specifies. Rather, the user is most interested in the hard scattering that initiates the parton shower. This user wants to have a calculation of the inclusive hard scattering cross section, including threshold corrections, that is as accurate as possible. Such a user might not want to have a cross section based on parton distributions $f^{\MSbar}_{a/A}(\eta_\La,Q^2)$ and $f^{\MSbar}_{b/B}(\eta_\La,Q^2)$ at the hard scale, since these parton distributions have been obtained by lowest order evolution from $m_\perp^2({\rm start})$. Instead, this user might prefer parton distributions $f^{\MSbar, {\rm NLO}}_{a/A}(\eta_\La,Q^2)$ and $f^{\MSbar, {\rm NLO}}_{b/B}(\eta_\La,Q^2)$ that have been obtained with next-to-leading order evolution. That is easily arranged by multiplying the default $d\sigma$ in eq.~(\ref{eq:defaultdsigma}) by a suitable weight factor:
\begin{equation}
\label{eq:newdsigma}
d\sigma^{\rm mod.} = d\sigma\
\frac{f^{\MSbar, {\rm NLO}}_{a/A}(\eta_\La,Q^2)}
{f^{\MSbar}_{a/A}(\eta_\La,Q^2)}\,
\frac{f^{\MSbar, {\rm NLO}}_{b/B}(\eta_\Lb,Q^2)}
{f^{\MSbar}_{b/B}(\eta_\Lb,Q^2)}
\;.
\end{equation}
We have, in fact, done this in our numerical comparisons in section \ref{sec:numerical}.

\section{Outlook}
\label{sec:outlook}

We have presented a formulation of parton shower event generators in which the ``threshold'' enhancements of the cross section at large hardness scale $Q^2$ are included within the parton shower. This has a disadvantage compared to analytical summations of threshold logs: as presented here, the calculation has not been systematically extended beyond the leading logarithm approximation, whereas many of the analytical results are for a much improved order of approximation. However, the parton shower formulation has the advantage compared to analytical calculations that the same algorithm works for a wide variety of physical observables. As long as the desired Born level hard process is included in the parton shower code, the user simply has to specify the observable that is to be measured at the end of the shower. 

Compared to standard parton shower formulations that do not include threshold effects, the methods presented here have the advantage that they make the parton shower more accurate in a base level approximation in which matching to an NLO calculation has not been applied. 

Every parton shower program is a little bit different. We have presented the threshold algorithms as needed for our program, \textsc{Deductor}. However, we believe that the methods presented here can be adapted with not much difficulty to other parton shower event generators.

%==========================================================================
\acknowledgments{ 
This work was supported in part by the United States Department of Energy and by the Helmholtz Alliance ``Physics at the Terascale." This project was begun while the authors were at the Munich Institute for Astro- and Particle Physics program ``Challenges, Innovations and Developments in Precision Calculations for the LHC.''   It was completed while one of us (DS) was at the Kavli Institute for Theoretical Physics at the University of California, Santa Barbara, program ``LHC Run II and the Precision Frontier'' which was supported by the U.\ S.\ National Science Foundation under Grant No.\ NSF PHY11-25915.  We thank the MIAPP and the KITP for providing stimulating research environments. We thank Thomas Becher,  Timothy Cohen, Hannes Jung, and George Sterman for helpful conversations. We thank Frank Petriello for advice about the Drell-Yan cross section and its perturbative calculation. We thank Jeff Owens for providing us with the Kidonakis-Owens code for threshold corrections to the one jet inclusive cross section. We thank Werner Vogelsang for providing us with the de Florian, Hinderer, Mukherjee, Ringer, Vogalsang code for threshold corrections to the jet cross section.}
%==========================================================================

%-------------------------------------------------------------------
\appendix

\section{Notation and kinematics}%%%%%%%%%%%%%%%%
\label{sec:notation}

In this appendix, we collect some notations used throughout the paper and put them in one place. We only sketch the notation that we need. Details can be found in refs.~\cite{NSI, NSII, NSspin, NScolor, Deductor, ShowerTime, PartonDistFctns, ColorEffects}.

In the parton shower, the partons are described by a state vector $\sket{\rho(t)}$ that represents the probability distribution of parton flavors and momenta and the density matrix in the quantum color space. (Recall that in this paper we effectively ignore quantum spin by summing over spins of the daughter partons after each splitting and averaging over the mother parton spin.) In this paper, all of the partons except top quarks are massless.  We expand $\sket{\rho(t)}$ in basis states $\sket{\{p,f,c',c\}_{m}}$. This basis state represents two initial state partons with labels ``a'' and ``b'' and $m$ final state partons with labels $l$. The partons have  momenta $p$, flavors $f$, and colors $c',c$.  The momentum fractions of the initial state partons are $\eta_\La$ and $\eta_\Lb$. In the context of a parton splitting, we generally use $p_l$ to denote the momentum of parton $l$ before the splitting and $\hat p_l$ to denote the momentum of parton $l$ after the splitting.

When a final state parton labelled $l$ splits into partons $l$ and $m+1$ with momenta $\hat p_l$ and $\hat p_{m+1}$, we characterize the splitting by its virtuality $(\hat p_l + \hat p_{m+1})^2$. Similarly, when the initial state parton with label ``a'' splits in backward evolution to a new initial state parton ``a'' and a new final state parton with label $m+1$, we characterize the splitting by its spacelike virtuality $(\hat p_\La - \hat p_{m+1})^2$.

The shower time that we use is related to the virtuality of a splitting:
\begin{equation}
\begin{split}
\label{eq:showertime}
e^{-t} ={}& \frac{(\hat p_l + \hat p_{m+1})^2}{2 p_l\cdot Q_0}
\hskip 1.5 cm {\rm final\ state},
\\
e^{-t} ={}&  -\frac{(\hat p_\La - \hat p_{m+1})^2}{2 p_\La \cdot Q_0}
\hskip 1 cm {\rm initial\ state}.
\end{split}
\end{equation}
Here $Q_0$ is a fixed vector equal to the total momentum of all of the final state partons just after the hard scattering that initiates the shower. With this notation, the virtuality in an initial state splitting is
\begin{equation}
\begin{split}
\label{eq:musqamusqbbis}
\mu_\La^2(t) ={}& 2 p_\La\cdot Q_0 \,e^{-t}
\;.
\end{split}
\end{equation}

It is convenient to use a dimensionless virtuality variable
\begin{equation}
\begin{split}
\label{eq:ydef}
y ={}& \frac{(\hat p_l + \hat p_{m+1})^2}{2 p_l\cdot Q}
\hskip 1.5 cm {\rm final\ state},
\\
y ={}& -\frac{(\hat p_\La - \hat p_{m+1})^2}{2 p_\La\cdot Q}
\hskip 1 cm {\rm initial\ state}.
\end{split}
\end{equation}
Here $Q = p_\La + p_\Lb$ is the total momentum of the final state partons just before the splitting. Then $y$ is related to shower time by
\begin{equation}
\begin{split}
\label{eq:yfromt}
y ={}& \frac{2 p_l\cdot Q_0}{2 p_l\cdot Q}\,e^{-t}
\hskip 1 cm {\rm final\ state},
\\
y ={}& \frac{2 p_\La\cdot Q_0}{2 p_\La\cdot Q}\,e^{-t}
\hskip 1 cm {\rm initial\ state}.
\end{split}
\end{equation}
Since $2 p_\La \cdot Q = Q^2$, we have a convenient identity for $\mu_\La^2(t)$:
\begin{equation}
\label{eq:yQsq}
\mu_\La^2(t) = y\,Q^2
\;.
\end{equation}

We sometimes use a squared transverse momentum variable $\bm k_\perp^2$ for a splitting of initial state parton ``a.'' With the exact kinematics and momentum mappings used in \textsc{Deductor}, the emitted parton has label $m+1$ and momentum $\hat p_{m+1}$. The part of $\hat p_{m+1}$ orthogonal to the momenta of both incoming partons after the splitting, $\hat p_\La$ and $\hat p_\Lb$, is $\hat {\bm p}_{m+1}^\perp$, whose square is
\begin{equation}
(\hat {\bm p}_{m+1}^\perp)^2  = y (1 - z - zy)\, 2 p_\La\cdot Q
\;.
\end{equation}
Here $z$ is defined by $z = \eta_\La/\hat \eta_\La$, where $\eta_\La$ is the momentum fraction of parton ``a'' before the splitting (in backward evolution) and $\hat \eta_\La$ is its momentum fraction after the splitting. We note that the condition $(\hat {\bm p}_{m+1}^\perp)^2 \ge 0$ implies
\begin{equation}
\label{eq:zlimit}
z < 1/(1+y)
\;.
\end{equation}

The quantity $(\hat {\bm p}_{m+1}^\perp)^2$ vanishes when the emitted parton is collinear to parton ``a,'' which corresponds to $y \to 0$ with fixed $(1-z)$. However, it also vanishes when the emitted parton is collinear to parton ``b,'' which corresponds to $(1 - z - zy) \to 0$ with fixed $y$. For our purposes, we prefer a variable that matches $(\hat {\bm p}_{m+1}^\perp)^2$ in the collinear limit, but does not vanish in the anticollinear limit $(1 - z - zy) \to 0$. We thus define a transverse momentum variable
\begin{equation}
\label{eq:kperpsqdef}
\bm k_\perp^2  = y (1 - z)\, 2 p_\La\cdot Q
\;.
\end{equation}
This discussion of $\bm k_\perp^2$ may be compared to that in section 2.3.1 of ref.~\cite{SjostrandSkands}.

We will need some standard flavor dependent constants. The number of flavors is $N_\Lf$. We have color factors $C_\LA = N_\Lc$ (with $N_\Lc = 3$) and $C_\LF = (N_\Lc^2 - 1)/(2 N_\Lc)$. We use $f$ for a parton flavor, $f \in \{\Lg, \Lu,  \bar \Lu, \Ld, \dots\}$. We define $C_f$ by
\begin{equation}
\begin{split}
\label{eq:Cf}
C_\Lg ={}& C_\LA
\;,
\\
C_q ={}& C_\LF\qquad q\in \{\Lu,  \bar \Lu, \Ld, \dots\}
\;.
\end{split}
\end{equation}
In the kernels for evolution of parton distributions, constants $\gamma_f$ appear, with
\begin{equation}
\begin{split}
\label{eq:gammaf}
\gamma_\Lg ={}& \frac{11 C_\LA}{6} - \frac{2 T_\LR N_\Lf}{3}
\;,
\\
\gamma_q ={}& \frac{3 C_\LF}{2} \qquad q\in \{\Lu,  \bar \Lu, \Ld, \dots\}
\;,
\end{split}
\end{equation}
where $T_\LR = 1/2$.

We will use three momenta $\vec p_l$ in a reference frame in which the total momentum $Q$ of the final state partons before the splitting has zero space components. (Thus $Q = p_\La + p_\Lb$.) In the case of an on-shell virtual particle, $p_l^0 = |\vec p_l|$. We then write $p_l =  |\vec p_l| v_l$ where
\begin{equation}
v_l = (1,\vec v_l)
\;,
\end{equation}
with $\vec v_l^{\,2} = 1$. We use $E_Q$ for the energy component of $Q$: $Q = (E_Q, \vec 0\,)$. We often use the convenient shorthand
\begin{equation}
\begin{split}
\label{eq:aldef}
a_l ={}& \frac{Q^{2}}{2p_{l}\!\cdot\! Q} = \frac{E_Q}{2 |\vec p_l|}
\;.
\end{split}
\end{equation}
In the case of an initial state parton, this is $a_\La = a_\Lb = 1$. We also sometimes use the shorthand
\begin{equation}
\label{eq:psikl}
\psi_{kl} = \frac{1 -  \cos\theta_{kl}}{\sqrt{8(1 +  \cos\theta_{kl})}}
\;,
\end{equation}
where $\theta_{kl}$ is the angle between a pair of partons with labels $k$ and $l$.

\section{Calculation of the probability conserving integrand $\Vreal(t)$}
\label{sec:CalculateV}

For each real splitting in a parton shower, we need a Sudakov factor associated with the evolution of the system from the time $t_1$ of the previous splitting and the time $t_2$ of the new splitting. The standard form for this factor is $\mathbb T \exp[-\int_{t_1}^{t_2}\! dt\ \Vreal(t)]$, where $\Vreal(t)$ is the probability per unit $dt$ to have a splitting at time $t$. Thus the Sudakov factor is the probability not to have had a splitting between $t_1$ and $t_2$. This is the structure of the Sudakov factor used in \textsc{Deductor} 1.0.

In this paper, we calculate a numerical factor $\int_{t_1}^{t_2}\! dt\ [\Vreal(t) - \Vvirt(t)]$, where $\Vvirt(t)$ gives the effect of parton evolution and virtual graphs. We are not able to calculate this factor exactly. Rather, we calculate $\Vreal(t)$ and $\Vvirt(t)$ for large t, or small $y \propto e^{-t}$ as defined in eq.~(\ref{eq:yfromt}). In this appendix, we calculate $\Vreal(t)$ for small y. Fortunately, although $\Vreal(t)$ in \textsc{Deductor} is quite complicated, it has a simple structure for small $y$. Our aim in this appendix is to exhibit enough details of the calculation to enable the reader to reproduce it.

The operator $\Vreal(t)$ is a sum over contributions from the partons in existence after time $t_1$, as in eq.~(\ref{eq:Vrealsum}) \cite{NSI}:
\begin{equation}
\label{eq:Vrealsumbis}
\Vreal(t)
= 
\Vreal_\La(t)
+ \Vreal_\Lb(t)
+ \sum_{l=1}^m \Vreal_l(t)
\;.
\end{equation}
We will first examine the case of final state partons, with labels $l = 1, \dots m$. Then we will turn to the initial state partons, with labels ``a'' and ``b.''

\subsection{Final state partons}
\label{sec:CalculateVFS}

We write $\Vreal_l(t)$ as \cite{NSI}
\begin{equation}
\label{eq:calRl}
\Vreal_l(t) = 
\sum_k \Vreal_{l k}(t)
\;.
\end{equation}
The sum includes all parton labels $k = \La, \Lb, 1, \dots, m$. When the operators $\Vreal_{l k}(t)$ act on a partonic state $\sket{\{p,f,c',c\}_{m}}$, they have the structure
\begin{equation}
\begin{split}
\label{eq:Rlk0}
\Vreal_{l k}(t)\sket{\{p,f,c',c\}_{m}}
={}& 
\overline\lambda^{\Vreal}_{l k}(\{p,f\}_{m},t)\,
\frac{1}{2}
\big([(\bm T_l\cdot \bm T_k)\otimes 1] + [1 \otimes (\bm T_l\cdot \bm T_k)]\big)
\\ &\times
\sket{\{p,f,c',c\}_{m}}
\;.
\end{split}
\end{equation}
As explained in section \ref{sec:Vreal}, the operators $[(\bm T_l\cdot \bm T_k)\otimes 1]$ act on the space of color density matrices for the parton state \cite{NSI, NScolor}, with basis elements $\ket{\{c\}_m}\bra{\{c'\}_m}$. A color generator matrix $T^a$ acting on parton $l$ in the ket state multiplies a generator matrix $T_a$ acting on parton $k$ in the ket state. The dot product indicates a sum over $a = 1,\dots, N_\Lc^2 - 1$. In $[1 \otimes (\bm T_l\cdot \bm T_k)]$, we have the same construction with the color generators acting on the bra state.

From eq.~(5.28) of ref.~\cite{NScolor}, we find that the functions $\overline\lambda^{\Vreal}_{l k}$ has the structure
\begin{equation}
\begin{split}
\label{eq:VresultF1}
\overline\lambda^{\Vreal}_{l k}(\{p,& f\}_{m},t)
\\
={}&
\frac{1}{m!}\int\!d\{\hat p,\hat f\}_{m+1}
\delta(t - T(\{\hat p,\hat f\}_{m+1}))\,
\sbra{\{\hat p,\hat f\}_{m+1}}{\cal P}_{l}\sket{\{p,f\}_m}
\\&\times \bigg[
\theta(k = l)\,\theta(\hat f_{m+1} \ne \Lg))\,
\frac{1}{2 C_\LA}\,\overline w_{ll}(\{\hat p,\hat f\}_{m+1})
\\&\qquad +
\theta(k = l)\,\theta(\hat f_{m+1} = \Lg))
[\overline w_{ll}(\{\hat p,\hat f\}_{m+1})
- \overline w_{ll}^{\rm eikonal}(\{\hat p,\hat f\}_{m+1})]
\\&\qquad -
\theta(k\ne l)\,\,\theta(\hat f_{m+1} = \Lg))\,
A'_{l k}(\{\hat p\}_{m+1})\overline w_{l k}^{\rm dipole}(\{\hat p,\hat f\}_{m+1})
\bigg]
\;.
\end{split}
\end{equation}

Consider the first line on the right hand side of eq.~(\ref{eq:VresultF1}). There is an integration over the variables that define the splitting of a mother parton with momentum $p_l$ into daughter partons with momenta $\hat p_l$ and $\hat p_{m+1}$. In this paper, we take all partons to be massless. We will use the auxiliary variables
\begin{equation}
\begin{split}
a_l ={}& \frac{Q^{2}}{2p_{l}\!\cdot\! Q}
\;,
\\
\lambda(y) ={}& 
\sqrt{(1 + y)^2 - 4 a_l y}
\;,
\\
h_\pm(y) ={}& \frac{1}{2}\,[1 + y \pm \lambda(y)]
\;.
\end{split}
\end{equation}
Here $Q$ is the total momentum of all the final state partons in $\sket{\{p,f,c',c\}_{m}}$. The dimensionless virtuality variable $y$ was defined in eqs.~(\ref{eq:ydef}) and (\ref{eq:yfromt}): 
\begin{equation}
y = \frac{2 \hat p_l\cdot \hat p_{m+1}}{2 p_l \cdot Q}
\;.
\end{equation}
We define the momentum fraction in the splitting by
\begin{equation}
\frac{\hat p_{m+1}\cdot \tilde n_l}{\hat p_{l}\cdot \tilde n_l}
= \frac{1-z}{z}
\;,
\end{equation}
where the auxiliary lightlike vector $\tilde n_l$ is 
\begin{equation}
\label{tildenldef}
\tilde n_l = \frac{1}{a_l}\, Q - p_l
\;.
\end{equation}
We also define an azimuthal angle $\phi$ of the splitting using the part $k_\perp$ of $\hat p_l$ that is orthogonal to $p_l$ and $\tilde n_l$. The choice of flavors of the daughters can be specified by giving the flavor $\hat f_{m+1}$ of daughter parton $m+1$. This gives us splitting variables $y,z,\phi,\hat f_{m+1}$. 

We can write the daughter parton momenta in terms of $y,z,\phi$ using
\begin{equation}
\begin{split}
\label{eq:daughtermomenta}
\hat p_l ={}& 
z\,h_+(y)\, p_l
+ (1 - z) h_-(y)\, \tilde n_l
+k_\perp
\;,
\\
\hat p_{m+1} ={}& 
(1-z)\, h_+(y)\, p_l
+ z  h_-(y)\, \tilde n_l
- k_\perp
\;.
\end{split}
\end{equation}
The magnitude of the transverse momentum $k_\perp$ is given by
\begin{equation}
- \frac{k_\perp^2}{2p_l\cdot Q} =
z (1-z)y
\;.
\end{equation}

Using eq.~(8.20) of ref.~\cite{NSI}, we find that integration over the splitting variables between shower times corresponding to $y$ values $y_1$ and $y_2$ is accomplished with
\begin{equation}
\begin{split}
\label{eq:integralF}
\frac{1}{m!}
\int \big[d\{\hat p,\hat f&\}_{m+1}\big]
\sbra{\{\hat p,\hat f\}_{m+1}}{\cal P}_{l}\sket{\{p,f\}_m}
\cdots
\\&
= \frac{p_{l}\!\cdot\! Q}{8\pi^2}\,
\int_{y_2}^{y_1}\! dy\
\lambda(y)\,
\int_{0}^{1}\! dz
\int_{-\pi}^\pi \frac{d\phi}{2\pi}\sum_{\hat f_{m+1}}\ \cdots
\;.
\end{split}
\end{equation}
The delta function that specifies the shower time is, from eqs.~(\ref{eq:showertime}) and (\ref{eq:ydef}),
\begin{equation}
\label{eq:yfromtbis}
\delta(t - T(\{\hat p,\hat f\}_{m+1}))
=
\delta\!\left(
\log y  - \log\!\left(\frac{p_l\cdot Q_0}{p_l\cdot Q}
e^{-t}\right)\right)
\;.
\end{equation}
Using eqs.~(\ref{eq:integralF}) and (\ref{eq:yfromtbis}) in eq.~(\ref{eq:VresultF1}) gives
\begin{equation}
\begin{split}
\label{eq:VresultF2}
\overline\lambda^{\Vreal}_{l k}(\{p,& f\}_{m},t)
\\
={}&
\frac{p_{l}\!\cdot\! Q}{8\pi^2}\,
y\,
\lambda(y)\,
\int_{0}^{1}\! dz
\int_{-\pi}^\pi \frac{d\phi}{2\pi}\ \sum_{\hat f_{m+1}}
\\&\times \bigg[
\theta(k = l)\,\theta(\hat f_{m+1} \ne \Lg))\,
\frac{1}{2 C_\LA}\,\overline w_{ll}(\{\hat p,\hat f\}_{m+1})
\\&\qquad +
\theta(k = l)\,\theta(\hat f_{m+1} = \Lg))
[\overline w_{ll}(\{\hat p,\hat f\}_{m+1})
- \overline w_{ll}^{\rm eikonal}(\{\hat p,\hat f\}_{m+1})]
\\&\qquad -
\theta(k\ne l)\,\,\theta(\hat f_{m+1} = \Lg))\,
A'_{l k}(\{\hat p\}_{m+1})\overline w_{l k}^{\rm dipole}(\{\hat p,\hat f\}_{m+1})
\bigg]
\;.
\end{split}
\end{equation}
The first line inside the square brackets here is for a gluon self-energy graph with a quark loop. The color factor is $T_\LR = 1/2$, but we follow the notation of eq.~(\ref{eq:Rlk0}), in which this term comes with a color operator $(\bm T_l\cdot \bm T_l)\otimes 1 = C_\LA 1\otimes 1$ or $1\otimes(\bm T_l\cdot \bm T_l)= C_\LA 1\otimes 1$. The factor $C_\LA$ does not really belong here, so we remove it by dividing $\overline w_{ll}$ by $C_\LA$. The second line of eq.~(\ref{eq:VresultF1}) covers gluon emission in a cut self-energy graph, while the third line covers gluon exchange between two lines, $l$ and $k$. 

For $k = l$ and $\hat f({m+1}) \ne \Lg$, we have a $\Lg \to q + \bar q$ splitting. From ref.~\cite{NSII}, eq.~(A.1), we have
\begin{equation}
\overline w_{ll}(\{\hat p,\hat f\}_{m+1})
= \frac{8\pi\as}{(\hat p_l + \hat p_{m+1})^2}
\left(
1 + \frac{2 \hat p_l\cdot D(p_l,Q) \cdot \hat p_{m+1}}{(\hat p_l + \hat p_{m+1})^2}
\right)
\;.
\end{equation}
Here $D(p_l,Q)$ is the Coulomb gauge numerator function, 
\begin{equation}
\label{eq:Dcoulomb}
D(q)^{\mu\nu} =
- g^{\mu\nu} 
-\frac{q^\mu \tilde q^\nu + \tilde q^\mu q^\nu - q^\mu q^\nu}{|\vec q\,|^2}
\;,
\end{equation}
where $\tilde q = (0,\vec q\,)$.  Since here $q^2 = 0$, this also equals
\begin{equation}
\label{eq:Daxial}
D(q)^{\mu\nu} =
- g^{\mu\nu} + \frac{q^\mu Q^\nu + Q^\mu q^\nu}{q\cdot Q}
- \frac{Q^2\ q^\mu q^\nu}{(q\cdot Q)^2}
\;.
\end{equation}
With the help of eq.~(\ref{eq:daughtermomenta}), we find
\begin{equation}
\hat p_l\cdot D(p_l,Q) \cdot \hat p_{m+1}
=
-2 z (1-z) \, y\, p_l \cdot Q
\;.
\end{equation}
This gives
\begin{equation}
\overline w_{ll}(\{\hat p,\hat f\}_{m+1})
= \frac{4\pi\as}{y\,p_l\cdot Q}
\left(
1 - 2 z (1-z)
\right)
\;.
\end{equation}

The function $\overline w_{ll}(\{\hat p,\hat f\}_{m+1}) - \overline w_{ll}^{\rm eikonal}(\{\hat p,\hat f\}_{m+1})$ is calculated in ref.~\cite{NSII}. For $\hat f_l$ equal to a quark or antiquark flavor, it is given in eq.~(2.23) of ref.~\cite{NSII} (for the quark mass equal to zero):
\begin{equation}
\begin{split}
\overline w_{ll}(\{\hat p,\hat f\}_{m+1})& - \overline w_{ll}^{\rm eikonal}(\{\hat p,\hat f\}_{m+1}) 
\\={}& 
\frac{4\pi\as}{y\, p_l\cdot  Q}
\bigg\{
\frac{(\lambda(y) - 1 + y)^2 + 4 y}{4\lambda(y)}\,
\left[
\frac{2x}{1-x} - \frac{2 a_l y}{(1-x)^2 (1 + y)^2}
\right]
\\&\qquad\qquad
+ \frac{1}{2}\,(1-z)(1 + y + \lambda(y))
\bigg\}
\;.
\end{split}
\end{equation}
Here 
\begin{equation}
\begin{split}
x ={}& \frac{\lambda(y)}{1 + y}\,z
+ \frac{2 a_l y}{(1+y)(1 + y + \lambda(y))}
\;,
\\
1-x ={}& \frac{\lambda(y)}{1 + y}\,(1-z)
+ \frac{2 a_l y}{(1+y)(1 + y + \lambda(y))}
\;.
\end{split}
\end{equation}
Here we have changed conventions compared to ref.~\cite{NSII} and exchanged $z \leftrightarrow (1-z)$ and $x \leftrightarrow (1-x)$. This expression is rather complicated, but we only need its $y \to 0$ limit:
\begin{equation}
\begin{split}
\label{eq:wllquark}
\overline w_{ll}(\{\hat p,\hat f\}_{m+1}) 
- \overline w_{ll}^{\rm eikonal}(\{\hat p,\hat f\}_{m+1}) 
\sim{}&
\frac{4\pi\as}{y\, p_l\cdot  Q}\,(1-z)
\;.
\end{split}
\end{equation}

We also need $\overline w_{ll}(\{\hat p,\hat f\}_{m+1}) - \overline w_{ll}^{\rm eikonal}(\{\hat p,\hat f\}_{m+1})$ for $\Lg \to \Lg + \Lg$ splittings, $\hat f_l = \Lg$. It is given in eq.~(2.50) of ref.~\cite{NSII}:
\begin{equation}
\begin{split}
\overline w_{ll}(\{\hat p,\hat f\}_{m+1})& - \overline w_{ll}^{\rm eikonal}(\{\hat p,\hat f\}_{m+1}) 
\\={}& 
\frac{4\pi\as}{y\, p_l\cdot  Q}\
\frac{z(1-z)}{2}
\left[
1 + \left(1 - \frac{2 a_l y}{x(1-x) (1 + y)^2}\right)^2
\right]
\;.
\end{split}
\end{equation}
We need only the $y\to 0$ limit of this:
\begin{equation}
\begin{split}
\label{eq:wllgluon}
\overline w_{ll}(\{\hat p,\hat f\}_{m+1}) 
- \overline w_{ll}^{\rm eikonal}(\{\hat p,\hat f\}_{m+1}) 
\sim{}&
\frac{4\pi\as}{y\, p_l\cdot  Q}\,z(1-z)
\;.
\end{split}
\end{equation}

Now we need the interference terms. Using eq.~(5.3) of ref.~\cite{NScolor} for $\overline w_{l k}$ and eq.~(7.12) of ref.~\cite{NSspin} for the partitioning function $A'_{l k}$, we have
\begin{equation}
\begin{split}
\label{eq:Awlk0}
A'_{l k}(\{\hat p\}_{m+1})&\, \overline w_{l k}^{\rm dipole}(\{\hat p,\hat f\}_{m+1})
\\={}& 4\pi\as\ \frac{2 \hat p_k\cdot \hat p_l}
{\hat p_{m+1}\cdot \hat p_ l}\
\frac{\hat p_l\cdot Q}
{\hat p_{m+1}\cdot \hat p_k\ \hat p_l\cdot Q
+ \hat p_{m+1}\cdot \hat p_l\ \hat p_k\cdot Q}
\;.
\end{split}
\end{equation}

To evaluate this, we write the spectator momentum after the splitting, $\hat p_k$, as
\begin{equation}
\hat p_k = 
A_k\left[
a_l e^{\xi + \omega(y)} p_l 
+ a_l e^{-\xi - \omega(y)}\tilde n_l
+ \ell_{\perp}
\right]
\;.
\end{equation}
where $\xi$ is a boost angle that is related to the angle $\theta_{lk}$ between $\vec p_l$ and $\vec p_k$ in the $\vec Q = 0$ frame by
\begin{equation}
e^{2\xi} = \frac{1+\cos\theta_{lk}}{1-\cos\theta_{lk}}
\;.
\end{equation}
The parameter $\omega(y)$ is an additional boost angle,
\begin{equation}
\begin{split}
\label{eq:omegadef}
e^{\omega(y)} ={}& 
\frac{a_l - h_+(y)}{a_l - h_+(0)}
\end{split}
\end{equation}
and in the case that $k = \La$ or $k = \Lb$, $\omega(y) = 0$. Since $\omega(y)$ is of order $y$ and we are interested in the limit of small $y$, we will replace $\omega(y) \to 0$. The normalization parameter $A_k$ drops out of our calculation.

The product in eq.~(\ref{eq:Awlk0}) is quite complicated in general. However, it is reasonably simple in the limit of small $y$. We find
\begin{equation}
\begin{split}
\label{eq:Awlk1}
A'_{l k}(\{\hat p\}_{m+1})&\, \overline w_{l k}^{\rm dipole}(\{\hat p,\hat f\}_{m+1})
\\\approx{}& 
\frac{4\pi\as}{y\, p_l\cdot Q}\,
\frac{2 z }
{ 
(1-z)
+ a_l y [2 e^{2\xi} + 1]
+ 2\sqrt{a_l y e^{2\xi}} \sqrt{1-z} \,
\cos \phi
}
\;.
\end{split}
\end{equation}
Here we have noted that the terms proportional to $y$ are negligible for small $y$ unless $(1-z)$ is small. Therefore in the coefficients of $y$ and $\sqrt{y(1-z)}$, we have replaced $z$ by 1. We can integrate this over $\phi$ with the result
\begin{equation}
\begin{split}
\label{eq:Awlk12}
\int\!\frac{d\phi}{2\pi}\ A'_{l k}(\{\hat p\}_{m+1})&\, \overline w_{l k}^{\rm dipole}(\{\hat p,\hat f\}_{m+1})
\\\approx{}& 
\frac{4\pi\as}{y\, p_l\cdot Q}\,
\frac{2 z}
{\left[[(1-z) + a_l y]^2 + 4 a_l^2 y^2 e^{2\xi}(1+e^{2\xi})\right]^{1/2}}
\;.
\end{split}
\end{equation}
We can then integrate this over $z$ and take the small $y$ limit of the result. We get
\begin{equation}
\begin{split}
\label{eq:Awlk13}
\int_0^1\!dz
\int\!\frac{d\phi}{2\pi}\ A'_{l k}(\{\hat p\}_{m+1})\, \overline w_{l k}^{\rm dipole}(\{\hat p,\hat f\}_{m+1})
\approx
\frac{8\pi\as}{y\, p_l\cdot Q}\bigg\{
\log\!
\left(
\frac{1 - \cos\theta_{kl}}
{2 a_l y}
\right)
- 1
\bigg\}
\;.
\end{split}
\end{equation}

We insert this result back into eq.~(\ref{eq:VresultF2}). The integrals over $\phi$ and $z$ of the other terms are simple, giving in the small $y$ limit,
\begin{equation}
\begin{split}
\label{eq:VresultF3}
\overline\lambda^{\Vreal}_{l k}(\{p,f\}_{m},t)
\approx{}&
\frac{\as}{2\pi}\,
\sum_{\hat f_{m+1}}
\bigg[
\theta(k = l)\,\theta(\hat f_{m+1} \ne \Lg))\,
\frac{1}{3 C_\LA}
\\&\qquad +
\theta(k = l)\,\theta(\hat f_{m+1} = \Lg))\theta(\hat f_l \ne \Lg))\
\frac{1}{2}
\\&\qquad +
\theta(k = l)\,\theta(\hat f_{m+1} = \Lg))\,\theta(\hat f_l = \Lg))\
\frac{1}{6}
\\&\qquad -
\theta(k\ne l)\,\,\theta(\hat f_{m+1} = \Lg))\,
2
\bigg\{
\log\!
\left(
\frac{1 - \cos\theta_{kl}}{2a_l y}
\right)
- 1
\bigg\}
\bigg]
\;.
\end{split}
\end{equation}
We can perform the sum over flavors $\hat f_{m+1}$. The first term applies only if $f_l = \Lg$. When $f_l = \Lg$, there are $N_\Lf$ equal terms. The third term applies if $f_l = \Lg$ and there is one term. The second term applies when $f_l \ne g$ and there is one term. The fourth term applies for any $f_l$. Thus
\begin{equation}
\begin{split}
\label{eq:VresultF4}
\overline\lambda^{\Vreal}_{l k}(\{p,f\}_{m},t)
\approx{}&
\frac{\as}{2\pi}\,
\bigg[
\theta(k = l)\,\theta(f_l = \Lg))\,
\frac{1}{C_\LA}\left[\frac{C_\LA}{6} + \frac{N_\Lf}{3}\right]
\\&\qquad +
\theta(k = l)\,\theta(f_l \ne \Lg)) \
\frac{1}{2}
\\&\qquad -
\theta(k\ne l)\,
2
\bigg\{
\log\!
\left(
\frac{1 - \cos\theta_{kl}}{2a_l y}
\right)
- 1
\bigg\}
\bigg]
\bigg]
\;.
\end{split}
\end{equation}
Using the constants $\gamma_f$ and $C_f$ from eqs.~(\ref{eq:gammaf}) and eq.~(\ref{eq:Cf}), this is
\begin{equation}
\begin{split}
\label{eq:VresultF5}
\overline\lambda^{\Vreal}_{l k}(\{p,f\}_{m},t)
\approx{}&
\frac{\as}{2\pi}\,
\bigg[
\theta(k = l)\,
\left[-\frac{\gamma_{f_l}}{C_{f_l}} + 2\right]
\\&\qquad
-
\theta(k\ne l)\,
2
\bigg\{
\log\!
\left(
\frac{1 - \cos\theta_{kl}}{2a_l y}
\right)
- 1
\bigg\}
\bigg]
\;.
\end{split}
\end{equation}

When we insert this result into eq.~(\ref{eq:Rlk0}), we have
\begin{equation}
\begin{split}
\label{eq:Rlk01}
\Vreal_{l}(t)\sket{\{p,f,c',c\}_{m}}\hskip - 1 cm {}&
\\={}& 
\frac{\as}{2\pi}\bigg\{
\left[-\frac{\gamma_{f_l}}{C_{f_l}} + 2\right]
\frac{1}{2}
\big([(\bm T_l\cdot \bm T_l)\otimes 1] + [1 \otimes (\bm T_l\cdot \bm T_l)]\big)
\\&\qquad -
\bigg[
\log\!
\left(
\frac{1}{a_l y}
\right)
- 1
\bigg]
\sum_{k \ne l}
\big([(\bm T_l\cdot \bm T_k)\otimes 1] + [1 \otimes (\bm T_l\cdot \bm T_k)]\big)
\\&\qquad -
\sum_{k \ne l}
\log\!
\left(
\frac{1 - \cos\theta_{kl}}{2}
\right)
\big([(\bm T_l\cdot \bm T_k)\otimes 1] + [1 \otimes (\bm T_l\cdot \bm T_k)]\big)
\bigg\}
\\ &\times
\sket{\{p,f,c',c\}_{m}}
\;.
\end{split}
\end{equation}
In the second line of the right hand side, we can use $\sum_{k \ne l} \bm T_l\cdot \bm T_k = -\bm T_l\cdot \bm T_l$. Then in lines 1 and 2 we can use $
\bm T_l\cdot \bm T_l = C_{f_l}$. This gives
\begin{equation}
\begin{split}
\label{eq:Rlk02}
\Vreal_{l}(t)\sket{\{p,f,c',c\}_{m}}\hskip - 1 cm {}&
\\={}& 
\frac{\as}{2\pi}\bigg\{
\left[2 C_{f_l}\log\!
\left(
\frac{1}{a_l y}
\right)
-\gamma_{f_l} \right]
(1\otimes 1)
\\&\qquad -
\sum_{k \ne l}
\log\!
\left(
\frac{1 - \cos\theta_{kl}}{2}
\right)
\big([(\bm T_l\cdot \bm T_k)\otimes 1] + [1 \otimes (\bm T_l\cdot \bm T_k)]\big)
\bigg\}
\\ &\times
\sket{\{p,f,c',c\}_{m}}
\;.
\end{split}
\end{equation}
We use this result with $a_l = E_Q/(2 | \vec p_l |)$ from eq.~(\ref{eq:alformulas}) in eq.~(\ref{eq:calV2}).

\subsection{Initial state partons}
\label{sec:CalculateVIS}

We write $\Vreal_\La(t)$ as \cite{NSI}
\begin{equation}
\label{eq:calRISl}
\Vreal_\La(t) = 
\sum_k \Vreal_{\La k}(t)
\;.
\end{equation}
The sum includes all parton labels $k = \La, \Lb, 1, \dots, m$. When the operators $\Vreal_{\La k}(t)$ act on a partonic state $\sket{\{p,f,c',c\}_{m}}$, they have the structure
\begin{equation}
\begin{split}
\label{eq:Rak0}
\Vreal_{\La k}(t)\sket{\{p,f,c',c\}_{m}}
={}& 
\overline\lambda^{\Vreal}_{\La k}(\{p,f\}_{m},t)\,
\frac{1}{2}
\big([(\bm T_\La\cdot \bm T_k)\otimes 1] + [1 \otimes (\bm T_\La\cdot \bm T_k)]\big)
\\ &\times
\sket{\{p,f,c',c\}_{m}}
\;,
\end{split}
\end{equation}
as in eq.~(\ref{eq:Rlk0}). We find from eq.~(5.28) of ref.~\cite{NScolor},
\begin{equation}
\begin{split}
\label{eq:lambdaRresultIS}
\overline\lambda^{\Vreal}_{\La k}(\{p,&f\}_{m},t)
\\
={}&
\frac{1}{m!}\int\!d\{\hat p,\hat f\}_{m+1}
\delta(t - T(\{\hat p,\hat f\}_{m+1}))\,
\sbra{\{\hat p,\hat f\}_{m+1}}{\cal P}_{l}\sket{\{p,f\}_m}
\\&\times
\frac
{n_\Lc(a) \eta_{\La}}
{n_\Lc(\hat a) \hat\eta_{\La}}\,
\frac{
f_{\hat a/A}(\hat \eta_{\La},\mu_\La^{2}(t))}
{f_{a/A}(\eta_{\La},\mu_\La^{2}(t)))}
\\&\times \bigg[
\theta(k = \La)\,\theta(\hat f_{m+1} \ne \Lg)\,\theta(a \ne \Lg)
\overline w_{\La \La}(\{\hat p,\hat f\}_{m+1})
\\&\qquad
+\theta(k = \La)\,\theta(\hat f_{m+1} \ne \Lg)\,\theta(a = \Lg)\,
\frac{T_\LR}{C_\LA}\,
\overline w_{\La \La}(\{\hat p,\hat f\}_{m+1})
\\&\qquad
+\theta(k = \La)\,\theta(\hat f_{m+1} = \Lg)\,
[\overline w_{\La \La}(\{\hat p,\hat f\}_{m+1})
- \overline w_{\La \La}^{\rm eikonal}(\{\hat p,\hat f\}_{m+1})]
\\&\qquad -
\theta(k\ne \La)\,\theta(\hat f_{m+1} = \Lg)\,
A'_{\La k}(\{\hat p\}_{m+1})\overline w_{\La k}^{\rm dipole}(\{\hat p,\hat f\}_{m+1})
\bigg]
\;.
\end{split}
\end{equation}

In an initial state splitting, parton ``a'' with momentum $p_\La$ becomes a new initial state parton with momentum $\hat p_\La$ and a final state parton with momentum $\hat p_{m+1}$. The momentum of the other initial state parton is unchanged: $\hat p_\Lb = p_\Lb$. The splitting kinematics is defined by
\begin{equation}
\begin{split}
\label{eq:momentaIS}
\hat p_\La ={}& \frac{1}{z}\,p_\La
\;,
\\
\hat p_\Lb ={}&  p_\Lb
\;,
\\
\hat p_{m+1} ={}& \left(\frac{1-z}{z} - y\right)\,p_\La + z y\, p_\Lb 
+ \sqrt{y(1-z-yz) Q^2}\ n_\perp
\;,
\end{split}
\end{equation}
where $n_\perp \cdot p_\La = n_\perp\cdot p_\Lb = 0$ and $n_\perp^2 = -1$. Note that $y = \hat p_\La \cdot \hat p_{m+1}/p_\La\cdot Q$, where $Q = p_\La + p_\Lb$, as in eq.~(\ref{eq:ydef}). This kinematics requires $y < (1-z)/z$, or
\begin{equation}
z < \frac{1}{1+y}
\;.
\end{equation}

Eq.~(\ref{eq:lambdaRresultIS}) contains a ratio of parton distributions, which are evaluated at momentum fractions defined by $p_\La = \eta_\La \, p_\LA$ and $\hat p_\La = \hat\eta_\La \, p_\LA$. We take the scales of the parton distributions to be the virtuality $\mu^2_\La(t)$ defined in eq.~(\ref{eq:musqamusqbbis}). There are also factors $n_\Lc(a)$ that count the number of colors (3 or 8) carried by partons of flavor $a$.

Using eq.~(8.20) of ref.~\cite{NSI} together with eqs.~(A.28) of ref.~\cite{PartonDistFctns}, we find that integration over the splitting variables between shower times corresponding to $y$ values $y_1$ and $y_2$ is accomplished with
\begin{equation}
\begin{split}
\label{eq:integralIS}
\frac{1}{m!}
\int \big[d\{\hat p,\hat f&\}_{m+1}\big]
\sbra{\{\hat p,\hat f\}_{m+1}}{\cal P}_{l}\sket{\{p,f\}_m}
\cdots
\\&
= \frac{Q^2}{16\pi^2}\,
\int_{y_2}^{y_1}\! dy\
\int_{0}^{1}\!\frac{dz}{z}
\int_{-\pi}^\pi \frac{d\phi}{2\pi}\sum_{\hat a}\ \cdots
\;.
\end{split}
\end{equation}
The delta function that specifies the shower time is, from eqs.~(\ref{eq:showertime}) and (\ref{eq:ydef}),
\begin{equation}
\label{eq:yfromtIS}
\delta(t - T(\{\hat p,\hat f\}_{m+1}))
=
\delta\!\left(
\log y  - \log\!\left(\frac{p_\La\cdot Q_0}{p_\La\cdot Q}
e^{-t}\right)\right)
\;,
\end{equation}
so that $\mu_\La^2(t) = y Q^2$. Using eqs.~(\ref{eq:integralIS}) and (\ref{eq:yfromtIS}) in eq.~(\ref{eq:lambdaRresultIS}) gives
\begin{equation}
\begin{split}
\label{eq:lambdaRresultIS2}
\overline\lambda^{\Vreal}_{\La k}(\{p,&f\}_{m},t)
\\
={}&
\frac{Q^2}{16 \pi^2}\,
\int_0^1\!\frac{dz}{z}\, \int_0^{2\pi}\frac{d\phi}{2\pi}\sum_{\hat a}\ y\,
\theta(z < 1/(1+y))
\\&\times \frac
{n_\Lc(a) z}
{n_\Lc(\hat a)}\,
\frac{
f_{\hat a/A}(\eta_{\La}/z,y Q^2)}
{f_{a/A}(\eta_{\La},y Q^2)}
\\&\times \bigg[
\theta(k = \La)\,\theta(\hat a \ne a)\,\theta(a \ne \Lg)\,
\overline w_{\La \La}(\{\hat p,\hat f\}_{m+1})
\\&\qquad
+\theta(k = \La)\,\theta(\hat a \ne a)\,\theta(a = \Lg)\,
\frac{T_\LR}{C_\LA}\,
\overline w_{\La \La}(\{\hat p,\hat f\}_{m+1})
\\&\qquad
+\theta(k = \La)\,\theta(\hat a = a)\,
[\overline w_{\La \La}(\{\hat p,\hat f\}_{m+1})
- \overline w_{\La \La}^{\rm eikonal}(\{\hat p,\hat f\}_{m+1})]
\\&\qquad -
\theta(k\ne \La)\,\theta(\hat a = a)\,
A'_{\La k}(\{\hat p\}_{m+1})\overline w_{\La k}^{\rm dipole}(\{\hat p,\hat f\}_{m+1})
\bigg]
\;.
\end{split}
\end{equation}
There are four terms here. The first three are for direct graphs. The first is for $(\hat a,a,\hat f_{m+1}) = (\Lg,q,\bar q)$ and $(\hat a,a,\hat f_{m+1}) = (\Lg,\bar q, q)$. The second is for $(\hat a,a,\hat f_{m+1}) = (q,\Lg, q)$ and $(\hat a,a,\hat f_{m+1}) = (\bar q,\Lg, \bar q)$. Here the color factor is $T_\LR = 1/2$, but this term multiplies $\bm T_\La \cdot \bm T_\La = C_A$, so we need to divide by $C_\LA$. The third term is for $(\hat a,a,\hat f_{m+1}) = (\Lg,\Lg,\Lg)$, $(\hat a,a,\hat f_{m+1}) = (\bar q,\bar q,\Lg)$ and $(\hat a,a,\hat f_{m+1}) = (q,q,\Lg)$. Here there is a soft gluon singularity, which is subtracted. The fourth term is for interference graphs, in which a gluon is exchanged between parton ``a'' and parton $k$.

Now we need the splitting functions. We take the limit $y \ll 1$. From ref.~\cite{PartonDistFctns}, eq.~(A.42), we have for $(\hat a,a,\hat f_{m+1}) = (\Lg,q,\bar q)$ and $(\hat a,a,\hat f_{m+1}) = (\Lg,\bar q, q)$,
\begin{equation}
\overline w_{\La \La}(\{\hat p,\hat f\}_{m+1})
\sim
\frac{8\pi \as}{Q^2}\,\frac{1}{yz}\
\frac{P_{a\hat a}(z)}{T_\LR}
\;.
\end{equation}
From ref.~\cite{PartonDistFctns}, eq.~(A.45), we have for $(\hat a,a,\hat f_{m+1}) = (q,\Lg, q)$ and $(\hat a,a,\hat f_{m+1}) = (\bar q,\Lg, \bar q)$,
\begin{equation}
\overline w_{\La \La}(\{\hat p,\hat f\}_{m+1})
\sim
\frac{8\pi \as}{Q^2}\,\frac{1}{yz}\
\frac{P_{a\hat a}(z)}{C_\LF}
\;.
\end{equation}
From ref.~\cite{PartonDistFctns}, eqs.~(A.35) and (A.39), we have for $(\hat a,a,\hat f_{m+1}) = (\Lg,\Lg,\Lg)$, $(\hat a,a,\hat f_{m+1}) = (\bar q,\bar q,\Lg)$ and $(\hat a,a,\hat f_{m+1}) = (q,q,\Lg)$,
\begin{equation}
\overline w_{\La \La}(\{\hat p,\hat f\}_{m+1})
\sim
\frac{8\pi \as}{Q^2}\,\frac{1}{yz}\
\left(\frac{P_{aa}(z)}{C_a}
- \frac{2y}{(1-z)^2}
\right)
\;.
\end{equation}
The eikonal function is defined in eq. (2.10) of ref.~\cite{NSII} as
\begin{equation}
\overline w_{\La \La}^{\rm eikonal}(\{\hat p,\hat f\}_{m+1})
=
4\pi\as \frac{\hat p_\La \cdot D(\hat p_{m+1},\hat Q) \cdot \hat p_\La}
{(\hat p_{m+1} \cdot \hat p_\La)^2}
\;,
\end{equation}
where $\hat Q = \hat p_\La + \hat p_\Lb$. That is,
\begin{equation}
\overline w_{\La \La}^{\rm eikonal}(\{\hat p,\hat f\}_{m+1})
=
4\pi\as 
\left\{
\frac{2\hat p_\La \cdot \hat Q}
{\hat p_{m+1} \cdot \hat p_\La\ \hat p_{m+1}\cdot \hat Q}
-
\frac{\hat Q^2}
{(\hat p_{m+1} \cdot \hat Q)^2}
\right\}
\;.
\end{equation}
Using eq.~(\ref{eq:momentaIS}) gives
\begin{equation}
\overline w_{\La \La}^{\rm eikonal}(\{\hat p,\hat f\}_{m+1})
=
\frac{16\pi\as}{Q^2}
\frac{1}{yz}
\left\{
\frac{z}
{(1-z)}
-
\frac{yz^2}
{(1-z)^2}
\right\}
\;.
\end{equation}
Thus
\begin{equation}
\begin{split}
\overline w_{\La \La}(\{\hat p,\hat f\}_{m+1})
-{}&\overline w_{\La \La}^{\rm eikonal}(\{\hat p,\hat f\}_{m+1})
\\
\sim{}&
\frac{16\pi \as}{Q^2}\,\frac{1}{yz}\
\left(\frac{P_{aa}(z)}{2C_a}
-\frac{z}{1-z}
- \frac{y (1+z)}{(1-z)}
\right)
\;.
\end{split}
\end{equation}
The term $z/(1-z)$ here removes the $1/(1-z)$ singularity from $P_{aa}(z)$. The third term, proportional to $y$ has a $1/(1-z)$ singularity. This can give a $\log(y)$ contribution to an integration over $z$ down to $(1-z) = y$. However, we can neglect $y \log(y)$. Thus we can throw this term away. This gives
\begin{equation}
\begin{split}
\overline w_{\La \La}(\{\hat p,\hat f\}_{m+1})
-{}&\overline w_{\La \La}^{\rm eikonal}(\{\hat p,\hat f\}_{m+1})
\sim
\frac{8\pi \as}{Q^2}\,\frac{1}{yz\, C_a}\
\left(P_{aa}(z)
-2C_a\,\frac{z}{1-z}
\right)
\;.
\end{split}
\end{equation}

Now we need the interference terms. From ref.~\cite{NScolor}, eq.~(5.3), we have
\begin{equation}
\overline w_{\La k}^{\rm dipole}(\{\hat p,\hat f\}_{m+1})
= 4\pi\as\ \frac{2 \hat p_k\cdot \hat p_\La}
{\hat p_{m+1}\cdot \hat p_k\ \hat p_{m+1}\cdot \hat p_\La}
\;.
\end{equation}
This multiplies $A'_{\La k}$. We use eq.~(7.12) of ref.~\cite{NSspin}:
\begin{equation}
A'_{\La k}(\{\hat p\}_{m+1}) = \frac{\hat p_{m+1}\cdot \hat p_k\ \hat p_\La\cdot \hat Q}
{\hat p_{m+1}\cdot \hat p_k\ \hat p_\La\cdot \hat Q
+ \hat p_{m+1}\cdot \hat p_\La\ \hat p_k\cdot \hat Q}
\;.
\end{equation}
The product is
\begin{equation}
\begin{split}
\label{eq:Aakwak1}
A'_{\La k}(\{\hat p\}_{m+1})\, \overline w_{\La k}^{\rm dipole}(\{\hat p,\hat f\}_{m+1})
={}&
\frac{16\pi\as\ }
{y\,Q^2}\
\frac{\hat p_k\cdot \hat p_\La}
{\hat p_{m+1}\cdot \hat p_k
+ yz\, \hat p_k\cdot \hat Q}
\;.
\end{split}
\end{equation}

To proceed further, we note that we need the function in eq.~(\ref{eq:Aakwak1}) in the limit of small $y$. The momentum $\hat p_k$ is related to the momentum $p_k$ by a Lorentz transformation that becomes the unit operator when $y \to 0$. Thus we can neglect the difference between $\hat p_k$, which varies as we integrate over splitting variables $(y,z,\phi)$, and $p_k$, which is fixed. For this reason, we substitute $p_k$ for $\hat p_k$ in eq.~(\ref{eq:Aakwak1}). Then we use
\begin{equation}
\label{eq:pkIS}
p_k = A_k\big(
(1+\cos\theta_{\La k})\,p_\La + (1-\cos\theta_{\La k})\,p_\Lb + 
\sqrt{Q^2}\sin\theta_{\La k}\ u_\perp
\big)
\;.
\end{equation}
Here $A_k$ is a normalization factor that cancels in eq.~(\ref{eq:Aakwak1}) and $\theta_{\La k}$ is the angle between the three-vector parts of $p_k$ and $p_\La$ in the frame in which $Q$ has only a time component, while $u_\perp$ is a vector transverse to $p_\La$ and $p_\Lb$ with $u_\perp^2 = -1$. Using the parameterizations in eqs.~(\ref{eq:momentaIS}) and (\ref{eq:pkIS}), we find
\begin{equation}
\begin{split}
&\frac{p_k\cdot \hat p_\La }
{\hat p_{m+1}\cdot p_k 
+ yz\,p_k\cdot \hat Q} 
\\&\ =
\frac{(1-\cos\theta_{\La k})}
{2(1 + \cos\theta_{\La k})yz^2
+(1 - \cos\theta_{\La k})(1-z)
-2z\sin\theta_{\La k}\cos\phi \sqrt{y(1-z-yz)}}
\;.
\end{split}
\end{equation}

We can perform the averaging of this over $\phi$ exactly:
\begin{equation}
\begin{split}
\int_0^{2\pi}&\frac{d\phi}{2\pi}\,
\frac{p_k\cdot \hat p_\La }
{\hat p_{m+1}\cdot \hat p_k 
+ yz\,p_k\cdot \hat Q} 
\\&=
\frac{1 - \cos\theta_{\La k}}
{\sqrt{(1 - \cos\theta_{\La k})^2 (1-z)^2 
+ 4 y^2 z^3[z(1 + \cos\theta_{\La k})^2 + \sin^2(\theta_{\La k})]}}
\;.
\end{split}
\end{equation}
We can write this in a suggestive form as
\begin{equation}
\label{eq:phiintegralak}
\int_0^{2\pi}\frac{d\phi}{2\pi}\,
\frac{\hat p_k\cdot \hat p_\La }
{\hat p_{m+1}\cdot \hat p_k 
+ yz\,\hat p_k\cdot \hat Q} 
=
\frac{1}
{\sqrt{(1-z)^2 +  y^2 z^2 /\Psi_{\La k}(z)^2}}
\;,
\end{equation}
where
\begin{equation}
\label{eq:Psiak}
\Psi_{\La k}(z) = \frac{1 -  \cos\theta_{\La k}}
{\sqrt{4 z[z(1 + \cos\theta_{\La k})^2 + \sin^2(\theta_{\La k})]}}
\;.
\end{equation}
We are interested in this in the small $y$ limit, in which $y^2 z^2 /\Psi_{\La k}(z)^2 \ll 1$. Then the second term in the denominator of eq.~(\ref{eq:phiintegralak}) is non-negligible only when $z$ is close to 1. Thus we can replace $\Psi_{\La k}(z)$ by $\psi_{\La k}$ defined in eq.~(\ref{eq:psikl}):
\begin{equation}
\label{eq:psiakdefbis}
\Psi_{\La k}(1) = \psi_{\La k} =  \frac{1 -  \cos\theta_{\La k}}
{\sqrt{8  (1 + \cos\theta_{\La k})}}
\;.
\end{equation}
Thus we use 
\begin{equation}
\label{eq:phiintegralak2}
\int_0^{2\pi}\frac{d\phi}{2\pi}\,
\frac{\hat p_k\cdot \hat p_\La }
{\hat p_{m+1}\cdot \hat p_k 
+ yz\,\hat p_k\cdot \hat Q} 
\approx
\frac{1}
{\sqrt{(1-z)^2 +  y^2 z^2 /\psi_{\La k}^2}}
\;.
\end{equation}

We may note that the angle $\theta_{\La k}$ is small when partons ``a'' and $k$ are the daughter partons of a previous initial state splitting that was nearly collinear. It is allowed in \textsc{Deductor} to have an initial state splitting with small momentum fraction $z_{\La k}$, so that the new initial state parton ``a'', which is the mother parton for the next splitting, has a much larger momentum fraction than the previous initial state parton. In this case, the virtualities of the initial state partons in successive splittings are not strongly ordered, so that one can have $y > 1 - \cos\theta_{\La k}$. This regime of multi-regge kinematics is discussed in section 5.4 of ref.~\cite{ShowerTime}.  This is the opposite kinematic regime from that of threshold logarithms, so we ignore this possibility in this paper. However, we still use $y^2 z^2$ instead of just $y^2$ in the denominator of eq.~(\ref{eq:phiintegralak2}) in order to keep the result reasonably accurate even when $y \gtrsim 1 - \cos\theta_{\La k}$.

We can now assemble our results:
\begin{equation}
\begin{split}
\label{eq:lambdaRresult3}
\overline\lambda^{\Vreal}_{\La k}(\{p,f\}_{m},t)
\approx{}&
\frac{\as}{2\pi}\,
\int_0^{1/(1+y)}\!\frac{dz}{z}\,
\sum_{\hat a}
\frac{
f_{\hat a/A}(\eta_{\La}/z,y Q^2)}
{f_{a/A}(\eta_{\La},y Q^2)}
\\&\times \Bigg[
\theta(k = \La)\,\theta(\hat a \ne a)\,\theta(a \ne \Lg)
\frac
{n_\Lc(a)}
{n_\Lc(\hat a)}\,
\frac{C_\LF}{T_\LR}
\frac{P_{a\hat a}(z)}{C_\LF}
\\&\qquad
+\theta(k = \La)\,\theta(\hat a \ne a)\,\theta(a = \Lg)\,
\frac
{n_\Lc(a)}
{n_\Lc(\hat a)}\,
\frac{T_\LR}{C_\LF}\,
\frac{P_{a\hat a}(z)}{C_\LA}
\\&\qquad
+\theta(k = \La)\,\theta(\hat a = a)\,
\frac{1}{C_a}
\left(P_{aa}(z)
-2C_a\,\frac{z}{1-z}
\right)
\\&\qquad -
\theta(k\ne \La)\,\theta(\hat a = a)\,
\frac{2z}
{\sqrt{(1-z)^2 + y^2 z^2/\psi_{\La k}^2}}\,
\Bigg]
\;.
\end{split}
\end{equation}
In the first term, for $(\hat a,a,\hat f_{m+1}) = (\Lg,q,\bar q)$ and $(\hat a,a,\hat f_{m+1}) = (\Lg,\bar q, q)$, we have $n_\Lc(a) = N_\Lc$ and $n_\Lc(\hat a) = (N_\Lc^2 - 1)$. Thus $[n_\Lc(a)/n_\Lc(\hat a)]\times C_\LF/T_\LR = 1$. Also, $C_\LF = C_a$. In the second term, for for $(\hat a,a,\hat f_{m+1}) = (q,\Lg, q)$ and $(\hat a,a,\hat f_{m+1}) = (\bar q,\Lg, \bar q)$, we have $n_\Lc(a) = (N_\Lc^2 - 1)$ and $n_\Lc(\hat a) = N_\Lc$. Thus $[n_\Lc(a)/n_\Lc(\hat a)]\times T_\LR/C_\LF = 1$. Also, $C_\LA = C_a$. After inserting $\overline\lambda^{\Vreal}_{\La k}$ into eqs.~(\ref{eq:calRISl}) and (\ref{eq:Rak0}), this gives
\begin{equation}
\begin{split}
\label{eq:Rresultak1}
\Vreal_{\La}(t)&\sket{\{p,f,c',c\}_{m}}
\\
={}&
\frac{\as}{2\pi}\,
\int_0^{1/(1+y)}\!dz\,\sum_{\hat a}\
\frac{
f_{\hat a/A}(\eta_{\La}/z, y Q^2)}
{z f_{a/A}(\eta_{\La}, y Q^2)}
\\&\times \Bigg\{
\frac{1}{C_a}
\left(P_{a\hat a}(z)
- \delta_{a \hat a} \frac{2C_a\,z}{1-z}
\right)
\frac{1}{2}
\big([(\bm T_\La\cdot \bm T_\La)\otimes 1] 
+ [1 \otimes (\bm T_\La\cdot \bm T_\La)]\big)
\\&\qquad -
\sum_{k\ne \La}\,\delta_{a \hat a}\,
\frac{2z}
{\sqrt{(1-z)^2 + y^2 z^2/\psi_{\La k}^2}}\,
\frac{1}{2}
\big([(\bm T_\La\cdot \bm T_k)\otimes 1] + [1 \otimes (\bm T_\La\cdot \bm T_k)]\big)
\Bigg\}
\\&\times
\sket{\{p,f,c',c\}_{m}}
\;.
\end{split}
\end{equation}

In the first term in eq.~(\ref{eq:Rresultak1}), we can replace $(\bm T_\La\cdot \bm T_\La)$ by $C_a$. We divide the second term  into three terms by defining $\Delta_{\La k}(z,y)$ according to
\begin{equation}
\begin{split}
\label{eq:wakdef}
\frac{1}
{\sqrt{(1-z)^2 + y^2 z^2/\psi_{\La k}^2}} ={}& 
\frac{1}{1-z}
-  \Delta_{\La k}(z,y)
\;.
\end{split}
\end{equation}
This gives
\begin{equation}
\begin{split}
\label{eq:Rresultak2}
\Vreal_{\La}(t)&\sket{\{p,f,c',c\}_{m}} 
\\={}&
\frac{\as}{2\pi}\,
\int_0^{1/(1+y)}\!dz
\\&\times \Bigg\{
\sum_{\hat a}
\frac{
f_{\hat a/A}(\eta_{\La}/z,y Q^2)}
{zf_{a/A}(\eta_{\La},y Q^2)}
\left(P_{a a}(z)
- \frac{2C_a\,z}{1-z}
\right)
[1\otimes 1] 
\\&\qquad -
\sum_{k\ne \La}\,
\left[
\frac{
f_{a/A}(\eta_{\La}/z,y Q^2)}
{f_{a/A}(\eta_{\La},y Q^2)}
-1
\right]
\frac{1}{1-z}\,
\big([(\bm T_\La\cdot \bm T_k)\otimes 1] + [1 \otimes (\bm T_\La\cdot \bm T_k)]\big)
\\&\qquad +
\sum_{k\ne \La}
\left[
\frac{
f_{a/A}(\eta_{\La}/z,y Q^2)}
{f_{a/A}(\eta_{\La},y Q^2)}
-1
\right]
\Delta_{\La k}(z,y)\,
\big([(\bm T_\La\cdot \bm T_k)\otimes 1] + [1 \otimes (\bm T_\La\cdot \bm T_k)]\big)
\\&\qquad -
\sum_{k\ne \La}\,
\frac{1}{\sqrt{(1-z)^2 + y^2 z^2/\psi_{\La k}^2}}\,
\big([(\bm T_\La\cdot \bm T_k)\otimes 1] + [1 \otimes (\bm T_\La\cdot \bm T_k)]\big)
\Bigg\}
\\&\times
\sket{\{p,f,c',c\}_{m}}
\;.
\end{split}
\end{equation}

In the last term, we can perform the $z$-integration approximately. For $2 y/(1 - \cos \theta_{\La k}) \ll 1$, one easily finds
\begin{equation}
\begin{split}
\int_0^{1/(1+y)}&\!dz
\frac{1}
{\sqrt{(1-z)^2 + y^2 z^2 /\psi_{\La k}^2}}
\approx
\log\!\left[
\frac{1 - \cos\theta_{\La k}}
{2y}
\right]
\;.
\end{split}
\end{equation}
We note that for $k = \Lb$, we have $\cos \theta_{\La \Lb} = -1$ and $1/\psi_{\La \Lb}^2 = 0$. Then the integral is just $- \log(y)$. This gives
\begin{equation}
\begin{split}
\label{eq:Rresultak3}
\Vreal_{\La}&(t)\sket{\{p,f,c',c\}_{m}}
\\
={}&
\bigg[
\frac{\as}{2\pi}\,
\int_0^{1/(1+y)}\!dz
\\&\times \Bigg\{
\sum_{\hat a}
\frac{
f_{\hat a/A}(\eta_{\La}/z,y Q^2)}
{z f_{a/A}(\eta_{\La},y Q^2)}
\left(P_{a\hat a}(z)
- \delta_{a \hat a} \frac{2C_a\,z}{1-z}
\right)
[1\otimes 1] 
\\&\quad - 
\sum_{k\ne \La}\,
\left[
\frac{
f_{a/A}(\eta_{\La}/z,y Q^2)}
{f_{a/A}(\eta_{\La},y Q^2)}
- 1
\right]
\frac{1}{1-z}\,
\big([(\bm T_\La\cdot \bm T_k)\otimes 1] + [1 \otimes (\bm T_\La\cdot \bm T_k)]\big)
\\&\quad +
\sum_{k\ne \La,\Lb}
\left[
\frac{
f_{a/A}(\eta_{\La}/z,y Q^2)}
{f_{a/A}(\eta_{\La},y Q^2)}
-1
\right]
\Delta_{\La k}(z,y)\,
\big([(\bm T_\La \cdot \bm T_k)\otimes 1] + [1 \otimes (\bm T_\La \cdot \bm T_k)]\big)
\Bigg\}
\\&\quad -
\frac{\as}{2\pi}\,
\sum_{k\ne \La}\,
\left(
\log\!
\left[
\frac
{1 - \cos\theta_{\La k}}{2}
\right]
- \log(y)
\right)
\big([(\bm T_\La\cdot \bm T_k)\otimes 1] + [1 \otimes (\bm T_\La\cdot \bm T_k)]\big)
\bigg]
\\&\times
\sket{\{p,f,c',c\}_{m}}
\;.
\end{split}
\end{equation}
Now there are three $k \ne \La$ terms. The first has the good feature that its only $k$ dependence is in the color factor, so that we can sum it over $k$. In the second term, we note that by its construction, $\Delta_{\La k}(z,y)$ has a $1/(1-z)$ singularity for $(1-z) \ll y/\psi_{\La k}$ but is suppressed compared to $1/(1-z)$ when $(1-z) \gg y/\psi_{\La k}$. Thus the second term has the good feature that, because of the structure of $\Delta_{\La k}(z,y)$, the only important integration region for the $z$-integration is $0 < (1-z) \lesssim y/\psi_{\La k}$. We also note that $1/\psi_{\La k} = 0$ when $\cos \theta_{ak} = -1$. In this limit, $\Delta_{\La k}(z,y) = 0$. We have $\cos \theta_{ak} = -1$ for $k = \Lb$, so $\Delta_{\La \Lb}(z,y) = 0$. This eliminates one term in our sum over $k$.  The third term has the good feature that we have been able to integrate it, at least approximately.

In the second term in eq.~(\ref{eq:Rresultak3}), we can sum over $k$ using $\sum_{k \ne \La} (\bm T_\La\cdot \bm T_k) = -(\bm T_\La\cdot \bm T_\La) \to -C_a$. In the last term, we can separate the $\log(1/y)$ contribution and perform the color sum in the same way. This gives
\begin{equation}
\begin{split}
\label{eq:calRa2}
\Vreal_{\La}(t)&\sket{\{p,f,c',c\}_{m}}
\\
={}& \bigg[
\frac{\as}{2\pi}\,
\int_0^{1/(1+y)}\!dz 
\\&\times \Bigg\{
\sum_{\hat a}
\left(\frac{
f_{\hat a/A}(\eta_{\La}/z,y Q^2)}
{zf_{a/A}(\eta_{\La},y Q^2)}
P_{a\hat a}(z)
- \delta_{a\hat a}\frac{2C_a}{1-z}
\right)
[1\otimes 1]
\\&\quad +
\sum_{k\ne \La,\Lb}
\left[\frac{
f_{a/A}(\eta_{\La}/z,y Q^2)}
{f_{a/A}(\eta_{\La},y Q^2)} - 1\right]
\Delta_{\La k}(z,y)
\big([(\bm T_\La\cdot \bm T_k)\otimes 1] + [1 \otimes (\bm T_\La\cdot \bm T_k)]\big)
\Bigg\}
\\& -
\sum_{k\ne \La,\Lb}
\frac{\as}{2\pi}\,
\log\!
\left[
\frac
{1 - \cos\theta_{\La k}}{2}
\right]\,
\big([(\bm T_\La\cdot \bm T_k)\otimes 1] + [1 \otimes (\bm T_\La\cdot \bm T_k)]\big)
\\&
+\frac{\as}{2\pi}\,2 C_a
\log\left[\frac{1}{y}\right]
[1\otimes 1]
\bigg]
\\&\times
\sket{\{p,f,c',c\}_{m}}
\;.
\end{split}
\end{equation}
We will use this result in eq.~(\ref{eq:calRa2encore}).

\section{Calculation of the virtual graphs}
\label{sec:CalculateS}

The Sudakov exponent $\Vvirt$ that appears in eq.~(\ref{eq:evolutionbis}) consists of two terms, as given in eq.~(\ref{eq:VfromVpertandF}): a term $\Vvirt^{\rm pert}(t)$ that comes from virtual graphs and a term that accounts for the evolution of the parton distribution functions. In this appendix, we outline the calculation of $\Vvirt^{\rm pert}(t)$.

Consider first an operator $\Vvirt^{\rm pert}_{\rm tot}$ that corresponds to the one loop virtual graphs that contribute to shower evolution. (More precisely, because of the minus sign in eq.~(\ref{eq:evolutionbis}), $\Vvirt^{\rm pert}_{\rm tot}$ corresponds to the negative of the one loop virtual graphs.) In these graphs, we integrate over a loop momentum $k$. There will be ultraviolet divergences that come from one loop corrections to QCD propagators and vertices. We suppose that these are removed by renormalization. There may be an additional ultraviolet divergence that arises from letting the scale of the loop momentum be much larger than the scale $Q^2$ of the hard interaction that initiates the shower. This can happen if we make the approximation $Q^2 \to \infty$ inside the graph. We should simply arrange to have a regulator in the integrand that eliminates such a divergence.

The operator $\Vvirt^{\rm pert}_{\rm tot}$ will also have infrared divergences, corresponding to the integration regions in which $k \to 0$ or $k$ becomes collinear with the momentum of one of the external lines of the graph. We regulate the infrared divergences with dimensional regulation or some other method. 

We will represent $\Vvirt^{\rm pert}_{\rm tot}$ as an integral over a shower time $t$, as we have done for real emission diagrams:
\begin{equation}
\label{eq:Vvirtsum}
\Vvirt^{\rm pert}_{\rm tot} = \int\! dt\ \Vvirt^{\rm pert}(t)
\;.
\end{equation}
The shower time $t$ corresponds to the negative of the logarithm of the hardness scale of the integrand, in analogy with the definition of $t$ in real emission graphs. Thus the infrared divergences are associated with $t \to \infty$. This general idea does not, however, tell us exactly how to define $t$. One way to proceed would be to use dimensional regulation and subtract the infrared poles. Then the result would depend on a parameter $\mu^2$. Then we could identify $e^{-t}$ with $\mu^2/Q^2$ and $\Vvirt^{\rm pert}$ with the derivative of the graphs with respect to $\log(Q^2/\mu^2)$. However, in eq.~(\ref{eq:ydefVlk}) below, we will make a more direct identification based on the physical meaning of our version of the shower time. 

Using the notation analogous to that of eq.~(\ref{eq:Vrealsum}) for real emission diagrams, we can write
\begin{equation}
\label{eq:VirtualTotal}
\Vvirt^{\rm pert}(t)
= \sum_{l=\La, \Lb, 1,\cdots, m} \Vvirt_l^{\rm pert}(t)
\;.
\end{equation}
The part associated with splitting of parton $l$ is $\Vvirt_l^{\rm pert}(t)$. As in eq.~(\ref{eq:calRl}) for real emission graphs, we write $\Vvirt_l^{\rm pert}(t)$ as
\begin{equation}
\label{eq:Virtuall}
\Vvirt_l^{\rm pert}(t) = \Vvirt_{ll}^{\rm pert}(t) + \sum_{k \ne l} \Vvirt_{l k}^{\rm pert}(t)
\;.
\end{equation}
The first term describes self-energy interactions. In the second term, a virtual gluon is exchanged between parton $l$ and parton $k$. The sum includes all parton labels $k = \La, \Lb, 1, \dots, m$ except for $k = l$. The operators $\Vvirt_{l k}(t)$ have the color structure
\begin{equation}
\begin{split}
\label{eq:Slkcolordef}
\Vvirt_{l k}^{\rm pert}(t)&\sket{\{p,f,c',c\}_{m}} 
\\={}& 
\big\{S^{\LL}_{l k}(\{p,f\}_{m};t)
[(\bm{T}_l\cdot \bm{T}_k)\otimes 1]
+ S^{\LR}_{l k}(\{p,f\}_{m};t)
[1 \otimes (\bm{T}_l\cdot \bm{T}_k)]
\big\}
\\
&\times \sket{\{p,f,c',c\}_{m}}
\;.
\end{split}
\end{equation}
As in appendix \ref{sec:CalculateV}, $T^a_k$ inserts a color generator matrix $T^a$ on line $k$. The functions $S^{\LR}_{l k}(t)$ are just the complex conjugates of the functions $S^{\LL}_{l k}(t)$, so we need only to define and analyze the functions $S^{\LL}_{l k}(t)$. 

We will begin with the case of interference diagrams: $k \ne l$. We start with the case that both $l$ and $k$ represent final state partons. Then we will look at the case in which both $l$ and $k$ represent initial state partons. Finally we let one of the partons be in the initial state while the other is in the final state. Once we have covered interference diagrams, we look at self-energy diagrams: $k = l$.

\subsection{Final state interference diagrams}
\label{sec:VirtualFSInterference}

Let's work with $S^{\LL}_{l k}(\{p,f\}_{m};t)$ in the case that $l$ and $k$ are (different) final state partons in the ket amplitude. We look at the contribution to $S^{\LL}_{l k}(\{p,f\}_{m};t)$ from gluon exchange between two partons. As we do throughout this paper, we take all partons to be massless. The momenta of partons $l$ and $k$ are $p_l$ and $p_k$. The gluon carries momentum $q$ from line $k$ to line $l$, so that, inside the loop, line $l$ carries momentum $p_l - q$ and line $k$ carries momentum $p_k + q$. 

In analyzing the virtual graphs, we follow as much as possible the treatment of the real graphs that we have used in refs.~\cite{NSI, NSII, NSspin}. In particular, this means that we calculate the virtual graphs in Coulomb gauge.

\subsubsection{Integral in Coulomb gauge}

We start with the full interference graph. We will want to identify the shower time, but we have not done that, so we start with the integral over $t$ of the functions that we ultimately want. Also, when a gluon is exchanged between partons $l$ and $k$, it is undefined which is the primary emitter and which is playing only a helping role. That is, we want $S^{\LL}_{l k}$ to represent an emission from parton $l$ with parton $k$ as helper, while $S^{\LL}_{kl}$ will represent an emission from parton $k$ with parton $l$ as helper. We have not yet defined how the total graph is partitioned into these two parts, so we begin with $S^{\LL}_{l k} + S^{\LL}_{kl}$. Thus we start with the definition (including the minus sign in eq.~(\ref{eq:evolutionbis}), so that we write the negative of the usual Feynman diagram),
\begin{equation}
\begin{split}
\label{eq:Glk0}
\int\!dt\  \big[
S^{\LL}_{l k}(\{p,f\}_{m};t) &
+ S^{\LL}_{kl}(\{p,f\}_{m};t)
\big]
\\
={}&
\mi\, \frac{\as}{(2\pi)^{3}}\,
\int\! d\vec q \int\! dE\
\frac{2 p_l\cdot D(q)\cdot p_k}
{(q\cdot p_l - \mi\epsilon)(q\cdot p_k + \mi \epsilon)(q^2 + \mi\epsilon)}\,
\theta(|\vec q\,| < M)
\;.
\end{split}
\end{equation}
We work in Coulomb gauge in a reference frame in which $\vec Q = 0$. The integration variables are defined by $q = (E,\vec q\,)$ in this frame. The integral is both ultraviolet and infrared divergent. For technical reasons have inserted an ultraviolet cutoff $|\vec q\,| < M$, where we will take $M$ to be large. The infrared divergence is associated with large positive values of the shower time $t$. We could imagine that the infrared divergences are regulated, but once we select a fixed value of $t$, the regulation is not needed. For this reason, we do not specify a regulation method. In Coulomb gauge at a fixed shower time $t$, only the soft integration region for $q$ is important. For this reason, it is appropriate to use the eikonal approximation (as in \textsc{Deductor}). We have applied the eikonal approximation in eq.~(\ref{eq:Glk0}). The numerator of the gluon propagator in Coulomb gauge is
\begin{equation}
D(q)^{\mu\nu} =
- g^{\mu\nu} 
-\frac{q^\mu \tilde q^\nu + \tilde q^\mu q^\nu - q^\mu q^\nu}{|\vec q\,|^2}
\;,
\end{equation}
where $\tilde q = (0,\vec q\,)$. 

Now, the integrand in eq.~(\ref{eq:Glk0}) is complicated because of the numerator $D(q)$ of the gluon propagator in Coulomb gauge. However, we can simplify it by writing
\begin{equation}
\begin{split}
\label{eq:CoulombSplitlk}
S^{\LL}_{l k}(\{p,f\}_{m};t) 
={}& 
S^{\LL}_{l k}(\{p,f\}_{m};t;{\rm dipole})  +
S^{\LL}_{ll}(\{p,f\}_{m};t;{\rm eikonal})
\end{split}
\end{equation}
and
\begin{equation}
\begin{split}
\label{eq:CoulombSplitkl}
S^{\LL}_{kl}(\{p,f\}_{m};t) 
={}& 
S^{\LL}_{kl}(\{p,f\}_{m};t;{\rm dipole})  +
S^{\LL}_{kk}(\{p,f\}_{m};t;{\rm eikonal})
\;.
\end{split}
\end{equation}
Here
\begin{equation}
\begin{split}
\label{eq:Glleik0}
\int\!dt\ S^{\LL}_{ll}(\{p,f\}_{m};&t;{\rm eikonal}) 
\\
={}&
\mi\, \frac{\as}{(2\pi)^{3}}\,
\int\! d\vec q \int\! dE\
\frac{p_l\cdot D(q)\cdot p_l}
{(q\cdot p_l - \mi\epsilon)^2 (q^2 + \mi\epsilon)}\,
\theta(|\vec q\,| < M)
\;,
\end{split}
\end{equation}
with an analogous definition of $S^{\LL}_{kk}(\{p,f\}_{m};t;{\rm eikonal})$. This is the eikonal approximation to the self-energy graph for parton $l$ in Coulomb gauge. Recall that we associate $S^{\LL}_{l k}$ with emissions from parton $l$ and $S^{\LL}_{kl}$ with emissions from parton $k$. In keeping with that interpretation, we count $S^{\LL}_{ll}(\{p,f\}_{m};t;{\rm eikonal})$ as contributing to $S^{\LL}_{l k}(\{p,f\}_{m};t)$ and we count $S^{\LL}_{kk}(\{p,f\}_{m};t;{\rm eikonal})$ as contributing to $S^{\LL}_{kl}(\{p,f\}_{m};t)$.

This defines $S^{\LL}_{l k}(\{p,f\}_{m};t;{\rm dipole})$ and $S^{\LL}_{kl}(\{p,f\}_{m};t;{\rm dipole})$. With simple algebra, we find
\begin{equation}
\begin{split}
\label{eq:Glk1}
\int\!dt\,  \big[
S^{\LL}_{l k}&(\{p,f\}_{m};t;{\rm dipole}) 
+ S^{\LL}_{kl}(\{p,f\}_{m};t;{\rm dipole})
\big]
\\
={}&
\mi\, \frac{\as}{(2\pi)^{3}}\,
\int\! d\vec q \int\! dE \
\frac{- P_{lk}\cdot D(q)\cdot P_{lk}}
{(q\cdot p_l - \mi\epsilon)^2(q\cdot p_k + \mi \epsilon)^2(q^2 + \mi\epsilon)}\,
\theta(|\vec q\,| < M)
\;,
\end{split}
\end{equation}
where
\begin{equation}
P_{lk} = q\cdot p_l\, p_k - q\cdot p_k\, p_l
\;.
\end{equation}
Since $q\cdot P_{lk} = 0$, none of the $q$ dependent terms in $D(q)$ contribute and we are left with
\begin{equation}
- P_{lk}\cdot D(q)\cdot P_{lk} = P_{lk}^2 
= - 2 q\cdot p_l\, q\cdot p_k\, p_l\cdot p_k
\;.
\end{equation}
Thus
\begin{equation}
\begin{split}
\label{eq:Glk2}
\int\!dt\  \big[
S^{\LL}_{l k}&(\{p,f\}_{m};t;{\rm dipole}) 
+ S^{\LL}_{kl}(\{p,f\}_{m};t;{\rm dipole})
\big]
\\
={}&
-\mi\, \frac{\as}{(2\pi)^{3}}\,
\int\! d\vec q \int\! dE \
\frac{2\,p_l\cdot p_k}
{(q\cdot p_l - \mi\epsilon)(q\cdot p_k + \mi \epsilon)(q^2 + \mi\epsilon)}\,
\theta(|\vec q\,| < M)
\;.
\end{split}
\end{equation}
That is, we get the familiar dipole formula for one gluon exchange in Feynman gauge.

\subsubsection{Dipole part}

We now analyze the Feynman gauge eikonal integral in eq.~(\ref{eq:Glk2}). We write 
\begin{equation}
\frac{1}{q^2 + \mi \epsilon} = \frac{1}{2 |\vec q|}
\left[
\frac{1}{E - |\vec q| + \mi \epsilon} - \frac{1}{E + |\vec q| - \mi \epsilon}
\right]
\;.
\end{equation}
In the first term, the gluon propagates forward in time from parton $k$ to parton $l$. In the second term, the gluon propagates forward in time from parton $l$ to parton $k$. In the second term, we redefine $q \to - q$, so that the direction of $q$ is the direction of propagation forward in time. Then
\begin{equation}
\begin{split}
\label{eq:Glk3}
\int\!dt\ & \big[S^{\LL}_{l k}(\{p,f\}_{m};t;{\rm dipole}) 
+ S^{\LL}_{kl}(\{p,f\}_{m};t;{\rm dipole})\big]
\\
={}&
-\mi\, \frac{\as}{(2\pi)^{3}}\,2 p_l\cdot p_k
\int\! \frac{d\vec q }{2 |\vec q|}\, \theta(|\vec q\,| < M)
\int\! dE\
\\&\times
\frac{1}{E - |\vec q| + \mi \epsilon}
\left[
\frac{1}
{(q\cdot p_l - \mi\epsilon)(q\cdot p_k + \mi \epsilon)}
+
\frac{1}
{(q\cdot p_l + \mi\epsilon)(q\cdot p_k - \mi \epsilon)}
\right]
\;.
\end{split}
\end{equation}

We let
$p_l =  |\vec p_l|\, v_l$
and
$p_k = |\vec p_k|\, v_k$, where
\begin{equation}
\begin{split}
v_l ={}& (1,\vec v_l)
\;,
\\
v_k ={}& (1,\vec v_k)
\;,
\end{split}
\end{equation}
with $\vec v_\La^{\,2} = \vec v_k^{\,2} = 1$. Also, we define $Q = E_Q(1,\vec 0)$. Then
\begin{equation}
\begin{split}
\label{eq:Glk4}
\int\!dt\ &\big[S^{\LL}_{l k}(\{p,f\}_{m};t;{\rm dipole}) 
+ S^{\LL}_{kl}(\{p,f\}_{m};t;{\rm dipole})\big]
\\
={}&
-\mi\, \frac{\as}{(2\pi)^{3}}\,2 v_l\cdot v_k
\int\! \frac{d\vec q }{2 |\vec q|}\, \theta(|\vec q\,| < M)
\int\! dE\
\frac{1}{E - |\vec q| + \mi \epsilon}
\\&\times
\left[
\frac{1}
{(E - \vec q\cdot \vec v_l - \mi\epsilon)(E - \vec q\cdot \vec v_k + \mi \epsilon)}
+
\frac{1}
{(E - \vec q\cdot \vec v_l + \mi\epsilon)(E - \vec q\cdot \vec v_k - \mi \epsilon)}
\right]
\;.
\end{split}
\end{equation}

We can immediately perform the $E$-integration to give
\begin{equation}
\begin{split}
\label{eq:Glk5}
\int\!dt\ &\big[S^{\LL}_{l k}(\{p,f\}_{m};t;{\rm dipole}) 
+ S^{\LL}_{kl}(\{p,f\}_{m};t;{\rm dipole})\big]
\\
={}&
-\frac{\as}{(2\pi)^{2}}\,2 v_l\cdot v_k
\int\! \frac{d\vec q }{2 |\vec q|}\, \theta(|\vec q\,| < M)
\\&\times
\left[
\frac{1}
{(|\vec q| - \vec q\cdot \vec v_l)
(\vec q\cdot \vec v_l - \vec q\cdot \vec v_k + \mi \epsilon)}
-
\frac{1}
{(|\vec q| - \vec q\cdot \vec v_k)
(\vec q\cdot \vec v_l - \vec q\cdot \vec v_k - \mi \epsilon)}
\right]
\;.
\end{split}
\end{equation}
Now, we can rewrite this using
\begin{equation}
\frac{1}{x \pm \mi \epsilon} = \frac{1}{[x]_\LP} \mp \mi \pi\, \delta(x)
\;.
\end{equation}
This gives
\begin{equation}
\begin{split}
\label{eq:Glk6}
\int\!dt\ \big[S^{\LL}_{l k} &(\{p,f\}_{m};t;{\rm dipole})
+ S^{\LL}_{kl}(\{p,f\}_{m};t;{\rm dipole})\big]
\\
={}&
-\frac{\as}{(2\pi)^{2}}\,2 v_l\cdot v_k
\int\! \frac{d\vec q }{2 |\vec q|}\,\theta(|\vec q\,| < M)
\\&\times
\bigg\{
\frac{1}{[\vec q\cdot \vec v_l - \vec q\cdot \vec v_k]_\LP}
\left[
\frac{1}
{|\vec q| - \vec q\cdot \vec v_l}
-
\frac{1}
{|\vec q| - \vec q\cdot \vec v_k}
\right]
\\& \qquad - \mi \pi\,
\delta(\vec q\cdot \vec v_k - \vec q\cdot \vec v_l)
\left[
\frac{1}
{|\vec q| - \vec q\cdot \vec v_l}
+
\frac{1}
{|\vec q| - \vec q\cdot \vec v_k}
\right]
\bigg\}
\;.
\end{split}
\end{equation}
That is
\begin{equation}
\begin{split}
\label{eq:Glk7}
\int\!dt\ \big[S^{\LL}_{l k}&(\{p,f\}_{m};t;{\rm dipole}) 
+ S^{\LL}_{kl}(\{p,f\}_{m};t;{\rm dipole})\big]
\\
={}&
-\frac{\as}{(2\pi)^{2}}\,2 v_l\cdot v_k
\int\! \frac{d\vec q }{2 |\vec q|}\,\theta(|\vec q\,| < M)
\bigg\{
\frac{1}
{(|\vec q| - \vec q\cdot \vec v_l)(|\vec q| - \vec q\cdot \vec v_k)}
\\&
- \mi \pi\,
\delta(\vec q\cdot \vec v_k - \vec q\cdot \vec v_l)\,
\left[
\frac{1}
{|\vec q| - \vec q\cdot \vec v_l}
+
\frac{1}
{|\vec q| - \vec q\cdot \vec v_k}
\right]
\bigg\}
\;.
\end{split}
\end{equation}

Now consider the first term. It has essentially the structure of the graphs for the emission of a real gluon with $q^2 = 0$, although it is not quite the same as our real emission factors because it does not contain momentum mappings that allow an on-shell parton to split into two on-shell partons. The integrand has poles at $|\vec q| = \vec q\cdot \vec v_l$ and at $|\vec q| = \vec q\cdot \vec v_k$. This reflects splittings both of parton $l$ and of parton $k$. In order to separate these splittings, we multiply the integrand by $1 = A'_{lk} + A'_{kl}$, where
\begin{equation}
\label{eq:Aprimelk}
A'_{lk} = \frac{|\vec q| - \vec q\cdot \vec v_k}
{(|\vec q| - \vec q\cdot \vec v_k) + (|\vec q| - \vec q\cdot \vec v_l)}
\;.
\end{equation}
The term containing $A'_{lk}$ is associated with $S^{\LL}_{l k}(\{p,f\}_{m};t)$, while the term containing $A'_{kl}$ is associated with $S^{\LL}_{kl}(\{p,f\}_{m};t)$. In the $\mi \pi$ term, the two contributions are actually equal, but we can associate the first with $S^{\LL}_{l k}(\{p,f\}_{m};t)$ and the second with $S^{\LL}_{kl}(\{p,f\}_{m};t)$. This gives
\begin{equation}
\begin{split}
\label{eq:Glk8}
\int\!dt\ S^{\LL}_{l k}(\{p,f\}_{m};t&;{\rm dipole})
\\
={}&
-\frac{\as}{(2\pi)^{2}}\,2 v_l\cdot v_k
\int\! \frac{d\vec q }{2 |\vec q|}\,\theta(|\vec q\,| < M)
\\&\times
\bigg\{
\frac{1}
{(|\vec q| - \vec q\cdot \vec v_l)}\,
\frac{1}
{(|\vec q| - \vec q\cdot \vec v_k) + (|\vec q| - \vec q\cdot \vec v_l)}
\\&\qquad
- \mi \pi\,
\delta(\vec q\cdot \vec v_k - \vec q\cdot \vec v_l)\,
\frac{1}
{|\vec q| - \vec q\cdot \vec v_l}
\bigg\}
\;.
\end{split}
\end{equation}

We can now identify the shower time with
\begin{equation}
\begin{split}
\label{eq:ydefVlk}
y = \frac{1}{E_Q}\
(|\vec q| - \vec q\cdot \vec v_l)
\;.
\end{split}
\end{equation}
This matches the definition (\ref{eq:ydef}) that we used for a real gluon emission.\footnote{In the case of real gluon emission, we take $\vec q$ and $\vec l$ to be the parton momenta after the splitting. The momentum of parton $l$ before the splitting is $\vec l + \vec q$, but here we use just $\vec l$ because $\vec q$ is small.}  We introduce this as a delta function, recognizing that $\int\!dt \cdots$ is equivalent to $\int\!d\log(y) \cdots$. Thus
\begin{equation}
\begin{split}
\label{eq:Glk9}
S^{\LL}_{l k}(\{p,f\}_{m};&t;{\rm dipole})
\\
={}&
-\frac{\as}{(2\pi)^{2}}\,2 v_l\cdot v_k
\int\! \frac{d\vec q }{2 |\vec q|}\,\theta(|\vec q\,| < M)\,
\delta\!\left[\log(E_Q y) - \log\left(
|\vec q| - \vec q\cdot \vec v_l\right)
\right]
\\&\times
\bigg\{
\frac{1}
{E_Q y}\,
\frac{1}
{(|\vec q| - \vec q\cdot \vec v_k) + E_Q y}
- \mi \pi\,
\delta(\vec q\cdot \vec v_k - \vec q\cdot \vec v_l)\,
\frac{1}
{E_Q y}
\bigg\}
\;.
\end{split}
\end{equation}
We will want to apply different methods for the two integrations.

\subsubsection{Dipole real part}

We examine first the real part of the integral (\ref{eq:Glk9}). We introduce transverse and longitudinal coordinates for $\vec q$:
\begin{equation}
\label{eq:qdef}
\vec q = (1 - z - 2 a_l y)\vec p_l + \vec q_\perp
\;,
\end{equation}
where $\vec q_\perp \cdot \vec p_l = 0$ and where we have defined
\begin{equation}
a_l = \frac{E_Q}{2 |\vec p_l|}
\;,
\end{equation}
as in eq.~(\ref{eq:aldef}). We denote by $\phi$ the azimuthal angle of $\vec q_\perp$ relative to the $(\vec p_k, \vec p_l)$ plane. We first need to find the value of $|\vec q_\perp|$. From eq.~(\ref{eq:ydefVlk}), we have
\begin{equation}
|\vec q|  = E_Q y + (1 - z - 2 a_l y) |\vec p_l|
\;.
\end{equation}
That is
\begin{equation}
|\vec q| = (1 - z) |\vec p_l|
\;.
\end{equation}
This implies that the cutoff $|\vec q\,| < M$ amounts to
\begin{equation}
(1-z) < M/|\vec p_l|
\;.
\end{equation}
We have
\begin{equation}
|\vec q|^2 = (1-z)^2 |\vec p_l|^2
\;.
\end{equation}
But $|\vec q|^2 = (1-z - 2 a_l y)^2 |\vec p_l|^2  + \vec q_\perp^{\,2}$. Thus
\begin{equation}
\begin{split}
\label{eq:qperppartonl}
\vec q_\perp^{\,2} ={}& |\vec p_l|^2 [ (1-z)^2 -  (1-z - 2 a_l y)^2]
\\
={}& |\vec p_l|^2 4 a_l y\,[(1-z) - a_l y]
\;.
\end{split}
\end{equation}
Note that there is a minimum possible value of $(1-z)$, corresponding to $q_\perp^2 = 0$:
\begin{equation}
\label{eq:zupperbound}
(1 - z) >  a_l y
\;.
\end{equation}

We can write the integration over $\vec q$ as
\begin{equation}
\label{eq:qperpjacobian}
d\vec q = \pi |\vec p_l|\, dz\ dq_\perp^2\ \frac{d\phi}{2\pi}
\;.
\end{equation}
To perform the $q_\perp^2$ integration against the delta function, we note that
\begin{equation}
\label{eq:qperpintegral}
\int\! dq_\perp^2\ \delta\!\left[\log(E_Q y) - \log\left( 
(|\vec q| - \vec v_l \cdot \vec q)\right)
\right]
= 2 |\vec q| E_Q y
\;.
\end{equation}
This gives
\begin{equation}
\begin{split}
\label{eq:Glk10}
{\rm Re}\, S^{\LL}_{l k}(\{p,f\}_{m};t;{\rm dipole})
={}&
-\frac{\as}{2\pi}\, v_l\cdot v_k \,
\int_{1 - M/|\vec p_l|}^{1 - a_l y}\!  dz\ \int\!\frac{d\phi}{2\pi}\
\frac{|\vec p_l|}
{(|\vec q| - \vec q\cdot \vec v_k) + E_Q y}
\;.
\end{split}
\end{equation}

We can integrate this over $\phi$:
\begin{equation}
\begin{split}
\int\!\frac{d\phi}{2\pi}\ \frac{|\vec p_l|}
{(|\vec q| - \vec q\cdot \vec v_k) + E_Q y} 
={}& 
\frac{1}{v_k\cdot v_l}\,
\frac{1}
{\sqrt{(1-z)^2 + a_l^2 y^2/\psi_{kl}^2}}
\;,
\end{split}
\end{equation}
where
\begin{equation}
\psi_{kl} = \frac{1 - \cos\theta_{kl}}{\sqrt{8 (1 + \cos\theta_{kl})}}
\;,
\end{equation}
as in eq.~(\ref{eq:psikl}). We now need to integrate this over $z$:
\begin{equation}
\label{eq:g3defstart}
\int_{1 - M/|\vec p_l|}^{1 - a_l y}\!\frac{dz}{\sqrt{(1-z)^2 + a_l^2 y^2/\psi_{kl}^2}}
= \log\left(
\frac{M/|\vec p_l| + \sqrt{M^2/|\vec p_l|^2 + a_l^2 y^2/\psi_{kl}^2}}{a_l y  \left(1 + \sqrt{1 + 1/\psi_{kl}^2}\right)}
\right)
\;,
\end{equation}
Neglecting terms that vanish like a power of $|\vec p_l|/M$ as $|\vec p_l|/M \to 0$, this becomes
\begin{equation}
\begin{split}
\label{eq:g3def1}
\int_{1 - M/|\vec p_l|}^{1 - a_l y}\!\frac{dz}{\sqrt{(1-z)^2 + a_l^2 y^2/\psi_{kl}^2}}
\sim{}&
\log\!\left(
\frac{2}{a_l y  \left(1 + \sqrt{1 + 1/\psi_{kl}^2}\right)}
\right)
+ \log(M/|\vec p_l|)
\\&={}
\log\!\left(
\frac{1 - \cos\theta_{kl}}{2 a_l y}
\right)
+ \log(M/|\vec p_l|)
\;.
\end{split}
\end{equation}
This gives
\begin{equation}
\begin{split}
\label{eq:GlkRe}
{\rm Re}\,S^{\LL}_{l k}(\{p,f\}_{m};t;{\rm dipole})
\approx{}&
-\frac{\as}{2\pi}\,\left[
\log\!\left(
\frac{1 - \cos\theta_{kl}}{2 a_l y}
\right)
+ \log(M/|\vec p_l|)
\right]
\;.
\end{split}
\end{equation}
The $\log(M/|\vec p_l|)$ term will cancel against an identical term in the integral that we subtracted and have to add back.

\subsubsection{Dipole imaginary part}

Now we examine the imaginary part of the integral (\ref{eq:Glk9}), which we rewrite slightly as
\begin{equation}
\begin{split}
\label{eq:GlkImdef}
{\rm Im}\,S^{\LL}_{l k}(\{p,f\}_{m}&;t;{\rm dipole})
\\ ={}& 
\pi
\frac{\as}{(2\pi)^{2}}\,
\frac{2 v_l\cdot v_k}{E_Q y}
\int\! \frac{d\vec q }{2 |\vec q|}\,\theta(|\vec q\,| < M)
\\&\times
\delta\!\left[\log(2E_Q y) - \log\left(
2|\vec q| - \vec q\cdot (\vec v_k + \vec v_l)\right)
\right]
\delta(\vec q\cdot (\vec v_k -  \vec v_l))
\;.
\end{split}
\end{equation}
We can immediately take the limit $M \to \infty$. It will be useful to choose coordinates $(\xi,\eta,\lambda)$ based on the orthogonal vectors $(\vec v_k + \vec v_l)$ and $(\vec v_k - \vec v_l)$:
\begin{equation}
\vec q = \left(
\xi + E_Q y \sqrt{\frac{1 + \cos\theta_{kl}}{1 - \cos\theta_{kl}}}
\right)
\frac{\vec v_k + \vec v_l}{\sin\theta_{kl}}
+ \eta\, \vec u
+ \lambda\,\frac{\vec v_k - \vec v_l}{2(1 - \cos\theta_{kl})}
\;,
\end{equation}
where $\vec u$ is a unit vector orthogonal to $\vec v_k$ and $\vec v_l$. We then have
\begin{equation}
d\vec q = \frac{1}{1 - \cos\theta_{kl}}\, d\xi\,d\eta\,d\lambda
\;.
\end{equation}
We can immediately perform the $\lambda$ integration using
\begin{equation}
\delta(\vec q\cdot (\vec v_k -  \vec v_l)) = \delta(\lambda)
\;.
\end{equation}
For the other delta function, define
\begin{equation}
f(\xi,\eta) = 2|\vec q| - \vec q\cdot (\vec v_k + \vec v_l)
\;.
\end{equation}
The delta function restricts $(\xi,\eta)$ to the surface $f(\xi,\eta) = 2 E_Q y$. On this surface, $2|\vec q| = \vec q\cdot (\vec v_k + \vec v_l) + 2E_Q y$, or
\begin{equation}
4 \vec q^{\,2} = [\vec q\cdot (\vec v_k + \vec v_l) + 2 E_Q y]^2
\;.
\end{equation}
After setting $\lambda = 0$, we find that the surface is a circle in our chosen coordinates:
\begin{equation}
\xi^2 + \eta^2 = \frac{2 E_Q^2 y^2}{1 - \cos\theta_{kl}}
\;.
\end{equation}

Consider what happens if $\xi \to \xi + \delta\xi$ and $\eta \to \eta + \delta\eta$. At the surface $f(\xi,\eta) = 2 E_Q y$, this gives
\begin{equation}
\begin{split}
\label{eq:deltaf1}
\delta f(\xi,\eta) ={}& 
\frac{2}{|\vec q|}
\left\{
\xi
\delta\xi
+ \eta\,\delta\eta
\right\}
\;.
\end{split}
\end{equation}
If we use polar coordinates $\xi = R\cos\theta$ and $\eta = R\sin\theta$ then the surface $f(\xi,\eta) = 2 E_Q y$ is at $R^2 = 2 E_Q^2 y^2/(1 - \cos\theta_{kl})$ and we have
\begin{equation}
\begin{split}
\label{eq:deltaf2}
\delta f(\xi,\eta) ={}& 
\frac{2}{|\vec q|}\,
R\,\delta R
\;.
\end{split}
\end{equation}
This gives
\begin{equation}
\begin{split}
\int\! \frac{d\xi\, d\eta}{2|\vec q|}\ 
\delta\!\left[\log(2E_Q y) - \log\left(
2|\vec q| - \vec q\cdot (\vec v_k + \vec v_l)\right)
\right]
={}& 2\pi\,\frac{E_Q y}{2}
\;.
\end{split}
\end{equation}

Inserting these results into eq.~(\ref{eq:GlkImdef}) and using $1 - \cos\theta_{kl} = v_k\cdot v_l$, we have
\begin{equation}
\begin{split}
\label{eq:GlkIm}
{\rm Im}\,S^{\LL}_{l k}(\{p,f\}_{m};t;{\rm dipole}) ={}& 
\pi\,
\frac{\as}{2\pi}
\;.
\end{split}
\end{equation}

\subsubsection{Dipole total}

Adding eqs.~(\ref{eq:GlkRe}) and (\ref{eq:GlkIm}), we have
\begin{equation}
\begin{split}
\label{eq:Glkdipole}
S^{\LL}_{l k}(\{p,f\}_{m};t;{\rm dipole})
\approx{}&
\frac{\as}{2\pi}\,\left[
-\log\!\left(
\frac{1 - \cos\theta_{kl}}{2 a_l y}
\right)
- \log(M/|\vec p_l|)
+ \mi \pi
\right]
\;.
\end{split}
\end{equation}

\subsubsection{Eikonal self-energy integral}

In order to construct $S^{\LL}_{l k}(\{p,f\}_{m};t)$ in eq.~(\ref{eq:CoulombSplitlk}), we add $S^{\LL}_{ll}(\{p,f\}_{m};t;{\rm eikonal})$, defined in eq.~(\ref{eq:Glleik0}), to $S^{\LL}_{l k}(\{p,f\}_{m};t;{\rm dipole})$.

We thus need to calculate $S^{\LL}_{ll}(\{p,f\}_{m};t;{\rm eikonal})$ in eq.~(\ref{eq:Glleik0}). The product $p_l\cdot D(q)\cdot p_l$ appears. This factor is
\begin{equation}
p_l\cdot D(q)\cdot p_l = 
\frac{q\cdot p_l\, |\vec p_l|}{|\vec q|^2}\
[E + \vec q\cdot \vec v_l]
\;.
\end{equation}
This gives us
\begin{equation}
\begin{split}
\label{eq:Glleik2}
\int\!dt\ &S^{\LL}_{ll}(\{p,f\}_{m};t;{\rm eikonal}) 
\\
={}&
\frac{\mi\, \as}{(2\pi)^{3}}\,
\int\! d\vec q\
\frac{\theta(|\vec q\,| < M)}{|\vec q|^2}
\int\! dE\
\frac{E + \vec q\cdot \vec v_l}
{(E - \vec q\cdot \vec v_l - \mi\epsilon) 
 (E - |\vec q| + \mi\epsilon)
 (E + |\vec q| - \mi\epsilon)}
\;.
\end{split}
\end{equation}
We can perform the $E$-integration to get
\begin{equation}
\begin{split}
\label{eq:Glleik3}
\int\!dt\ S^{\LL}_{ll}(\{p,f\}_{m};t;{\rm eikonal}) 
={}&
\frac{\as}{(2\pi)^{2}}\,
\int\! d\vec q\
\frac{\theta(|\vec q\,| < M)}{2|\vec q|^3}\
\frac{|\vec q| + \vec q\cdot \vec v_l}
{|\vec q| - \vec q\cdot v_l}
\;.
\end{split}
\end{equation}
We recognize the denominator as defining the shower time, so
\begin{equation}
\begin{split}
\label{eq:Glleik4}
S^{\LL}_{ll}(\{p,&f\}_{m};t;{\rm eikonal}) 
\\={}&
\frac{\as}{(2\pi)^{2}}\,
\int\! d\vec q\ \theta(|\vec q\,| < M)\,
\delta\!\left[\log(E_Q y) - \log\left(
|\vec q| - \vec q\cdot \vec v_l\right)
\right]
\frac{E_Q y + 2\vec q\cdot \vec v_l}{2|\vec q|^3\,E_Q y}\
\;.
\end{split}
\end{equation}
Introducing variables $q_\perp, z, \phi$ with the aid of eqs.~(\ref{eq:qdef}), (\ref{eq:zupperbound}), (\ref{eq:qperpjacobian}), and (\ref{eq:qperpintegral}), we have
\begin{equation}
\begin{split}
\label{eq:Glleik6}
S^{\LL}_{ll}(\{p,f\}_{m};t;{\rm eikonal}) 
={}&
\frac{\as}{2\pi}\,
\int_{1- M/|\vec p_l|}^{1 - a_l y}\! dz\ 
\frac{(1-z) - a_l y}
{(1-z)^2}\
\;.
\end{split}
\end{equation}
We can perform the integration to obtain
\begin{equation}
\begin{split}
\label{eq:Glleik7}
S^{\LL}_{ll}(\{p,f\}_{m};t;{\rm eikonal}) 
={}&
\frac{\as}{2\pi}\,
\left[
\log\left(
\frac{M/|\vec p_l|}{a_l y}
\right)
- \frac{M/|\vec p_l| - a_l y}{M/|\vec p_l|}
\right]
\;.
\end{split}
\end{equation}
We want the limit of this for large $M/|\vec p_l|$:
\begin{equation}
\begin{split}
\label{eq:Glleik8}
S^{\LL}_{ll}(\{p,f\}_{m};t;{\rm eikonal}) 
={}&
\frac{\as}{2\pi}\,
\left[
\log\left(
\frac{1}{a_l y}
\right)
- 1
+
\log\left(
\frac{M}{|\vec p_l|}
\right)
\right]
\;.
\end{split}
\end{equation}
\subsubsection{Total $l$-$k$ interference graph}

We put our contributions back together, inserting $G^{\LL}_{lk}(\{p,f\}_{m};t;{\rm dipole})$ from  eq.~(\ref{eq:Glkdipole}) and $G^{\LL}_{ll}(\{p,f\}_{m};t;{\rm eikonal})$ from eq.~(\ref{eq:Glleik8}) into eq.~(\ref{eq:CoulombSplitlk}):
\begin{equation}
\begin{split}
\label{eq:Glktotal}
S^{\LL}_{l k}(\{p,f\}_{m};t)
\approx{}&
\frac{\as}{2\pi}\,\left[
-\log\!\left(
\frac{1 - \cos\theta_{kl}}{2}
\right)
- 1
+ \mi \pi
\right]
\;.
\end{split}
\end{equation}
Here the cutoff dependent terms proportional to $\log(M/|\vec p_l|)$ have cancelled. We will use this result in eq.~(\ref{eq:calGl2}).

\subsection{Initial state interference diagram}%%%%%%%%%%%%%%%%

Let's now look at the case that the active parton is one of the initial state partons, $l = \La$, and the helper parton $k$ is the other, $k = \Lb$. Thus, we examine $S^{\LL}_{\La \Lb}(\{p,f\}_{m};t)$, corresponding to gluon exchange between the two initial state partons. The gluon carries momentum $q$ from line ``a'' to line ``b,'' so that, inside the loop, line ``a'' carries momentum $p_\La - q$ and line ``b'' carries momentum $ p_\Lb + q$. 

We start with the eikonal approximation to the exchange in Coulomb gauge,
\begin{equation}
\begin{split}
\label{eq:Gab0}
\int\!dt\  [S^{\LL}_{\La\Lb}&(\{p,f\}_{m};t)
+ S^{\LL}_{\Lb\La}(\{p,f\}_{m};t)]
\\
={}&
\mi\, \frac{\as}{(2\pi)^{3}}\,
\int\! d\vec q \int\! dE\
\frac{2 p_\La\cdot D(q)\cdot p_\Lb}
{(q\cdot p_\La - \mi\epsilon)(q\cdot p_\Lb + \mi \epsilon)(q^2 + \mi\epsilon)}\,
\theta(|\vec q\,| < M)
\;.
\end{split}
\end{equation}
There is an ultraviolet cutoff $|\vec q\,| < M$ that we eventually remove.

This integral is exactly the same as we had in eq.~(\ref{eq:Glk0}) for the final state case, with $p_l \to p_\La$ and $p_k \to p_\Lb$. We can apply the same treatment, partitioning the integral into two terms using the partitioning function~(\ref{eq:Aprimelk}) and identifying the shower time using eq.~(\ref{eq:ydefVlk}). Thus we can simply use the result in eq.~(\ref{eq:Glktotal}), noting that $\cos\theta_{kl} = -1$:
\begin{equation}
\begin{split}
\label{eq:Gabtotal}
S^{\LL}_{\La\Lb}(\{p,f\}_{m};t)
\approx{}&
\frac{\as}{2\pi}\,\left[
- 1
+ \mi \pi
\right]
\;.
\end{split}
\end{equation}
We will use this result in eq.~(\ref{eq:calGa2}).

\subsection{Initial state - final state interference}%%%%%%%%%%%%%%%

We now examine the case of gluon exchange between one of the initial state partons, say $l = \La$, and a final state parton $k$. Thus, we examine $S^{\LL}_{\La k}(\{p,f\}_{m};t)$ and $S^{\LL}_{k \La}(\{p,f\}_{m};t)$. An exchanged gluon carries momentum $q$ from line ``a'' to line $k$, so that, inside the loop, line ``a'' carries momentum $p_\La - q$ and line $k$ carries momentum $p_k - q$.

We start with the eikonal approximation to the exchange in Coulomb gauge,
\begin{equation}
\begin{split}
\label{eq:Gak0}
\int\!dt\ [S^{\LL}_{\La k}&(\{p,f\}_{m};t) 
+ S^{\LL}_{k \La}(\{p,f\}_{m};t)]
\\
={}&
\mi\, \frac{\as}{(2\pi)^{3}}\,
\int\! d\vec q \int\! dE\
\frac{2 p_\La\cdot D(q)\cdot p_k}
{(q\cdot p_\La - \mi\epsilon)(q\cdot p_k - \mi \epsilon)(q^2 + \mi\epsilon)}\,
\theta(|\vec q\,| < M)
\;.
\end{split}
\end{equation}
The integrand in eq.~(\ref{eq:Gak0}) is complicated, but we can simplify it by writing
\begin{equation}
\begin{split}
\label{eq:CoulombSplitak}
S^{\LL}_{\La k}(\{p,f\}_{m};t) 
={}& 
S^{\LL}_{\La k}(\{p,f\}_{m};t;{\rm dipole})  +
S^{\LL}_{\La\La}(\{p,f\}_{m};t;{\rm eikonal})
\;,
\\S^{\LL}_{k\La}(\{p,f\}_{m};t) 
={}& 
S^{\LL}_{k\La}(\{p,f\}_{m};t;{\rm dipole})  +
S^{\LL}_{kk}(\{p,f\}_{m};t;{\rm eikonal})
\;.
\end{split}
\end{equation}
Here
\begin{equation}
\begin{split}
\label{eq:Gaaeik0}
\int\!dt\ S^{\LL}_{\La\La}(\{p,f\}_{m};&t;{\rm eikonal}) 
\\
={}&
\mi\, \frac{\as}{(2\pi)^{3}}\,
\int\! d\vec q \int\! dE\
\frac{p_\La\cdot D(q)\cdot p_\La}
{(q\cdot p_\La - \mi\epsilon)^2 (q^2 + \mi\epsilon)}\,
\theta(|\vec q\,| < M)
\;,
\end{split}
\end{equation}
with an analogous definition of $S^{\LL}_{kk}(\{p,f\}_{m};t;{\rm eikonal})$.

This defines $S^{\LL}_{\La k}(\{p,f\}_{m};t;{\rm dipole})$ and $S^{\LL}_{k\La}(\{p,f\}_{m};t;{\rm dipole})$. After some algebra, we find
\begin{equation}
\begin{split}
\label{eq:Gak2}
\int\!dt\  \big[
S^{\LL}_{\La k}&(\{p,f\}_{m};t;{\rm dipole}) 
+ S^{\LL}_{k\La}(\{p,f\}_{m};t;{\rm dipole})
\big]
\\
={}&
-\mi\, \frac{\as}{(2\pi)^{3}}\,
\int\! d\vec q \int\! dE \
\frac{2\,p_\La\cdot p_k}
{(q\cdot p_\La - \mi\epsilon)(q\cdot p_k + \mi \epsilon)(q^2 + \mi\epsilon)}\,
\theta(|\vec q\,| < M)
\;.
\end{split}
\end{equation}
As with eq.~(\ref{eq:Glk3}) we write this using $E$ and $\vec q$: 
\begin{equation}
\begin{split}
\label{eq:Gak3}
\int\!dt\  \bigg[
S^{\LL}_{\La k}(\{p,f\}_{m};&t;{\rm dipole}) 
+ S^{\LL}_{k \La}(\{p,f\}_{m};t;{\rm dipole})
\bigg]
\\
={}&
-\mi\, \frac{\as}{(2\pi)^{3}}\,
\int\! d\vec q \int\! dE \
\frac{2\,v_\La\cdot v_k}
{(E - \vec q\cdot \vec v_\La - \mi\epsilon)(E - \vec q\cdot \vec v_k - \mi \epsilon)}
\\ & \times
\frac{1}{2 |\vec q|}
\left[
\frac{1}{E - |\vec q| + \mi \epsilon} - \frac{1}{E + |\vec q| - \mi \epsilon}
\right]
\theta(|\vec q\,| < M)
\;.
\end{split}
\end{equation}

We can immediately perform the energy integration, noting that the term $1/(E + |\vec q| - \mi \epsilon)$ does not contribute:
\begin{equation}
\begin{split}
\label{eq:Gak4}
\int\!dt\  \bigg[
S^{\LL}_{\La k}(\{p,f\}_{m};&t;{\rm dipole}) 
+ S^{\LL}_{k \La}(\{p,f\}_{m};t;{\rm dipole})
\bigg]
\\={}&
-\frac{\as}{(2\pi)^{2}}\,
\int\!\frac{d\vec q}{2 |\vec q|}\
\frac{2\,v_\La\cdot v_k}
{(|\vec q| - \vec q\cdot \vec v_\La)
(|\vec q| - \vec q\cdot \vec v_k)}\,
\theta(|\vec q\,| < M)
\;.
\end{split}
\end{equation}
Notice how this compares to the equivalent result for virtual gluon exchange between two  final state partons, as in eq.~(\ref{eq:Glk5}), or two initial state partons. Here there is no imaginary part.

The integrand has poles at $|\vec q| = \vec q\cdot \vec v_\La$ and at $|\vec q| = \vec q\cdot \vec v_k$. This reflects splittings both of parton ``a'' and of parton $k$. In order to separate these splittings, we multiply the integrand by $1 = A'_{\La k} + A'_{k\La}$, where
\begin{equation}
\label{eq:Aprimeak}
A'_{\La k} = \frac{|\vec q| - \vec q\cdot \vec v_k}
{(|\vec q| - \vec q\cdot \vec v_k) + (|\vec q| - \vec q\cdot \vec v_\La)}
\;.
\end{equation}
The term containing $A'_{\La k}$ is associated with $G^{\LL}_{\La k}(\{p,f\}_{m};t)$, while the term containing $A'_{k\La}$ is associated with $G^{\LL}_{k\La}(\{p,f\}_{m};t)$. Thus we define
\begin{equation}
\begin{split}
\label{eq:Gak5}
\int\!dt\ 
S^{\LL}_{\La k}&(\{p,f\}_{m};t;{\rm dipole}) 
\\={}&
-\frac{\as}{(2\pi)^{2}}\,
\int\!\frac{d\vec q}{2 |\vec q|}\
\frac{2\,v_\La\cdot v_k}
{(|\vec q| - \vec q\cdot \vec v_\La)
[(|\vec q| - \vec q\cdot \vec v_k) + (|\vec q| - \vec q\cdot \vec v_\La)]}\;
\theta(|\vec q\,| < M)
\;.
\end{split}
\end{equation}

We can now identify the shower time with
\begin{equation}
\begin{split}
\label{eq:ydefInitialFinal}
y = \frac{1}{E_Q}\
(|\vec q| - \vec q\cdot \vec v_\La)
\end{split}
\end{equation}
as we did for a final state splitting. We introduce this as a delta function, recognizing that $\int\!dt \cdots$ is equivalent to $\int\!d\log(y) \cdots$. Thus
\begin{equation}
\begin{split}
\label{eq:Gak6}
S^{\LL}_{\La k}(\{p,f\}_{m};&t;{\rm dipole}) 
\\
={}&
-\frac{\as}{(2\pi)^{2}}\,2v_\La\cdot v_k\,
\int\!  \frac{d\vec q}{2|\vec q|}\,
\theta(|\vec q\,| < M)\,
\delta\!\left[\log(E_Q y) - \log\left(
|\vec q| - \vec q\cdot \vec v_\La\right)
\right]
\\&\times
\frac{1}
{E_Q y\,
(|\vec q| - \vec q\cdot \vec v_k + E_Q y)}\;
\;.
\end{split}
\end{equation}

This is precisely the real part of the integral in eq.~(\ref{eq:Glk9}) with $l \to \La$. To calculate $S^{\LL}_{\La k}(\{p,f\}_{m};t)$ using eq.~(\ref{eq:CoulombSplitak}), we also need $S^{\LL}_{\La\La}(\{p,f\}_{m};t;{\rm eikonal})$, which is the same as $S^{\LL}_{ll}(\{p,f\}_{m};t;{\rm eikonal})$ with $l \to \La$. Thus $S^{\LL}_{\La k}(\{p,f\}_{m};t)$ is simply the real part of $S^{\LL}_{l k}(\{p,f\}_{m};t)$, eq.~(\ref{eq:Glktotal}), with $l \to \La$: 
\begin{equation}
\begin{split}
\label{eq:Gaktotal}
S^{\LL}_{\La k}(\{p,f\}_{m};t)
\approx{}&
\frac{\as}{2\pi}\,\left[
-\log\!\left(
\frac{1 - \cos\theta_{\La k}}{2}
\right)
- 1
\right]
\;.
\end{split}
\end{equation}
We will use this result in eq.~(\ref{eq:calGa2}). For $S^{\LL}_{k \La}(\{p,f\}_{m};t)$, essentially the same calculation gives the same result,
\begin{equation}
\begin{split}
\label{eq:Gkatotal}
S^{\LL}_{k \La }(\{p,f\}_{m};t)
\approx{}&
\frac{\as}{2\pi}\,\left[
-\log\!\left(
\frac{1 - \cos\theta_{\La k}}{2}
\right)
- 1
\right]
\;.
\end{split}
\end{equation}
We will use this result in eq.~(\ref{eq:calGl2}).

\subsection{Self-energy diagrams}
\label{sec:VirtualSelfEnergy}

In this section, we look at self-energy graphs. As with the interference graphs, we use Coulomb gauge in the rest frame of the total momentum $Q$ of the final state partons. Consider a gluon that enters the final state or comes from the initial state. The gluon has momentum $p_l$, with $l \in \{1,2,\dots,m\}$ for a final state gluon and with $l \in \{\La,\Lb\}$ for the initial state case. We combine the self-energy subgraph $-\mi\Pi(q)^{\alpha\beta}$ with the adjoining virtual propagator and the adjoining cut propagator. Such a graph really represents a field strength renormalization for the gluon field. We interpret the graph together with the propagators as
\begin{equation}
\label{eq:Gllintegralgluon}
\int\!dt\ S^{\LL}_{l l}(\{p,f\}_{m};t;{\rm gluon})\, D(p_l)^{\mu\nu} = 
-\frac{1}{2}\left[\frac{1}{p_l^2}D(p_l)^\mu_\alpha\,
\Pi(p_l)^{\alpha\beta}\,
D^\nu_\beta(p_l)\right]_{p_l^2 = 0}
\;.
\end{equation}
The minus sign is from eq.~(\ref{eq:evolutionbis}), so that we write the negative of the usual Feynman diagram. We will also consider a quark that enters the final state or comes from the initial state. The analogous definition is
\begin{equation}
\label{eq:Gllintegralquark}
\int\!dt\ S^{\LL}_{l l}(\{p,f\}_{m};t;{\rm quark})\, \s{p}_l = 
-\frac{1}{2}\left[\frac{\s{p}_l}{p_l^2}
\Sigma(p_l)\,
\s{p}_l\right]_{p_l^2 = 0}
\;.
\end{equation}

We can use the results of ref.~\cite{KSCoulomb}. For the gluon case, these results include the contributions from a gluon loop, a quark loop, and a ghost loop. For the quark case, we have a quark-gluon loop.

In the loop integral, we integrate over the energy going around the loop, giving a result in time-ordered perturbation theory, with on-shell partons and energy denominators. We write the result using the three-momenta $\vec k_\pm$ of the two partons in the loop, with $\vec k_+ + \vec k_- = \vec p_l$. Inside this integral, we need to identify the shower time or, equivalently, the dimensionless virtuality variable $y$. The definition (\ref{eq:ydef}) for a real splitting is
\begin{equation}
\begin{split}
\label{eq:ydefbis}
y ={}& \frac{(k_+ + k_-)^2}{2 p_l\cdot Q}
\;,
\end{split}
\end{equation}
where $k_\pm$ are the (on-shell) parton momenta after the splitting. The virtuality $(k_+ + k_-)^2$ is $(|\vec k_+| + |\vec k_-|)^2 - (\vec k_+ + \vec k_-)^{2}$. This is normalized by dividing by $2 p_l\cdot Q = 2 p_l^0 E_Q$. In a real emission, a small amount of momentum is taken from elsewhere in the event to put $p_l$ on shell. For the virtual graph, we simply replace $p_l^0 \to |\vec p_l| = |\vec k_+ + \vec k_-|$. Thus we identify
\begin{equation}
\begin{split}
\label{eq:ydefSelfEnergy}
y ={}& \frac{(|\vec k_+| + |\vec k_-|)^2 - \vec p_l^{\,2}}{2|\vec p_l|E_Q}
\;.
\end{split}
\end{equation}

In the result from ref.~\cite{KSCoulomb}, the integral is written as an integration over $y$, the azimuthal angle $\phi$ of $\vec k_+$ around the direction of $\vec p_l$, and and a variable\footnote{In the quark case, $k_-$ is the momentum of the quark line inside the loop, so $x$ is the momentum fraction of the gluon. However, we present the results symmetrized over $x \leftrightarrow (1-x)$, so that it does not matter whether $x$ or $(1-x)$ is identified with the gluon in the loop. Ref.~\cite{KSCoulomb} explains how to remove the symmetrization, but we do not need to do that here.}
\begin{equation}
\begin{split}
x ={}& \frac{|\vec k_+| - |\vec k_-| + |\vec p_l|}{2|\vec p_l|}
\;.
\end{split}
\end{equation}
The variable $\bar q^2 = 2|\vec p_l|E_Q\,y$ appears instead of $y$ in ref.~\cite{KSCoulomb}. We include here the factor 1/2 in eqs.~(\ref{eq:Gllintegralgluon}) and (\ref{eq:Gllintegralquark}). Ref.~\cite{KSCoulomb} gives directly the right hand side of these equations without the factor 1/2. Also, we maintain the notation of eq.~(\ref{eq:Slkcolordef}) in which $S^{\LL}_{l l}$ multiplies a factor $[(\bm T_l\cdot \bm T_l)\otimes 1]$. For the gluon case, this is a factor $C_\LA [1\otimes 1]$ and for the quark case, it is a factor  $C_\LF [1\otimes 1]$. For this reason, we remove a factor $C_\LA$ or $C_\LF$ from the results as given in ref.~\cite{KSCoulomb}. As in previous sections, we will use the convenient abbreviation $a_l = E_Q/(2|\vec p_l|)$ from eq.~(\ref{eq:aldef}).

The variable $y$ is proportional to $e^{-t}$, where $t$ is the shower time. Thus integrating over $t$ is the same as integrating over $y$, with $dt = d\log y$. We are interested in $S^{\LL}_{l l}(\{p,f\}_{m};t)$, the integrand of the $\log y$ integration. Integrations over $x$ and $\phi$ will remain. 

We now turn to the gluon and quark cases separately.

\subsubsection{Gluon self-energy}%%%%%%%%%%%%%%%%

For the gluon self energy, we find from ref.~\cite{KSCoulomb}
\begin{equation}
S^{\LL}_{l l}(\{p,f\}_{m};t;{\rm gluon}) = 
\int_0^1\!dx \int_{-\pi}^\pi\!\frac{d\phi}{2\pi}\
\frac{\as}{8\pi}\,\frac{1}{C_\LA}
\sum_{J=0}^2
\frac{A_{\LT,J}}{[4 a_l y + 4 x (1-x)]^J}
\;.
\end{equation}
The coefficients $A_{\LT,J}$ are
\begin{equation}
\begin{split}
A_{\LT,0} ={}& 
- (2 C_\LA - N_\Lf)\,\frac{e^2 \mu_\LR^2}{2 |\vec p_l| E_Q y + e^2 \mu_\LR^2}
+ 2 C_\LA\,x(1-x) \,\frac{e^{5/3} \mu_\LR^2}
  {2 |\vec p_l| E_Q y + e^{5/3} \mu_\LR^2}
\\&
+ (4 C_\LA - 2 N_\Lf)\,x(1-x) \,\frac{e^{8/3} \mu_\LR^2}
  {2 |\vec p_l| E_Q y + e^{8/3} \mu_\LR^2}
\;,
\\
A_{\LT,1} ={}& 2 C_\LA \left[
12 x(1-x) - 24 x^2 (1-x)^2
\right]
\;,
\\
A_{\LT,2} ={}& 16 C_\LA x(1-x) \left[
2 - 8 x(1-x) + 8 x^2 (1-x)^2
\right]
\;.
\end{split}
\end{equation}
Here we have renormalized the graphs according to the $\MSbar$ prescription, but have subtracted a numerical function that gives the same result as subtracting a pole. The parameter $\mu_\LR^2$ is the $\MSbar$ renormalization scale.

We can perform the integrations. The $\phi$-integral is trivial. Performing the $x$-integral gives
\begin{equation}
S^{\LL}_{l l}(\{p,f\}_{m};t;0;{\rm gluon}) = 
\frac{\as}{8\pi}\,\frac{1}{C_\LA}
\sum_{J=0}^2 I^{\rm g}_J
\;,
\end{equation}
where
\begin{equation}
\begin{split}
I^{\rm g}_0 ={}& 
- (2 C_\LA - N_\Lf)\,\frac{e^2 \mu_\LR^2}{2 |\vec p_l| E_Q y + e^2 \mu_\LR^2}
+ \frac{1}{3} C_\LA\,\frac{e^{5/3} \mu_\LR^2}
  {2 |\vec p_l| E_Q y + e^{5/3} \mu_\LR^2}
\\&
+ \frac{1}{3}(2 C_\LA -  N_\Lf)\,\frac{e^{8/3} \mu_\LR^2}
  {2 |\vec p_l| E_Q y + e^{8/3} \mu_\LR^2}
  \;,
\\
I^{\rm g}_1 ={}& 2 C_\LA \left[
2 + 6 a_l y  
-\frac{6 a_l y (1+ 2 a_l y)}{\sqrt{1 + 4 a_l y}}\,
\log\!\left(\frac{(1 + \sqrt{1 + 4 a_l y}\,)^2}{4 a_l y}\right)
\right]
\;,
\\
I^{\rm g}_2 ={}& 4 C_\LA \bigg[
- \frac{4(1 + 3 a_l y)(2 + 5 a_l y)}{3 ( 1 + 4 a_l y)} 
\\&\quad
+\frac{(1 + 2 a_l y) (1 + 8 a_l y + 20 a_l^2 y^2)}{(1 + 4 a_l y)^{3/2}}\,
\log\!\left(\frac{(1 + \sqrt{1 + 4 a_l y})^2}{4 a_l y}\right)
\bigg]
\;.
\end{split}
\end{equation}

We are interested in the small $y$ limit of this. There is a constant term and a term proportional to $\log(y)$:
\begin{equation}
\label{eq:Gllgluon}
S^{\LL}_{l l}(\{p,f\}_{m};t;{\rm gluon}) = 
\frac{\as}{2\pi}\,\frac{1}{C_\LA}
\left(
-\frac{23 C_\LA}{12} + \frac{N_\Lf}{6}
-  C_\LA \log\left(\frac{E_Q y}{2 |\vec p_l|} \right)
\right)
\;.
\end{equation}

\subsubsection{Quark self-energy}%%%%%%%%%%%%%%%%

For the quark self energy, we find from ref.~\cite{KSCoulomb}
\begin{equation}
S^{\LL}_{l l}(\{p,f\}_{m};t;{\rm quark}) = 
\int_0^1\!dx \int_{-\pi}^\pi\!\frac{d\phi}{2\pi}\
\frac{\as}{8\pi}\,\frac{1}{C_\LF}
\sum_{J=0}^2
\frac{B_{\LL,J}}{[4 a_l y + 4 x (1-x)]^J}
\;.
\end{equation}
The coefficients $B_{\LT,J}$ are
\begin{equation}
\begin{split}
B_{\LL,0} ={}& 
C_\LF\left[
-  \frac{e^3 \mu_\LR^2}{2 |\vec p_l| E_Q y + e^3 \mu_\LR^2}
+ 12  x(1-x) \,\frac{e^{5/3} \mu_\LR^2}
  {2 |\vec p_l| E_Q y + e^{5/3} \mu_\LR^2}
  \right]
\;,
\\
B_{\LL,1} ={}& 2 C_\LF \left[
20 x(1-x) - 56 x^2 (1-x)^2
\right]
\;,
\\
B_{\LL,2} ={}& 32 C_\LF x(1-x) \left[
1 - 6 x(1-x) + 8 x^2 (1-x)^2
\right]
\;.
\end{split}
\end{equation}
Here again $\mu_\LR^2$ is the $\MSbar$ renormalization scale.

We can perform the integrations. The $\phi$-integral is trivial. Performing the $x$-integral gives
\begin{equation}
S^{\LL}_{l l}(\{p,f\}_{m};t;{\rm quark}) = 
\frac{\as}{8\pi}\,\frac{1}{C_\LF}
\sum_{J=0}^2 I^{\rm q}_J
\;,
\end{equation}
where
\begin{equation}
\begin{split}
I^{\rm q}_0 ={}& 
C_\LF\left[
-\frac{e^3 \mu_\LR^2}{2 |\vec p_l| E_Q y + e^3 \mu_\LR^2}
+ \frac{2 e^{5/3} \mu_\LR^2}{2 |\vec p_l| E_Q y + e^{5/3} \mu_\LR^2}
\right]
\;,
\\
I^{\rm q}_1 ={}& C_\LF \left[
\frac{16}{3} + 28 a_l y  
-\frac{4 a_l y (5 + 14 a_l y)}{\sqrt{1 + 4 a_l y}}\,
\log\!\left(\frac{(1 + \sqrt{1 + 4 a_l y})^2}{4 a_l y}\right)
\right]
\;,
\\
I^{\rm q}_2 ={}& C_\LF \bigg[
- \frac{40}{3} (1 + 3 a_l y) 
+\frac{4(1 + 10 a_l y + 20 a_l^2 y^2)}{\sqrt{1 + 4 a_l y}}\,
\log\left(\frac{(1 + \sqrt{1 + 4 a_l y})^2}{4 a_l y}\right)
\bigg]
\;.
\end{split}
\end{equation}

We are interested in the small $y$ limit of this. There is a constant term and a term proportional to $\log(y)$:
\begin{equation}
\label{eq:Gllquark}
S^{\LL}_{l l}(\{p,f\}_{m};t;{\rm quark}) = 
\frac{\as}{2\pi}\,
\left(
-\frac{7}{4}
-  \log\left(\frac{E_Q y}{2|\vec p_l|} \right)
\right)
\;.
\end{equation}

\subsubsection{Self-energy in general}%%%%%%%%%%%%%%%%

We can combine the results (\ref{eq:Gllgluon}) and (\ref{eq:Gllquark}) using the notation from eqs.~(\ref{eq:Cf}) and (\ref{eq:gammaf}):
\begin{equation}
\label{eq:Gllnet}
S^{\LL}_{l l}(\{p,f\}_{m};t) = 
-\frac{\as}{2\pi}\,
\left(
\frac{\gamma_{f_l}}{2 C_{f_l}}
+\log\left(\frac{E_Q y}{2|\vec p_l|} \right)
+ 1
\right)
\;.
\end{equation}
We will use this result in eqs.~(\ref{eq:calGl2}) and (\ref{eq:calGa2}).

%-------------------------------------------------------------------

%-------------------------------------------------------------------
\end{document}